\documentclass[fleqn,usenatbib]{mnras}

\usepackage{newtxtext,newtxmath}
\usepackage[T1]{fontenc}

\DeclareRobustCommand{\VAN}[3]{#2}
\let\VANthebibliography\thebibliography
\def\thebibliography{\DeclareRobustCommand{\VAN}[3]{##3}\VANthebibliography}

\usepackage{graphicx}	% Including figure files
\usepackage{amsmath}	% Advanced maths commands
\usepackage{array}
\usepackage{siunitx}
\usepackage{anyfontsize}
\usepackage{subcaption} %Subfigures
\usepackage{makecell} % 
\usepackage[dvipsnames]{xcolor}
\usepackage{braket}

\title[The Catalogue of virtual-ETGs from IllustrisTNG]{The Catalogue of Virtual Early-Type Galaxies from IllustrisTNG: 
Validation and Real Observation Consistency}

\author[de Araujo Ferreira  et al.]{
Pedro de Araujo Ferreira$^{1}$\thanks{E-mail: pedroitalo96@academico.ufs.br},
Nicola R. Napolitano$^{2,3,4}$,
Luciano Casarini$^{1}$,
Crescenzo Tortora$^{3}$,
\newauthor~Rodrigo von Marttens$^{5}$
and Sirui Wu$^{6}$
\\
$^{1}$Physics Department, Federal University of Sergipe, 49100-000, São Cristovão-SE, Brazil\\
$^{2}${Department of Physics E. Pancini, University Federico II, Via Cinthia 21, I-80126, Naples, Italy}\\
$^{3}${INAF – Osservatorio Astronomico di Capodimonte, Salita Moiariello 16, I-80131 Napoli, Italy}\\
$^{4}${INFN, Sez. di Napoli, via Cintia, 80126, Napoli, Italy}\\
$^{5}$Physics Institute, Federal University of Bahia, 40210-340, Salvador-BA, Brazil\\
$^{6}$Dark, Niels Bohr Institute, Univeristy of Copenhagen, Jagtvej 155, 2200 Copenhagen, Denmark}

\date{Accepted 2025 April 15. Received 2025 April 15; in original form 2025 January 8}

\pubyear{2024}

\begin{document}
\label{firstpage}
\pagerange{\pageref{firstpage}--\pageref{lastpage}}
\maketitle

\begin{abstract}
Early-type galaxies (ETGs) are reference systems to understand galaxy formation and evolution processes. The physics of their %collapse
{formation}
and internal dynamics are codified in well-known scaling relations. %Cosmological hydrodynamical simulations play an important role, providing insights into the 3D distribution of matter and galaxy formation mechanisms, as well as validating methods to infer the properties of real objects. 
{In this context, cosmological hydrodynamical simulations play an important role in probing the physical origins of scaling relations by providing a controlled environment to study the formation and evolution of galaxies, linking their internal dynamics to underlying physical processes, and testing the robustness of observational inference methods}. In this work, we present the closest-to-reality sample of ETGs from the IllustrisTNG100-1 simulation, dubbed "virtual-ETGs", based on an observational-like algorithm that combines standard projected and three-dimensional galaxy structural parameters. We extract 2D photometric information by projecting the galaxies' light into three planes and modelling them via Sérsic profiles. Aperture velocity dispersions, corrected for softened central dynamics, are calculated along the line-of-sight orthogonal to the photometric projection plane. Central mass density profiles assume a power-law model, while 3D masses remain unmodified from the IllustrisTNG catalogue. The final catalogue includes $10121$ galaxies at redshifts $z \leq 0.1$. By comparing the virtual properties with observations, we find that the virtual-ETG scaling relations (e.g., size-mass, size-central surface brightness, and Faber-Jackson), central density slopes, and scaling relations among total density slopes and galaxy structural parameters are generally consistent with observations. We make the virtual-ETG publicly available for galaxy formation studies and plan to use this sample as a training set for machine learning tools to infer galaxy properties in future imaging and spectroscopic surveys.
\end{abstract}

\begin{keywords}
galaxies: photometry -- galaxies: structure -- galaxies: elliptical and lenticular, cD
\end{keywords}

%%%%%%%%%%%%%%%%%%%%%%%%%%%%%%%%%%%%%%%%%%%%%%%%%%

%%%%%%%%%%%%%%%%% BODY OF PAPER %%%%%%%%%%%%%%%%%%

\section{Introduction}
\label{sec:intro}
The prevailing  {cosmological} model,
%of the universe, 
%the $\Lambda$CDM, 
 {based on a cosmological constant plus a Cold Dark Matter, 
the $\Lambda$CDM, predicts}
%proposes 
a hierarchical  {growth of the large scale} structure  {of the universe}, 
%formation scheme, 
where galaxy formation is driven from the assembly of dark halos composed of collisionless dark matter. % {within which gas cools to form stars}. 
  {Inside these dark matter haloes, the baryonic matter clumps and  {triggers the}
%, subsequently, 
star formation via effective gas cooling} 
%Via effective gas cooling, the baryonic matter inside the dark matter halos clumps and  {triggers the}
%%, subsequently, 
%star formation 
%%is triggered 
\citep{White+1978, White+1991, Cole+1994, Blumenthal+1984}.  {The complex} interplay 
%of 
 {between} baryonic and dark matter {, commonly  referred to as the galaxy-halo connection \citep{Wechsler},} shapes the evolution history and structure of galaxies. 
%In contrast, 
 {Observationally,} while the baryonic component of galaxies can be studied from its luminous properties, the dark component can only be traced by its gravitational effects on the baryonic matter, for instance, via the study of the motions (i.e., the kinematics) of stars and gas.
%, and this relation is commonly  referred to as the galaxy-halo connection \citep{Wechsler}. 
%In particular, 
%A common tracer of dark matter is its effect on the kinematics of the baryonic component. 
%This connection can be explored for 
 {The photometry and kinematics of galaxies have been deeply investigated for} a wide variety of galaxies 
%thanks to the advent of 
by numerous observational programs  {either targeting 
%surveys dedicated to collect good quality photometric and spectroscopic data  
galaxy centers}, such as 
SAURON \citep{Zeew+2002},
SPIDER \citep{LaBarbera+2010}, 
ATLAS$^{\text{3D}}$ \citep{Atlas3D}, 
SAMI \citep{Bryant+2015},
%Frontier Fields\citep{FrontierFields}, 
and MaNGA \citep{Zhu}, or galaxy outskirts, e.g., using globular clusters 
%\citep{use some SLUGGS paper} 
\citep{Usher+2019} and planetary nebulae 
%\citep{use some PN.S paper}. 
\citep{Douglas+2007, Arnaboldi+2016, Pulsoni+2020}. 
%The data product of these samples plus theoretical considerations incorporated into dynamical models allows the investigation of the matter content and distribution within galaxies. One typical approach is, for instance, the Jeans Anisotropic Modelling (JAM) (see, e.g., \citealt{Cappellari+2013, cappellari+2016}) which is used to investigate the mass distribution of galaxies by combining two-dimensional kinematic information with dynamical modelling.
In particular, early-type galaxies (ETGs), being the %the gas-poor products of the 
final stage of galaxy evolution process, are crucial objects to understand
%on the study of structure 
the formation processes (see, e.g., \citealt{Costantin2021ApJ...913..125C}) and the dark matter assembly (\citealt{Tortora_2018MNRAS}) across cosmic time. The internal dynamics of ETGs, including their stellar kinematics and velocity dispersion profiles, offer important insights about their formation history, assembly processes and chemical evolution \citep{Coccato+2009, cappellari+2016, Lapi+2018, Cannarozzo+2023}.
%and probe of the high-redshift universe. 
Indeed, the manner ETGs form and evolve shapes their internal structure as well as their central dark matter fraction and central density profiles
%, which have been a source of research in a variety of works  
\citep{Tortora+2014,D.Xu,Sharma+2021,Tortora+2022, C.Wang}. In particular, mergers play a fundamental role in ETGs' structure evolution. These interactions can alter the distribution of baryonic matter, size evolution and redistribute dark matter within the merging systems \citep{Naab, Oogi, López-Sanjuan}. The cumulative effects of several mergers over time can result in the formation of the diverse population of ETGs observed in the local universe. 

If, on one hand, photometry and spectroscopy serve as the cornerstone to the development of methods and models able to unveil the physics and structure of galaxies,
%, such as their geometry, dark and baryonic content distribution, properties of the environment where they are embedded, etc. However, once 
on the other hand, they provide a limited view of these systems, as we can only observe a galaxy through its projection on the sky, leaving their  three-dimensional shape and matter distribution 
%can only be inferred 
accessible only via assumptions on their equilibrium and internal geometry.
%, e.g., assume that the galaxies are in a steady-state and geometrical considerations, as that objects have spherical, triaxial shapes or are axis-symmetric. 
Here, it is where hydrodynamical simulations (e.g., \citealt{Schaye+2014,  Vogelsberger+2014,Crain+2015, Nelson_et_al_2019, Wetzel+2023}) can play a crucial role, 
%State-of-the-art simulations 
offering a powerful framework to explore galaxy formation in a cosmological context and validating the methods typically used to infer the internal structure of galaxies, including their dark matter content.
%and combined with theoretical studies, these tools can help to further refine the models and recipes of galaxy formation and evolution. 
%For instance, given the simulated data, one can derive the three-dimensional distribution (baryonic and/or dark) of the objects. Thus, based on the 3D simulated quantities, that represent the \textit{ground truth} (i.e., the exact results expected by theoretical predictions), models can be validated and possibly refined to better describe 
%the properties of 
%real world samples. 
%E.g.,
{One such example is the validation of the Jeans Anisotropic Modelling (JAM for short, e.g., \citealt{Cappellari+2013, cappellari+2016}), which is one typical manner to investigate the mass distribution of galaxies by combining two-dimensional kinematic information with dynamical modelling. By applying the Multi-Gaussian Expansion (MGE) formalism \citep{Emsellem+1994} to the projected light distribution of galaxies from the Illustris simulations \citep{Vogelsberger+2014}, it was demonstrated by \citet{Li+2015} that the JAM approach is a high accuracy method, being able to reproduce the distribution of mass in galaxies with negligible bias. }
%JAM for short, \citealt{Cappellari+2013, cappellari+2016})
%and find these are 
%able to  reproduce the distribution of galaxy masses with negligible bias for oblate galaxies, while the JAM-recovered stellar masses are, on average, $18\%$ higher for prolate systems \citep{Li+2015}.

%The Jeans Anisotropic Modelling (JAM) (see, e.g., \citealt{Cappellari+2013, cappellari+2016}) is one typical manner to investigate the mass distribution of galaxies by combining two-dimensional kinematic information with dynamical modelling. For instance, by applying the Multi-Gaussian Expansion (MGE) formalism \citep{Emsellem+1994} to the projected light distribution of galaxies from the Illustris simulations \citep{Vogelsberger+2014}, it has been demonstrated by \citet{Li+2015} that the JAM approach is a high accuracy method, being able to reproduce the distribution of mass in galaxies with negligible bias.  
{Unfortunately,
%, as pointed by \citet{vonMarttens+2022}, 
techniques involving dynamical modelling, such as JAM, often need rather long computational times. 
%large computational resources for efficient results. 
A consequence is that such approaches are more feasible for samples of thousands of galaxies (see, e.g., \citealt{Tortora+2014, Tortora+2022, Zhu}) while 
%thousand-sized  have recently been preferably used on relatively small datasets, as in \citet{Cappellari+2013}, for example.
the upcoming large galaxy surveys will observe millions to billions of galaxies. }In this context,
%requires the development of techniques able to deal with large amounts of data. In particular, 
machine Learning (ML) models present a promising avenue for addressing this challenge.
%, once they are suitable for dealing with a large amount of data and with superior computational speed when compared to regular approaches. 
One example of its use is as an alternative approach to inference of the total and dark matter content in galaxies (e.g., \citealt{vonMarttens+2022,wu2023total,Chu_machine_learning_IFU_2024}) or the emulation of dynamical models such as JAM with neural networks \citep{Gomer+2023}. In particular, in \citet{wu2023total}, we have provided a first application of these techniques to real samples and shown that machine learning can be as accurate as JAM or similar Jeans analyses, but providing total and dark masses of thousands of galaxies in a timescale of a few seconds.

In this context, though, the use of simulation datasets as training sample needs extreme caution (see again \citealt{wu2023total} for a discussion). Indeed,  %limitation of using 
the parameters that are the direct product of simulations %is that their definitions 
are, in general, not compatible with their observational counterparts as the former are computed from the three-dimensional distribution of particles, while the latter rely on measurements of projected quantities.
%However, a number of these methods are still maturing, given that a limitation in some simulations and works using their product is the reliance on internal quantities, which in general are not directly comparable to observations that, on the other hand, rely on projected properties, such as those obtained from photometric surveys.
 { Thus, the 
conduction of 
analysis or 
%the construction of 
models based on parameters that are the direct product of the simulations can 
%make the analysis 
result in being} either too much idealized or even biased. 
%in the sense that these 
 {For instance,} galaxy parameters will not carry the same scatter and/or systematic errors as their observational equivalents. Hence, appropriate methods for the extraction of galaxy structural parameters have to be explored in order to make the determination of the physical parameters of simulated galaxies 
%less idealized. 
more realistic.
Possible approaches to incorporate this ``observational realism'' into the simulated data include: 1) approximating Petrosian apertures \citep{Strauss_2002} for simulated galaxies by considering only the projected light and matter within $30$ kpc from the galactic centre (e.g., \citealt{Pillepich+2017, D.Xu, Y.Wang, Lu+2020}) 2) using a surface brightness (SB) limit, compatible with the typical depth of imaging surveys used to measure structural parameters of large galaxy samples with automatic tools (see, e.g., \citealt{Tang+2021}). In particular, \citet{Pillepich+2017} have shown that the former method consents to the construction of a galaxy stellar mass function in agreement with observations. However, \citet{Tang+2021} have shown that the 30 kpc cut is a too rough approximation to reproduce the size-mass relation, showing that the SB cut is a more appropriate procedure to realistically reproduce both stellar mass function and the size-mass relation. 
Dust attenuation effects to the measured light \citep{D.Xu,Nelson+2018} is another piece of observational realism to add to the mock data. As shown by \citet{Nelson+2018} this can indeed produce an improved agreement between the observed galaxy colour bimodality and the simulated result. More recently, \citet{Tang+2021} used light distribution information rather than mass to approach the mock galaxies  and obtained good agreement between the observed and simulated galaxy luminosity function. Therefore, by leveraging control over the 2D and the fiducial 3D structural properties of the objects, the state-of-the-art simulations can help us address the efficacy and limitations of existing methodologies and possibly to develop new approaches. 

In this work, we introduce a catalogue composed of   {$10121$ low-redshift ETGs ($z \leq 0.1$)} for which we have re-analysed 
the Snapshots' data\footnote{https://www.tng-project.org/data/docs/specifications/\#sec1}
%data product of the 
from the high resolution TNG100-1 simulation by the IllustrisTNG project.
%(see \S\ref{the_simulation} for details). 
 {The choice of TNG100-1 has been motivated by the best compromise between statistics, as in a $(100 \rm Mpc)^3$ we expect to collect a catalogue of the order of 10k galaxies with stellar masses larger than $10^{10.5} \rm M_{\sun}$, and resolution, as the softening length of the TNG100 for stars and galaxies is sub-kpc (see \S\ref{the_simulation} for details). 
} {The aim of our new catalogue is to serve as a resource for a wide variety of applications, by providing a sample of ETGs that closely matches those found in large scale surveys in terms of the methodology used to derive their ``observed'' properties. The implemented {\it observational realism} consists of 1) the conversion of stellar mass particles into ``light'' particles using stellar population models for all the Snapshot data used (TNG provides these data only for a few redshifts); 2) the 
%One of its notable features is the 
derivation of typical projected ``on sky'' photometric and ``along the line-of-sight'' spectroscopic quantities; 3) the use of standard surface brightness fitting techniques to derive the structural parameters of galaxies, paying maximum attention to minimize the impact of the regions affected by the central softening length, and 4) the derivation, for the first time, of a physically motivated correction to the central velocity dispersion that allows the full recovery of the Faber-Jackson relation and the Fundamental Plane (de Araujo Ferreira et al. in preparation). This is the closest-to-reality TNG galaxy sample ever made, eventually enabling a fully unbiased comparisons with observational data. In the following, we will refer to this as the {\it virtual-ETG} sample.} 
Three-dimensional quantities are also included, as masses and density profiles, allowing for comparison between the 2D quantities that mocks observational procedures  and the fiducial 3D quantities, that, apart from the density profiles, are standard products of the simulation. A similar approach was adopted by  {\citet{deGraaff_eagle} for mock images of $z = 0.1$ galaxies in the EAGLE simulations \citep{Schaye+2014}} and \citet{D.Xu} for synthetic galaxies from the Illustris-1 simulation \citep{Vogelsberger+2014}. Here, we follow most of this latter paper methodology, with some relevant refinements in the photometric analysis and fundamental adds in the kinematics (see \S\ref{pararaph:projected_quantities}) and resolution (for comparison, the resolution in \citealt{deGraaff_eagle} is $2$ kpc).
%We will use the IllustrisTNG simulation with the second-highest resolution, namely TNG100-1 (see \S\ref{the_simulation}). 
% {The choice of TNG100-1 has been motivated by the best compromise between statistics, as in a $(100 \rm Mpc)^3$ we expect to collect a catalogue of the order of 10k galaxies with stellar masses larger than $10^{10.5} \rm M_{\sun}$, and resolution, as the softening length of the TNG100 for stars and galaxies is sub-kpc (see \S\ref{the_simulation} for details). For comparison, the resolution in \citet{deGraaff_eagle} is $2$ kpc.}
%to construct a large sample of local early-type galaxies (with redshifts $z\leq 0.1$) with small impact of outliers.
%Moreover, the robust morphological classification of ETGs employed here (see \S\ref{classification}) enables one to skip the commonly tedious task of separating galaxy types and make direct use of the galaxies provided in our sample. 

The final catalogue will comprehend projected quantities of ETG simulated galaxies derived using non-parametric methods and Sérsic photometry, including: axis ratio, $r$-band luminosity, sizes and line-of-sight velocity dispersion.  The three-dimensional parameters include total and central masses of the various components (i.e., gas, stars and dark matter), central density slope derived assuming a simple power-law description $\rho(r) \propto r^{-\gamma}$ (\S\ref{density_profiles}), star formation rate, 3D $r$-band luminosity derived from stellar population synthesis (\S\ref{fsps}) and 3D sizes. To validate the collected sample, we investigate the  {most relevant scaling relation of ETGs,}  
%the fundamental plane \citep{Mo, Binney},
%of ETGs 
and study the slopes of the central density profiles of stellar, gas, dark matter and total contribution. For the total density slope, we explore its correlations with typical   {galaxy} parameters, such as stellar mass, effective radius and line-of-sight velocity dispersion.  {This is particularly relevant, due to the claimed discrepancies between simulations and observation \citep{D.Xu,Y.Wang}.}   {Finally, the relation between 3D parameters and their projected counterparts is also explored}. The fundamental plane \citep{Mo, Binney} will be discussed in a separate paper (de Araujo Ferreira et al., in preparation).
In all cases we provide comparison of these relations with observations, showing a consensus.

The catalogue constructed in the paper will be used for training machine learning tools for predicting the dark matter content of real galaxies, as previously done (see, e.g., \citealt{wu2023total}), but using a more realistic variety of features, allowing us to push the dark matter studies to their density slopes, for the first time. The catalogue will be made publicly available to become a shared resource of local (low redshift) early-type galaxies extracted from the IllustrisTNG100-1 simulation to be used in future researches concerning galaxy formation and evolution or to develop and test ML techniques further.

%To further verify the reliability of the catalogue, we compare the properties of the mock sample with observational samples from different surveys and methodologies. Hence, in addition to the observation-driven catalogue construction, this article is a review of various observational results concerning the central density trends of galaxies and some scaling relations involving ETGs' structural parameters.
%,   {and projection effects}.

The structure of the paper is organized as follows: in  \S\ref{methodology} we describe the methods introduced in this work: in particular, the stellar population synthesis to derive the luminosity of the galaxies, the morphological classification criteria, the radial light decomposition to fit 1D Sérsic models and the construction of density profiles.  {This allows us to define 
%After this, we present 
the main observational-like quantities that we use to collect a catalogue of ``virtual-ETG observables''.}
%in the catalogue along with their description and definitions}
%, and explain a data cleaning procedure applied to remove possible outliers added during its construction. 
In \S\ref{results_discussion} we discuss the scaling relations of the galaxies within the catalogue, the properties of the density slopes, the correlations between the total density slopes and some parameters of the catalogue and a comparison between the projected and the three-dimensional quantities. Finally, in \S\ref{conclusions} we summarize the main results obtained in this work and further steps involving possible applications of the sample.

\section{Methodology}\label{methodology}
In this section, we detail the procedures undertaken in our study, starting from the data gathering, then definitions of parameters, and, lastly, the catalogue content. 
%We detail the underlying assumptions and methodologies employed at each stage of the process.

\subsection{The simulation}\label{the_simulation}
%The data used to make this work possible comes from 
 {This work is based on the data from} IllustrisTNG, a set of state-of-the-art magnetohydrodynamic cosmological simulations of galaxy formation \citep{Nelson_et_al_2019}. The simulations consist of three cubic volumes of side $51.7$, $110.7$ and $302.6$ Mpc,  dubbed TNG50, TNG100 and TNG300, respectively.   {The assumed cosmology is the Planck 2015 \citep{planck15}, with a total matter density parameter $\Omega_{m} = 0.3089$, dark energy density parameter $\Omega_{\Lambda} = 0.6911$, baryonic matter density parameter $\Omega_b = 0.0486$ and reduced Hubble constant $h=0.6774$.}

Specifically, we use the {highest resolution simulation} from the TNG100 volume, namely TNG100-1 \citep{Nelson+2018, Pillepich+2018, Marinacci+2018, Springel+2018, Naiman+2018}, which contains a baryonic and dark matter mass resolution of $1.4\times 10^6~M_{\odot}$ and $7.5\times10^6~M_{\odot}$, respectively. The gas softening length is adaptive and has the minimum value $\epsilon_{\rm gas,\min} = 0.185$ kpc. For the stellar and dark matter content, the softening length is fixed in comoving units to $\epsilon_{\rm DM, \star} = 0.74$ kpc until $z=1$, and become fixed in physical units for $0<z<1$, such that at $z=0$ it has half the softening length it would have had if the comoving evolution had been kept.

The galaxies are identified as gravitationally bound structures of stellar, gas and dark matter particles using the \texttt{SUBFIND} algorithm \citep{Springel_et_al_2001, Dolag_et_al_2009}. For this study, we selected only central galaxies that are the most massive objects identified by \texttt{SUBFIND} in each friend-of-friend (FoF) group and are of cosmological origin (\texttt{SubhaloFlag=True}), in the sense that they have formed
due to the process of structure formation and collapse \citep{Nelson_et_al_2019}. We filtered galaxies in nine snapshots corresponding to the redshift values $z=0,  0.01, 0.02, 0.03, 0.05, 0.06, 0.07, 0.08, 0.1$. We assume that in this redshift window there is no significant galactic evolution to impact the studies of this paper and, for now, we do not try to inspect the impact of redshift in the analysis. We also consider only the galaxies with total stellar masses {(extracted from the \texttt{SUBFIND} catalogue)} in the range $10^{10.3} \leq M_{\star}/M_{\odot} \leq 10^{11.9}$,  {to guarantee a sufficiently large number of stellar particles per galaxy to minimize the Poissonian noise on the average quantities derived by the binning of the particles (see below).} This first selection results in a sample of $23161$ objects for $z\leq 0.1$. Further filters to this sample (such as galaxy morphological classification) are discussed in detail in the next sections.

\subsection{External observation data}
\label{sec:external_data}
In this paper, we will make extensive comparisons with external observational datasets providing the same structural parameters that we extract for the TNG100-1 simulation described in the previous paragraph. For sake of space, we list the major datasets that we will discuss later in the paper, without giving all details of these catalogues, but only providing the necessary references describing the original dataset used. In particular, we use data from the SAGES Legacy Unifying Globulars and GalaxieS Survey (SLUGGS, \citealt{Brodie+2014, Forbes+2016}), the Sloan Lens ACS (SLACS) Survey \citep{SLACS}, the SL2S Galaxy-Scale Gravitational Lens Sample \citep{Gavazzi_2012}, the Dynamics and Stellar Population (DynPop) from the MaNGA survey \citep{Zhu}, the SPIDER sample \citep{LaBarbera+2010} and ATLAS$^\text{3D}$ \citep{Atlas3D}. Relevant data used in this paper come from \citet{Auger+2010, Grillo} for SLACS, \citet{Sonnenfeld_2013} for SL2S, \citet{Zhu} for DynPop\footnote{\url{https://manga-dynpop.github.io}}, \citet{Krajnovic-2013,Cappellari_2015, Poci,Derkenne} for ATLAS$^\text{3D}$, and \citet{Tortora+2014} for SPIDER. More details on the quantities used in the paper will be given throughout the paper, when needed.

\subsection{Light-based properties }
In this section, we discuss the methods used to infer the $r$-band magnitudes for each stellar particle
%associated with 
of a given galaxy, the estimations of projected quantities, such as ellipticity, centroid of luminosity and orientation angle, and, finally, 
%in the remaining part, we discuss 
the 
%derivation of Sérsic 
 {surface brightness profile fitting using the Sérsic model \citep{Sersic}, to derive the related Sérsic} parameters,  {namely the effective radius, $R_{\rm e}$, the Sérsic index, $n$, and the total luminosity, $L_{\rm tot}$}.
%from these \textit{projected} galaxies and the classification criteria  used to filter our sample of ETGs. 
In order to  {check the variance of the} 
%maintain the 
observational constraints in our analysis, we consider the 2D projection of each 3D simulated galaxy along the $X$, $Y$ and $Z$ axes of the simulation box as one realization of the galaxy as seen by an observer whose line of sight lies parallel to that particular axis of projection.

\subsubsection{Stellar light measurement}
\label{fsps}
To derive the luminosity of the stars within each TNG galaxy, we use the \texttt{FSPS} (Flexible Stellar Population Synthesis) code\footnote{\url{https://github.com/cconroy20/fsps}} \citep{C.Conroy,C.ConroyII,D.Foreman-Mackey} to compute simple stellar populations. The method applied in this section is similar to that in \citet{Nelson+2018}, with minor modifications. 
%The steps are described in the next paragraphs. 

Each 
  {star} 
 {particle} in a TNG galaxy is assumed to be a single-burst 
  {simple} 
stellar population (SSP). Using the information of birth time, metallicity and initial mass, which are recorded 
%respectively stored 
in the simulations as the  \texttt{GFM\_StellarFormationTime}, \texttt{GFM\_Metallicity} and \texttt{GFM\_InitialMass} keywords, respectively, the configurations we use in \texttt{FSPS} assume a \citet{Chabrier2003} initial mass function, along with the MILES stellar library \citep{Sanchez+2006, Falcon-Barroso+2011} and Padova isochrones (see, e.g., \citealt{Marigo+2007, Marigo+2008}). We also incorporate dust effects by the adoption of the simple two-component power-law extinction model of \citet{S.Charlot}, which is defined as:
\begin{equation}
     \tau(t) \equiv \begin{cases}
                    \tau_1\left(\lambda/5500~\mathring{A}\right)^{-0.7}, \quad t\leq 10^7~\text{yr}\\
                    \tau_2\left(\lambda/5500~\mathring{A}\right)^{-0.7}, \quad t > 10^7~\text{yr},
                    \end{cases}
\end{equation}
where $\tau$ is the optical depth. In this model, the $\tau$ for young stellar systems is associated with dust in molecular clouds, in which the young stars are embedded. Following the disruption of molecular clouds, which occurs on a timescale of {$t\sim 10^7$ yr} \citep{Blitz}, the optical depth is associated with a uniform screen across the galaxy. Following \citet{S.Charlot}, we use $\tau_1 \sim 1.0$ and $\tau_2 \sim 0.3$, that have been found to adequately describe a wide range of observations.
%found that values of $\tau_1 \sim 1.0$ and $\tau_2 \sim 0.3$ adequately described an array of observational data. The same values for  $\tau_1$ and $\tau_2$ are adopted for our computations. 

  {Based on the usual range of metallicities in the selected galaxies, we computed isochrones composed of $22$ metallicity steps in the range $-2 \leq \log(Z_i/Z_{\odot}) \leq 0.5$ (where $Z_{\odot}= 0.127$, \citealt{Wiersma+2009}), and $94$ age steps for $-3.5 \leq \log(t_{\rm ssp}/\text{Gyrs}) \leq 1.15 $}. For each point on the 2D grid of metallicities and ages, we convolved the resulting population spectrum with the observed SDSS $r$-band filter (airmass 1.3), and saved the total magnitude in each. We also included the models for dust emission and nebular line and continuum emission. Hence, for each star particle, we applied a bicubic interpolation in the ($Z_i$, $t_{\rm ssp}$) plane to obtain $m_1$, the $r$-band magnitude of an $1 M_{\odot}$ equivalent population. Multiplying by the initial mass $M_i$, we obtain the total $r$-band magnitude of the actual population by $m_r^{\rm abs} = m_1 - 2.5 \times \log(M_i/M_{\odot})$. In the AB system, the solar absolute magnitude in this pass-band is $m_{\odot,r}^{\rm abs} = 4.64$\footnote{A table about the \texttt{FSPS} filters' details, including the solar magnitude for most of them, is found in \url{https://dfm.io/python-fsps/current/filters/}.}. 
%The  magnitudes can be easily  converted to luminosity values in solar units.

\subsubsection{Galaxy geometry: ellipticity, centroid of luminosity, and orientation angle}\label{geometric_params}

With the luminosity information obtained 
%from the procedure described 
in the previous section, we now are able to discuss the construction of the light profiles. In doing this we need to introduce %along with 
the main geometrical properties of galaxies, such as ellipticity, centroid of luminosity and position angle, which we will use throughout the paper. 

 {Before we move to detail this procedure,} it is important to recall that when a smaller halo is accreted onto a bigger structure, its dark and baryonic matter at the outskirts can be tidally stripped while sinking to the centre of the host. This stellar mass is still ``weakly'' gravitationally bound to the galaxy, and hence normally included in the raw measurement of \texttt{SUBFIND}. In the context of galaxy clusters, this stellar material is usually referred as \textit{intracluster} light, and it generally accounts for a significant fraction of the total luminosity of a galaxy it overlaps with \citep{Lin+2004, Zibetti+2005, Puchwein+2010}.  {Despite its importance in the context of the galaxy and group/cluster formation theories, this component represents a contaminant of the individual galaxies, as it does not constitute a fully bound component of any specific individual galaxy, but rather a component eventually relaxing in the cluster potential (see, e.g., \citealt{pota+2018, spiniello+2018})}, with a phase space distribution of the stars which is different from the individual galaxy populations. {To mitigate the impact of this component on the stellar mass and derived light profiles, we applied a two-step selection process, following \citet{D.Xu}. First, we employed a commonly used technique: a 3D radial cut at 30 kpc from the galactic centre (e.g., \citealt{Pillepich+2017, Lu+2020}). The galactic centre is defined as the position of the particle with the minimum gravitational potential, as determined by the \texttt{SUBFIND} algorithm.   Next, the selected stellar particles were projected onto a 2D plane, where we defined the galaxy’s centroid of light $(X_{gc}, Y_{gc})$. This centre is computed as the luminosity-weighted average position of the projected stellar particles:  
\begin{equation}
X_{gc} = \frac{\sum_i X_i L_i}{\sum_{i}L_i}, \quad Y_{gc} =  \frac{\sum_i Y_i L_i}{\sum_{i}L_i},
\end{equation}  
where $L_i$ represents the dust-attenuated luminosity of each stellar particle. Finally, disregarding the previous 3D cut, we applied a 2D radial cut that selects all projected stellar particles within a radius of 30 kpc from $(X_{gc}, Y_{gc})$.}

 %for a given galaxy projection we define a 2D radial cut of $30$ kpc  from the centre of light $(X_{gc}, Y_{gc})$}, defined as the average projected $(X_i,Y_i)$ position of each stellar particle weighted by their dust-attenuated luminosity value $L_i$:

%\pf{In addition to this 3D cut, we follow \citet{D.Xu} and define a 2D radial cut of 30 kpc from the centre of light for each projected galaxy. The centre of light $(X_{gc}, Y_{gc})$, computed as:}
%\begin{equation}
%X_{gc} = \frac{\sum_i X_i L_i}{\sum_{i}L_i}, \quad Y_{gc} =  \frac{\sum_i Y_i L_i}{\sum_{i}L_i},
%\end{equation}
%where $\sum_{i}L_i$ is the total luminosity within this $30$ kpc aperture. 
%\pf{where $\sum_i L_i$ is the dust-attenuated luminosity within a given aperture. }
%The iterative refinement begins with a projected aperture of $30$ kpc (see above) and decreases in steps of $1$ kpc until reaching a minimum radius of $1$ kpc, consistent with the simulation's softening length. At each step, only particles within the current aperture contribute to the centroid calculation. This ensures that the final centre of light is robust against extended low surface brightness features, while remaining consistent across galaxies of different sizes.

Given the values of $(X_{gc}, Y_{gc})$, the measurement of the axis ratio $K = b/a$ and orientation angle $\phi$ of a galaxy 
%were achieved 
were finally obtained through the computation of the second moments of luminosity, given by the set of equations:
\begin{equation}
\begin{aligned}
M_{XX} &=  \frac{\sum_{i} L_i(X_i - X_{gc})^2}{\sum_i L_i},\\
 M_{YY} &=  \frac{\sum_{i} L_i(Y_i - Y_{gc})^2}{\sum_i L_i},\\
M_{XY} &=  \frac{\sum_{i} L_i(X_i - X_{gc})(Y_i - Y_{gc})}{\sum_i L_i}.
\end{aligned}
\end{equation}
%Here, the axis ratio is given by:
  {Hence,}
\begin{equation}
K = \frac{b}{a} = \left(\frac{M_{XX} +M_{YY}-\sqrt{(M_{XX}-M_{YY})^2 + 4M_{XY}^2}}{M_{XX} +M_{YY}+\sqrt{(M_{XX}-M_{YY})^2 + 4M_{XY}^2}}\right)^2,
\end{equation}
%and the orientation angle by:
\begin{equation}
    \phi = \frac{1}{2}\arctan\left(\frac{2M_{XY}}{M_{XX}-M_{YY}}\right).
\end{equation}

\begin{figure*}
    \centering
    \begin{subfigure}{0.3\textwidth}
        \centering
        \includegraphics[width=\textwidth]{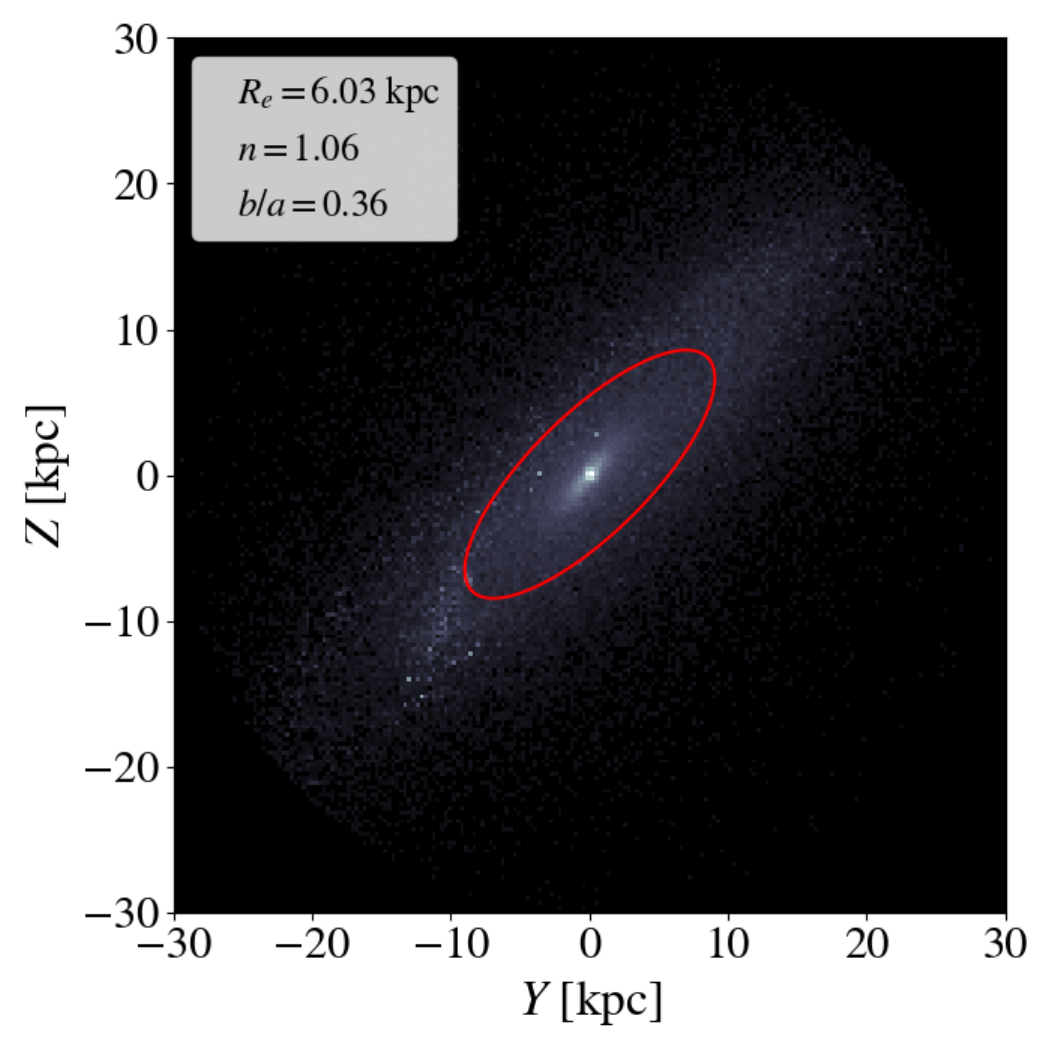}
    \end{subfigure}
    \begin{subfigure}{0.3\textwidth}
        \centering
        \includegraphics[width=\textwidth]{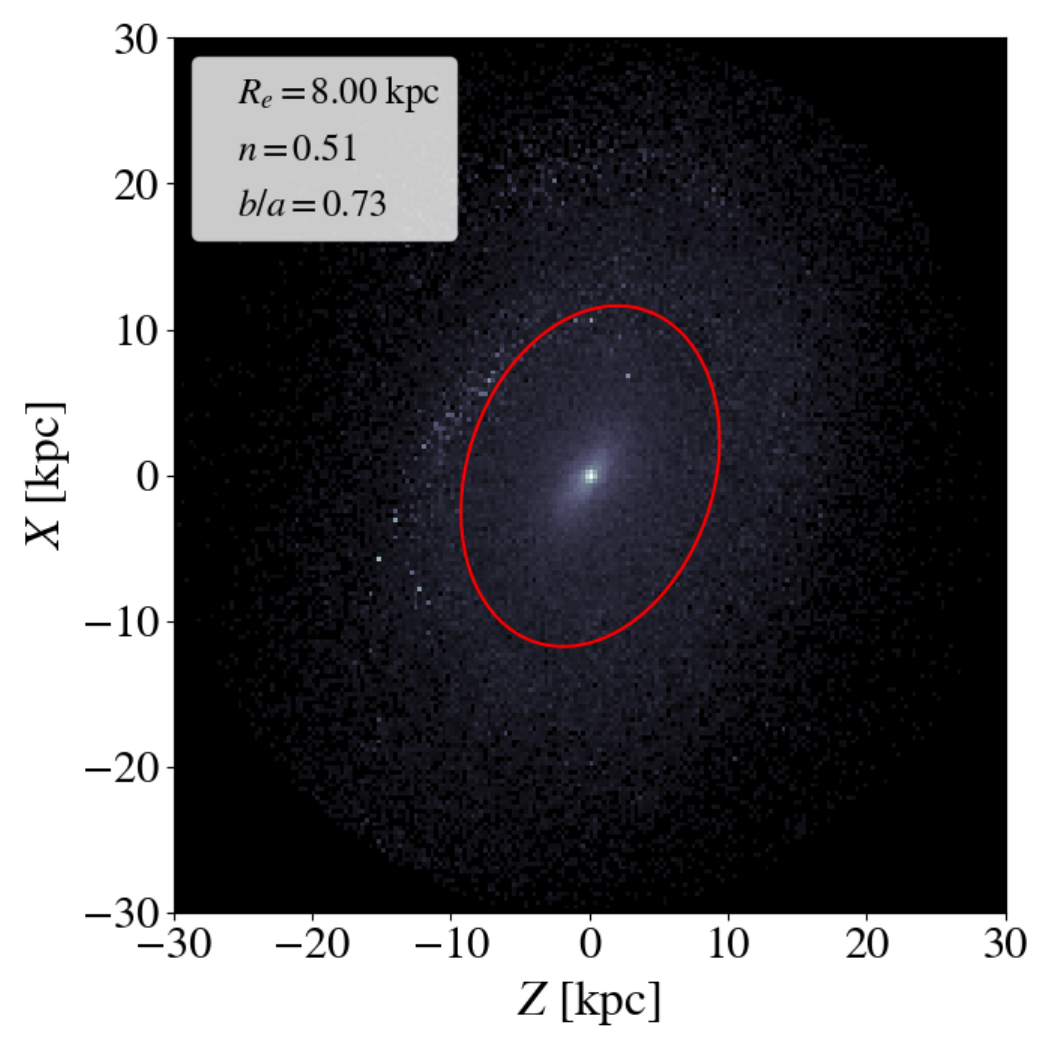}

    \end{subfigure}
    \begin{subfigure}{0.3\textwidth}
        \centering
        \includegraphics[width=\textwidth]{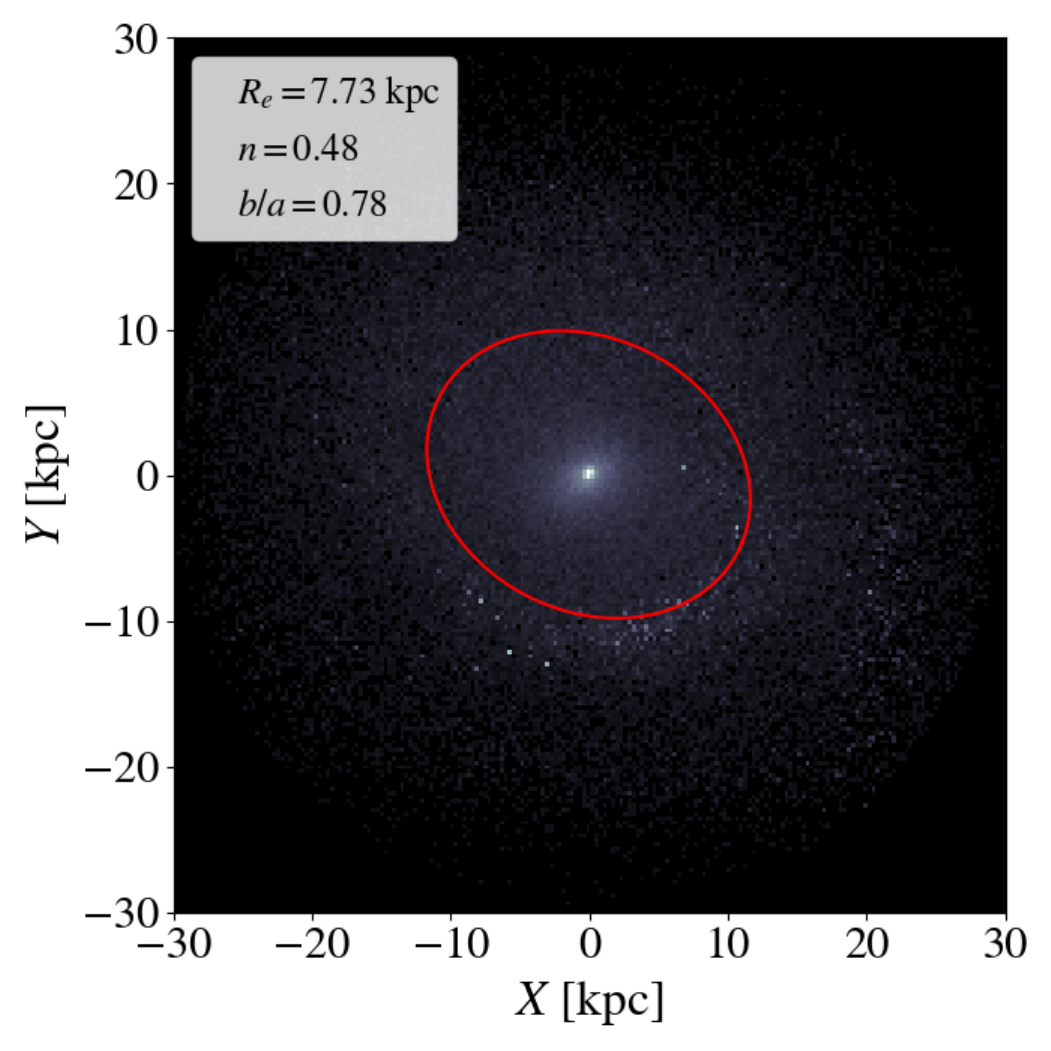}
    \end{subfigure}

    \begin{subfigure}{0.3\textwidth}
        \centering
        \includegraphics[width=\textwidth]{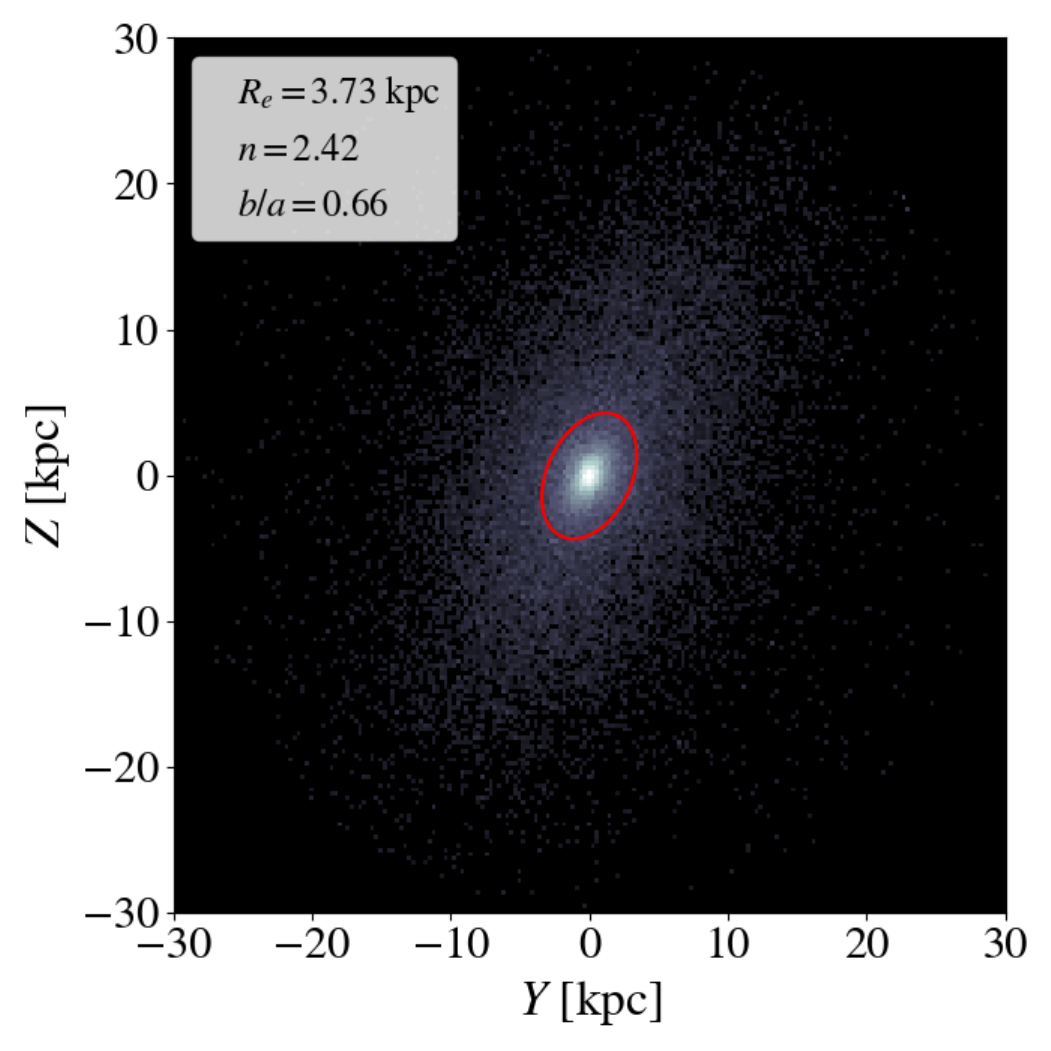}
        \caption{$X$ projection}
    \end{subfigure}
    \begin{subfigure}{0.3\textwidth}
        \centering
        \includegraphics[width=\textwidth]{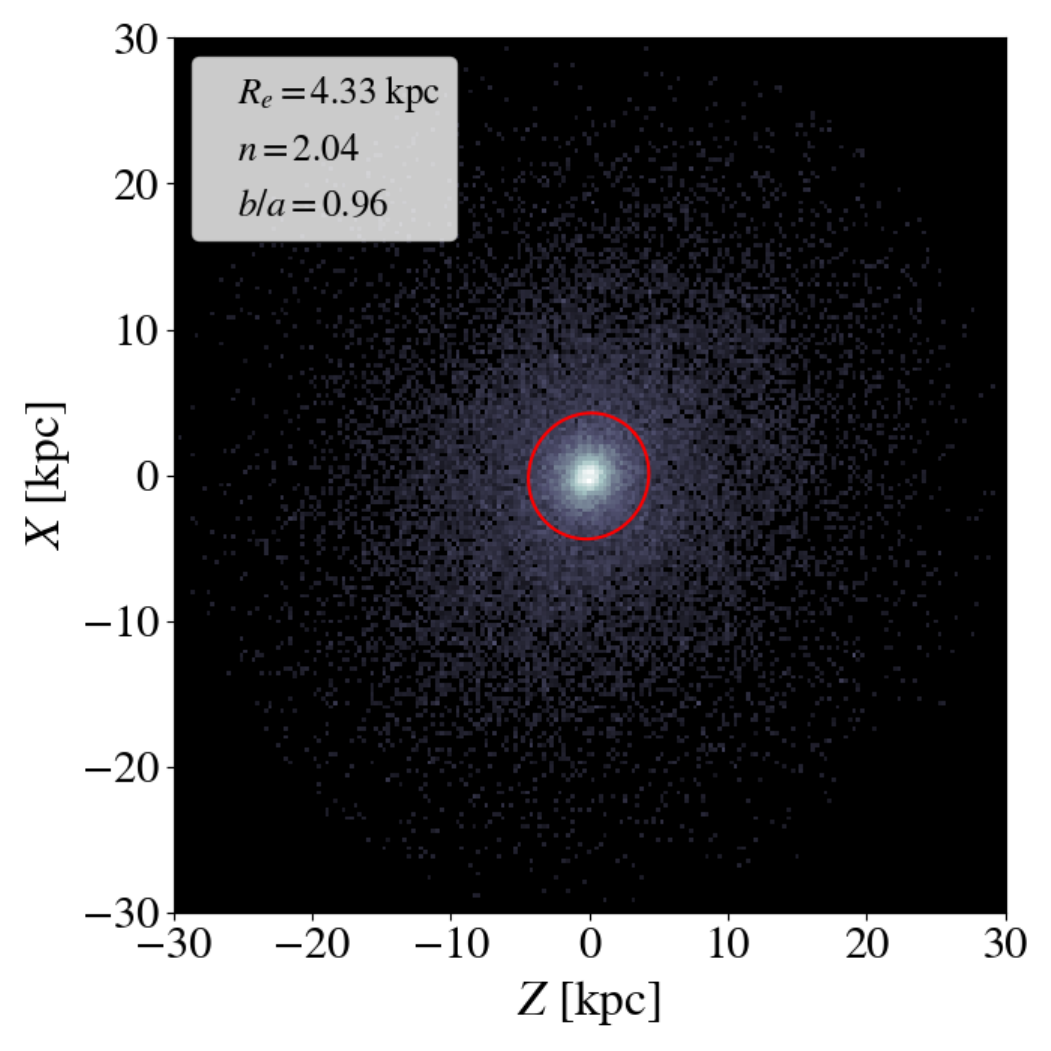}
        \caption{$Y$ projection}
    \end{subfigure}
    \begin{subfigure}{0.3\textwidth}
        \centering
        \includegraphics[width=\textwidth]{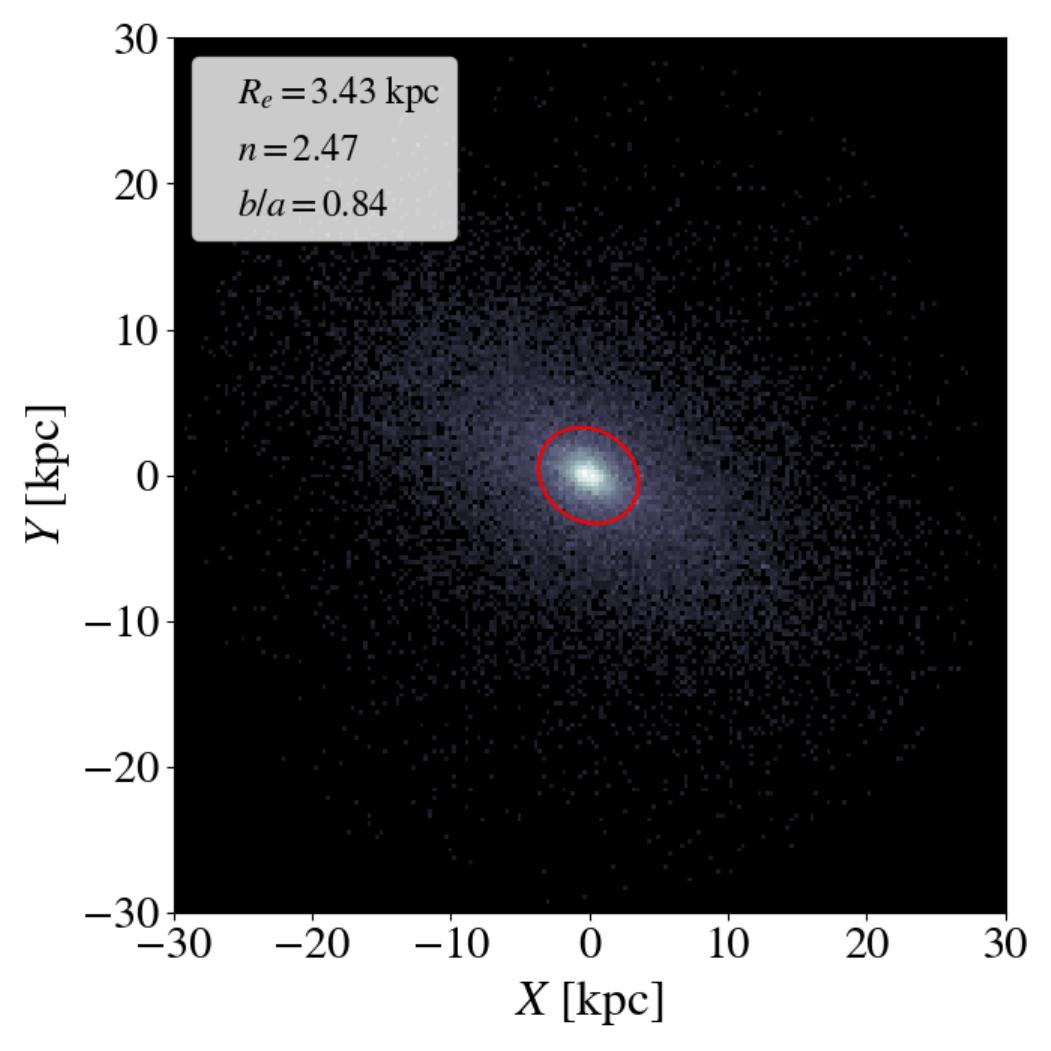}
        \caption{$Z$ projection}
    \end{subfigure}
    \caption{Light density distribution of three independent projections of an LTG (upper row) and an ETG (lower row) along with their structural parameters. The red ellipses represent the half-light ellipse of circularized radius $R_{\rm e}$. The TNG ID for the LTG is $431201$ and for the ETG is $524400$, both at $z=0$.}
    \label{fig:mock_images}
\end{figure*}

\begin{figure}
    \hspace{-0.5cm}
    \includegraphics[width=0.535\columnwidth]{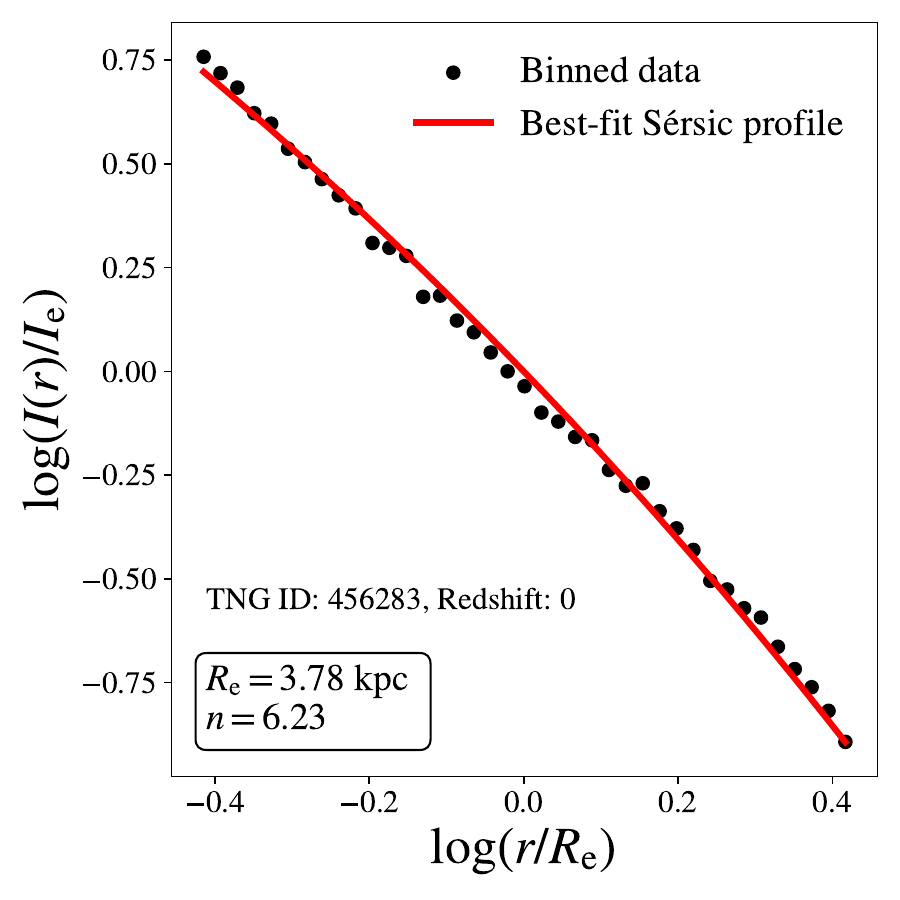}
    \hspace{-0.18cm}
        \includegraphics[width=0.535\columnwidth]{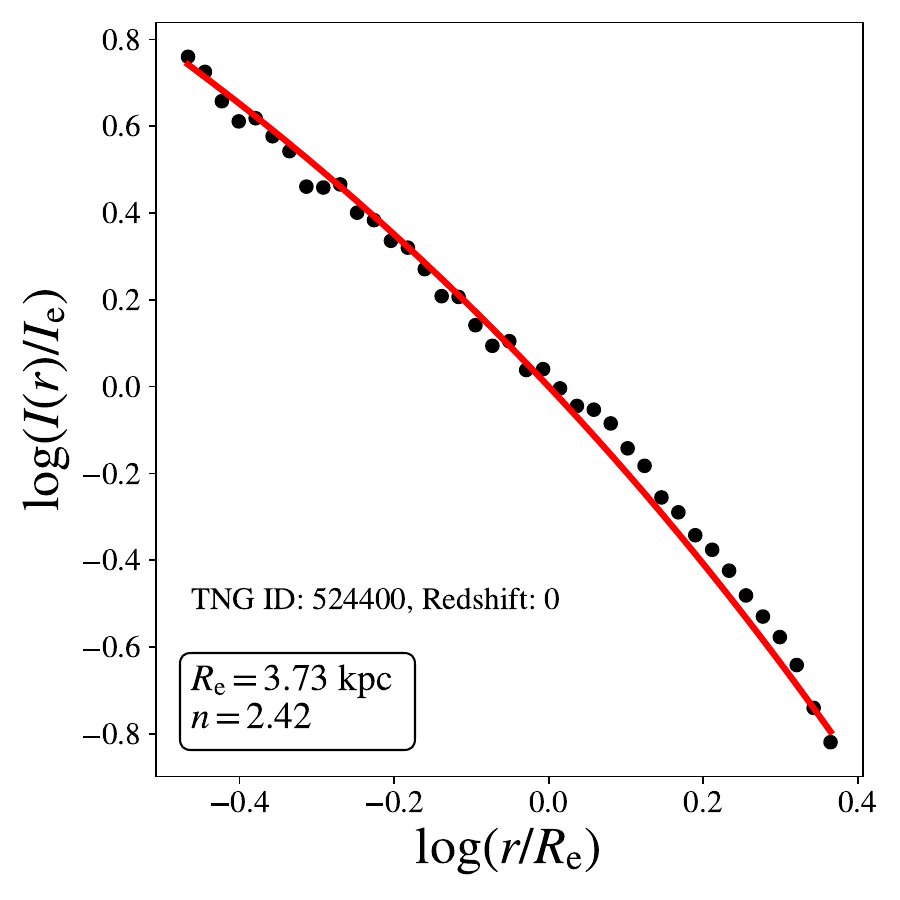}
        
    \hspace{-0.5cm}
    \includegraphics[width=0.535\columnwidth]{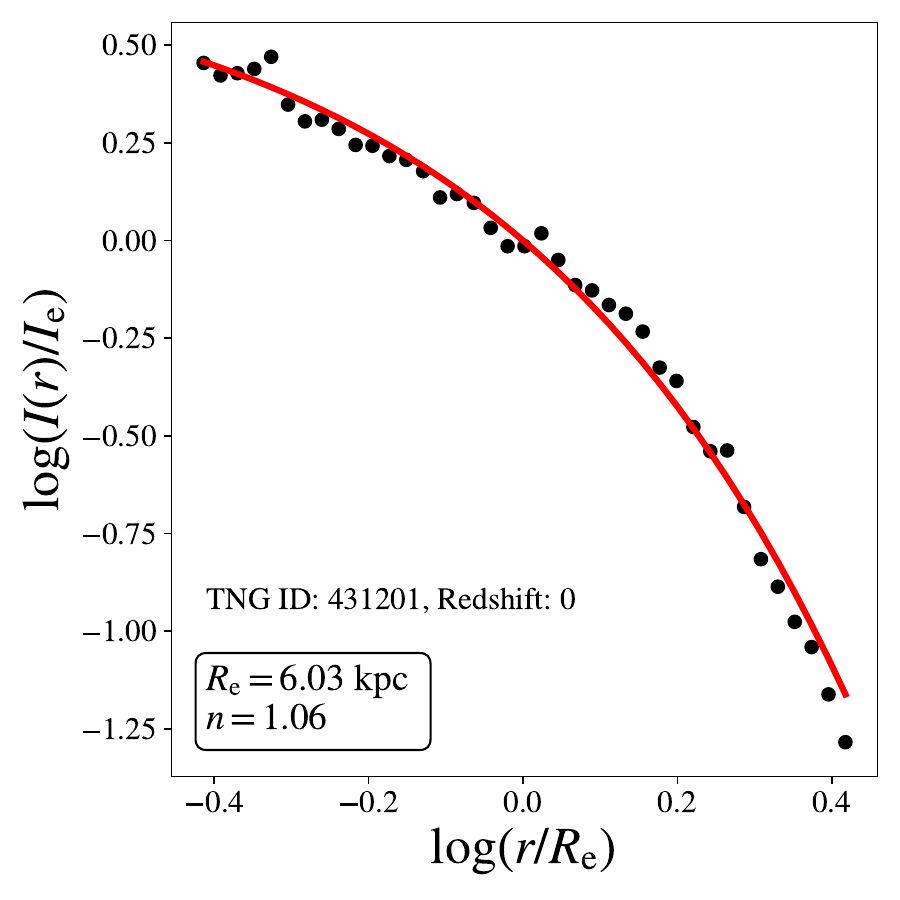}
    \hspace{-0.18cm}
    \includegraphics[width=0.535\columnwidth]{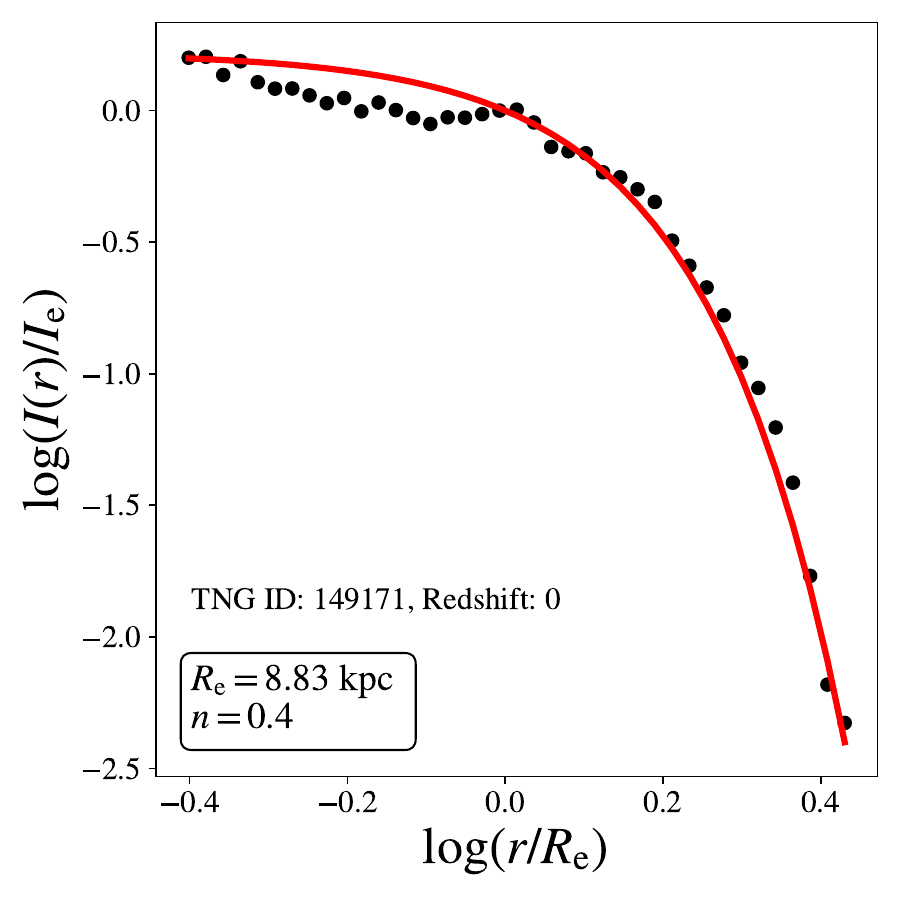}
    
    \caption{Surface brightness profiles for a pair of ETGs (upper panels) and a pair of LTGs (bottom panels). For each plot, the TNG ID is reported together with the best-fit S\'ersic parameters for the projection along the $X$-axis. The black dots are the binned light distribution, and the red line represents the best fit Sérsic profile.}
    \label{fig:sb_plots}
\end{figure}

\subsubsection{Surface brightness profiles}
\label{sbp}
To calculate the galaxy luminosity and effective radius within non-circular apertures, we assumed that the 2D surface brightness (SB) distribution in a given projection follows a series of isophotal ellipses that are described by a constant axis ratio $K$ and orientation angle $\phi$, measured according to the former definitions. We transform the positions of each star to a \textit{canonical} system of coordinates by applying the linear transformation  $\mathbf{P}' = R(\phi) \mathbf{P}$, where $\mathbf{P} = (X-X_{gc}, Y-Y_{gc})$, $\mathbf{P}' = (X', Y')$ and $R(\phi)$ is the two-dimensional rotation matrix \citep{arfken}, defined as:
\begin{equation}
    R(\phi) = \begin{bmatrix}
        \cos\phi & \sin\phi \\
        -\sin\phi & \cos\phi
    \end{bmatrix}.
\end{equation}
The new position $(X'_i, Y_i')$ of each star is given by:
\begin{equation}
\begin{aligned}
X'_i &= (X_i - X_{gc})\cos{\phi} + (Y_i - Y_{gc})\sin\phi\\
Y'_i &= -(X_i - X_{gc})\sin\phi + (Y_i - Y_{gc})\cos\phi.
\end{aligned}
\end{equation}
In this system, the elliptical isophotes are concentric, %with the origin 
and their major axes lay parallel to the horizontal axis $X'$.
From the canonical equation of an ellipse, we write:
\begin{equation}
a^2 = X'^2 + \frac{Y'^2}{K^2}.
\end{equation}
Based on the light distribution, we selected 
%$a$ such that it is 
the semi-major axis $a_{\rm e, dir}$ of the ellipse enclosing half of the total luminosity of a galaxy. 
%Denoting this \textit{half-light} semi-major axis by $a_e$ and 
Using the assumption that $K = b/a =$ constant, it is straightforward to find the \textit{half-light} semi-minor axis $b_{\rm e, dir}=K a_{\rm e, dir}$. With these quantities, we estimated a direct half-light radius (or effective radius) as being the geometrical mean of $a_{\rm e, dir}$ and $b_{\rm e, dir}$, i.e., $R_{\rm e, dir}= \sqrt{a_{\rm e, dir} b_{\rm e, dir}}$, such that the area enclosing half of a galaxy's light is $A = \pi R_{\rm e, dir}^2 = \pi a_{\rm e, dir} b_{\rm e, dir}$. We remark that  {the constant axis ratio and position angle assumptions are not the most realistic approaches to the surface brightness analysis, as galaxies show generally variations of these quantities as a function of the distance from the center. However, this assumption still allows us to catch the average geometrical properties of galaxies (e.g., average ellipticity), which are usually used in scaling relations (e.g., the ellipticity -- anisotropy relation; see, e.g., \citealt{Binney, Emsellem+2011}) }

With the $r$-band luminosity and position information for   {the stellar} particles, we derived the binned SB distribution
(in units of $L_{\odot}/$kpc$^2$)
%the projected 
 {of each of the three projections for each} galaxy in circularized concentric annuli, radially ordered by $r = \sqrt{KX'^2 + Y'^2/K}$. This binning procedure consists of $40$ bins equally spaced in logarithmic scale, from $xR_{\rm e, dir}$ to $3 R_{\rm e, dir}$\footnote{We tried to extend the range of the upper limit of these intervals up to $4R_{\rm e}$, $5R_{\rm e}$ and $6R_{\rm e}$, and found that, due to the fewer number of particles in the outer bins, both the individual and the overall parameter distribution becomes noisier, i.e., on average, the scatter increased by $0.02\%$ for $n$, $3.31\%$ for $R_{\rm e}$ and $1.66\%$ for $I_{\rm e}$. Hence, we decided to maintain the  upper limit used by \cite{D.Xu} for a more homogeneous comparison.
}, where $x$ is an adaptive fraction of $R_{\rm e, dir}$, such that $0.1 \leq x\leq 0.3$ for $R_{\rm e, dir} \geq 10$ kpc and  $0.3 < x \leq 0.4$ for $R_{\rm e, dir} < 10$ kpc. 
%We added 
 {This adaptive partitioning is an additional step adopted by us with respect} to the procedure used by \citet{D.Xu}, which ensures a balanced treatment of light profiles  {for small and large galaxies, in particular by preventing}
%and for both large and small galaxies. Indeed, it prevents 
profiles from staying too far from the centre for larger galaxies and too close for smaller galaxies.
We also impose the condition that $x R_{\rm e} \geq 1$ kpc to ensure that all the profiles have a lower-bound value considerably above the simulation softening length (see, e.g., \citealt{Du_2020}).  {We believe this is the best workaround to minimize the impact of the softening length on galaxies of all sizes and masses, but we cannot exclude the possibility that it may introduce biases in the Sérsic parameters (see below).} 

{As for the maximum radius to adopt to include stellar particles in the fit, we have used a $30$ kpc aperture  (see, e.g., \citealt{D.Xu,Pillepich+2018}) centred at $(X_{gc}, Y_{gc})$, according to the previous discussion}. We have also tried a more observationally motivated SB cut, corresponding to 26.5 mag/arcsec$^2$ in $g$-band
%\footnote{To make the selection consistent, we have used the $r$-band light for the stellar particles  (see \S\ref{fsps}) and converted the $g$-band cut into a 25.7 mag/arcsec$^2$ in this $r$-band, assuming typical $g-r=0.8$ for galaxies in the red sequence (see, e.g., \citealt{LaBarbera+2010b}).} 
(see \citealt{Tang+2021}, and also \S\ref{sec:intro}). However, since we have found no significant differences in the best-fit quantities that we will discuss later in this paper (in particular the effective radius and the total luminosity of the virtual-ETG sample), we have decided to discard the quantities based on the SB cut. The reason for the low sensitivity to the outer cut resides on the fact that this does not directly impact the inferred quantities that are always the result of a fitting procedure. This is different from the reason motivating the 26.5 mag/arcsec$^2$ in $g$-band in \citet{Tang+2021}, who use as total quantities the ones obtained by integrating all particles within that limit, with neither fit or extrapolation.

%For brevity, we will refer to these different radii as $r_{\rm 30kpc}$ (with lower case indicating this is a 3D selection) and $R_{\rm 26.5g}$ (with upper case indicating this is a 2D selection), for the two ``radius cut'' above, respectively. In the following, we will generally use the results based on the ``standard'' $r_{\rm 30kpc}$ to make the comparison with previous literature easier, unless explicitly specified.} 
%, where they excluded the central $1$ kpc region of the IllustrisTNG100 galaxies used to fit the Sérsic profiles, in order to avoid the smoothing effects in scales below the softening length). 
%Using $R_{e}$, we estimate $I_e$, which is the surface brightness value at $r=R_{e}$. 

Once selected the range of radii to analyse, we fit them with a single-component Sérsic profile \citep{Sersic}
\begin{equation}
I(r) = I_{\rm e, mod} \exp\left\{-b_n\left[\left(r/R_{\rm e, mod}\right)^{1/n}-1\right]\right\}. \label{sersic_profile}
\end{equation}
 {This choice is dictated by the necessity to adopt an approach that is close to typical observational-like analyses (see, e.g., \citealt{Shen+2003,Lange+2015, Roy+2018}) and easily interpretable parameters, that are routinely used for scaling relations, like the size-mass \citep{Hyde+2009} or the fundamental plane (e.g., \citealt{LaBarbera+2010, Bernardi+2020}). Despite its generality, though, the single-Sérsic profile could not capture the full complexity of the virtual-ETG sample, as well as it does not for real galaxies (see, e.g., \citealt{Conselice_2005,Lang+2014, Gao_2017}). We will return on the impact of this simplified assumption, and the adaptive partitioning discussed above in \S\ref{data_cleaning}.}

The Sérsic model depends on the free parameters $n$, $R_{\rm e, mod}$, $I_{\rm e, mod}$, where $n$ is the Sérsic index, $R_{\rm e, mod}$ and $I_{\rm e, mod}$ are the model half-light radius and surface brightness, respectively. The parameter $b_n$ is a function of $n$ and satisfies $2\gamma(2n; b_n) =\Gamma(2n)$, where $\Gamma$ and $\gamma$ are respectively the Gamma function and the  (lower) non-regularized incomplete Gamma function \citep{arfken}. For our computations, we used an approximation for $b_n$ up to fourth order in $\nu=1/n$ \citep{MacArthur+2003}:
\begin{equation}
b_n \approx \frac{2}{\nu}-\frac{1}{3}+\frac{4\nu}{405}+\frac{46\nu^2}{25515}+\frac{131\nu^3}{1148175}-\frac{2194697\nu^4}{30690717750}. 
\end{equation}

\begin{figure}
    \centering
    \includegraphics[width=\columnwidth]{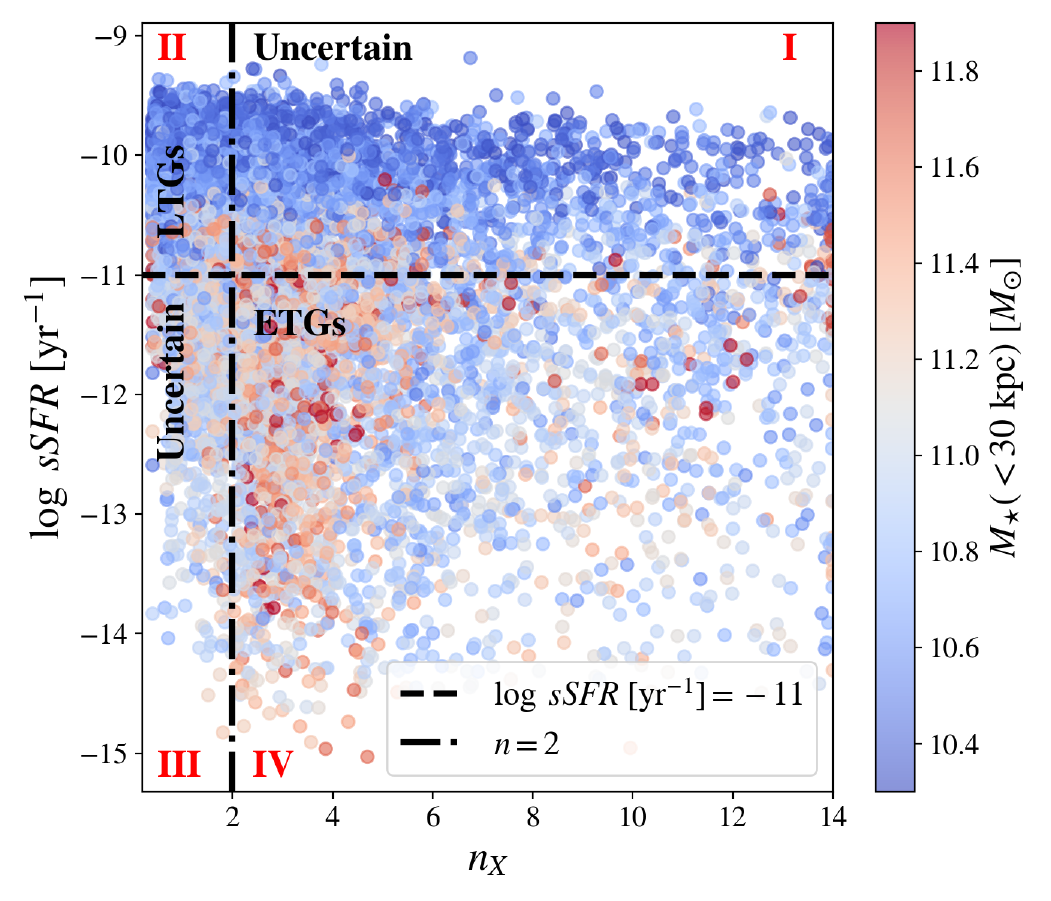}
    \caption{Illustration of classification scheme of galaxies based on the criteria described in the main text. The horizontal axis represents the Sérsic index with respect to the galaxy projections along the $X$ axis and the vertical axis is the specific star formation rate (which is a 3D quantity; see text for details). The blueish dots represent galaxies with smaller stellar mass, whilst reddish dots are galaxies with larger stellar mass.   The second (II) and fourth (IV) quadrants respectively correspond to the galaxies classified as LTGs and ETGs in this specific projection, while the first (I) and third (III) quadrants correspond to galaxies that did not satisfy the criteria to be either LTG or ETG.}
    \label{ssfr_classification}
\end{figure}

We chose the best-fit parameters based on a minimum $\chi^2$ approach, having used  $R_{\rm e, dir}$ and $I_{\rm e, dir}$ for each galaxy as initial guesses for the fitting procedure, while $n=2.5$ is assumed as initial values for all galaxies. 
Using a least-squares algorithm\footnote{\url{https://docs.scipy.org/doc/scipy/reference/generated/scipy.optimize.least_squares.html}}, we fitted several profiles considering the different allowed values of $x$ (with $5$ steps between each sub-range) and chose the profile in which the fitted parameters give the minimum value of $\chi^2$. In short, ten profiles with different lower radial limits were fitted to each galaxy, and the best-fit was chosen based on the minimum $\chi^2$ value among them. After the fitting step, we took $R_{\rm e, mod}$, $I_{\rm e, mod}$ and $n$ as being the \textit{final} half-light radius, half-light surface brightness and Sérsic index for each galaxy. 
%The estimated parameters $R_{\rm e, dir}$ and $I_{\rm e, dir}$ for each galaxy are used as first guesses, while $n=2.5$ is assumed as a starting point to all the profiles. 
 {For these ones, from now on, we drop the subscript \textit{\rm mod}, and refer to them as $R_{\rm e}$, $I_{\rm e}$ and $n$, respectively.} 
%are going to refer to the best-fit parameters without mentioning the superscript \textit{\rm mod}. 
{Particular values of $n$ in Eq.~(\ref{sersic_profile}) include $n=1$ that is the so-called exponential profile \citep{Mo, Binney}, and characterizes the surface brightness distribution of disky or other light-diffuse systems. For $n=4$ we have the De Vaucouleurs profile \citep{DeVaucouleurs, Mo, Binney}, that is better suited for the description of spheroidal systems, such as bulges and elliptical galaxies.}   {In Fig. \ref{fig:mock_images}, we show the visual representations of two example galaxies and their respective projections. The one in the upper row has a low Sérsic index value ($n \lesssim 1$), and a disky appearance, which characterizes a Late-Type Galaxy (LTG). The galaxy in the lower row has a higher Sérsic index ($n \geq 4$) and has a smooth, featureless appearance, which are characteristics of an ETG. Furthermore, in Fig.~\ref{fig:sb_plots} {we present the fitted surface brightness profiles and their mock measruments for a pair of ETGs and a pair of LTGs, projected along the simulation's $X$-axis\footnote{{While we initially omitted uncertainties in the light profiles, we acknowledge that they could be estimated from the brightness scatter within each bin. However, our tests show that including these does not significantly affect the Sérsic fits or the scaling relations examined (see \S\ref{results_discussion}), and thus does not impact our discussion.}}.} More details on the galaxy morphological classification are presented in \S\ref{classification}.}
\begin{figure}
    \centering
    \includegraphics[width=0.9\columnwidth]{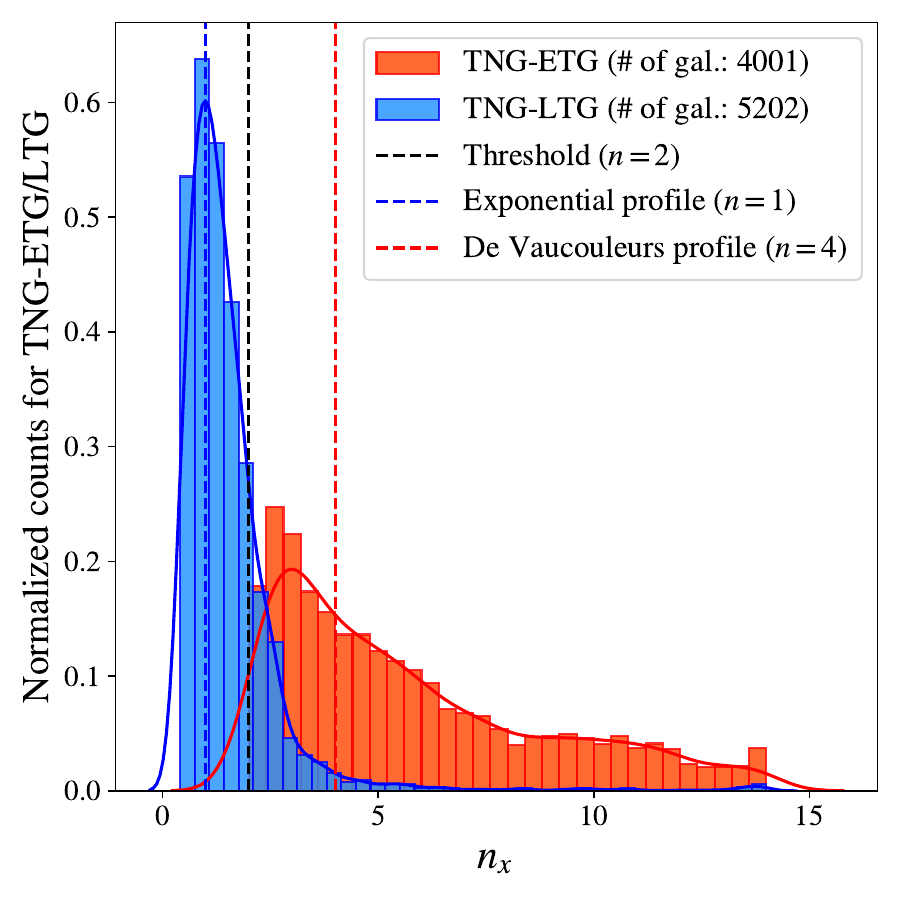}
    \caption{Sérsic index distribution of the galaxies for $z$ between $0$ and $0.01$ projected along the $X$ axis of the simulation box. All the galaxies in this distribution are in accordance to the classification criteria described in the text. For reference, we also include the values of $n$ equivalent to the exponential and the De Vaucouleurs profile.}
    \label{sersic_distro}
\end{figure}

One can compute the luminosity inside a radius $r$ by integrating equation (\ref{sersic_profile}) over a projected area $A=\pi r^2$ \citep{Graham}:
\begin{equation}
L(r) =\int_{0}^{r} I(r') 2\pi r' dr',
\end{equation}
which can be written in the explicit form:
%The explicit form is written in the following manner:
\begin{equation}
L(r) = I(R_{e}) R_{\rm e}^2 \exp(b_n)\frac{2\pi n}{b_n^{2n}}\gamma(2n; b_n(r/R_{\rm e})^{1/n}),\label{luminosity}
\end{equation}
where, again, $\gamma(2n; x)$ is the (lower) non-regularized incomplete Gamma function, which we here explicitly report to be
%expressed by
\begin{equation}
\gamma(2n;x) = \int_{0}^{x} t^{2n-1} e^{-t}dt.
\end{equation}
In practice, galaxies' light are assumed to have finite boundaries that are typically determined by growth curves of the luminosity, where the maximum extent of a galaxy $r_{\rm max}$ is defined as the radius where the growth curve starts to approach a nearly constant value, and its value will typically depend on the exposure time of the outer parts of the galaxy image \citep{Trujillo}. However, there are other ways one can define the total luminosity of a galaxy. For instance, by choosing a sufficiently large multiple of $R_{\rm e}$ (see, e.g., \citealt{D.Xu}) as an upper bound to integrate Eq. (\ref{luminosity}), or, more formally, integrating the surface brightness (\ref{luminosity}) to infinity, which is a straightforward way of defining a galaxy total luminosity from the Sérsic model.  {This will be particularly useful} in \S\ref{corr_stellar_mass}, where we use the integral of the Sérsic profile to infinite and stellar mass-to-light ratios (see \S\ref{pararaph:projected_quantities}) to estimate the galaxy's total stellar mass. 
  {We include this 2D total luminosity, $L_{2D}(R\to\infty)$, in our catalogue, as well as two other apertures: the first one corresponding to a circular aperture of radius $R_{\rm e}$, namely $L_{2D}(R_{\rm e})$, and the second one corresponding to $2R_{\rm e}$, denoted by $L_{2D}(2R_{\rm e})$. %We will also include the means surface brightness inside the effective radius, $\langle I_{\rm e} \rangle = {L_{\rm 2D}(<R_{\rm e})}/{(\pi R_{\rm e}^2)}$, as this is an observable that is used to describe the ETG fundamental plane.
%(see also \S\ref{sec:FP}). 
In \S\ref{projected_vs_3d_light}, we validate these central luminosity definitions by comparing them  with their corresponding three-dimensional counterparts. Furthermore, the projected central stellar masses may be obtained from these central luminosity definitions and stellar mass-to-light ratios (see \S\ref{pararaph:projected_quantities} and \S\ref{projected_versus_3d_stellar_mass}). }

\subsubsection{Galaxy classification}\label{classification}
The galaxies in our sample are separated in ETGs and LTGs, using two criteria. The first criterion is based on their Sérsic index ($n$) {: if $n\geq2$ for all three 2D projections a galaxy is classified as ETG, while it is classified as LTG, if one of the projections has $n<2$.
%. If the Sérsic index in all three 2D projections is greater than or equal to $2$, the galaxy is classified as an Early-Type Galaxy. Conversely, if the Sérsic index is less than $2$ in at least one of the 2D projections, the galaxy is classified as a Late-Type Galaxy (LTG). 
This criterion is purely morphological and, despite being rather widely used for observed galaxies, having the advantage of being applicable to single band images, it might not be robust enough (see, e.g., \citealt{Roy+2018} and references therein).}
The second criterion 
%involves checking 
is based on the specific star formation rate ($sSFR$), defined as the ratio between the star formation rate ($SFR$) and the stellar mass of the galaxy, $M_{\star}$, i.e., $sSFR = SFR/M_{\star}$ [yr$^{-1}$]. If the logarithm of $sSFR$ is greater than $-11$, the galaxy is likely to be an LTG, otherwise, it is likely an ETG. The classification of galaxies from their specific star formation rate follows from observational results and has been widely used in previous works to separate star forming from quiescent systems (e.g., \citealt{Fontanot, Donnari, paspaliaris} and references therein). We note that the stellar mass used to apply this criterion is that defined within a spherical aperture of $30$ kpc, according to the discussion in \S\ref{geometric_params}.

 {As we possess both the morphological and the $sSFR$ information, we combine these into a single classification scheme:}
%In summary, 
1) for a galaxy to be classified as an ETG, it must satisfy the condition: $n_i \geq 2$ and $\log(sSFR~[\text{yr$^{-1}$}]) < -11 $, where $n_i$ is the Sérsic index corresponding to the projection along the $i~(=X,~Y,~Z)$ axis; 2) 
%On the other hand, 
for a galaxy to be classified as an LTG, 
%it must meet the requirement: 
we impose that $\log(sSFR[\text{yr$^{-1}$}]) > -11$ and have at least one of the projected Sérsic indices smaller than $2$. 

In Fig. \ref{ssfr_classification} we show an illustration of our classification scheme taking as reference galaxies projected along the $X$ axis of the simulation box. The second quadrant represents the galaxies with typical Sérsic indices less than two and with higher specific star formation rate, while the fourth quadrant represents galaxies with Sérsic indices greater than two and with low specific star formation rate. The first and third quadrants are galaxies that do not satisfy 
%any of the two 
 {both the} criteria we imposed, making their %morphological 
classification 
%inconclusive, 
 {more uncertain, hence} we chose to not include them in any of the two categories. {Notice that galaxies classified as "LTG" in this particular projection illustrated in the Figure, may be classified as "Uncertain" in any of the other projections and the same applies to the "ETGs" in the figure}. 
 
 By applying  the above-mentioned criteria, we established a reliable and robust sample of ETGs, as demonstrated in  Fig. \ref{sersic_distro} for the Sérsic index distribution of galaxies for $0 \leq z \leq 0.1$, also taking as reference the projections along the $X$ axis of the simulation box.  {From this figure, we see that it still remains an overlap between ETGs and LTGs, if looking at the $n-$index in only one direction, which comes from the fact that the ``true'' class is based on the ``morphological'' criterion in the three directions. Hence, this shows that LTGs might still produce a misleading $n-$index in one of the directions, while ETGs, that are the main target of this work, being based on $n>2$ criterion on the three directions has a more robust morphological selection. The final sample of selected ETGs contains 4\,001 galaxies, seen in the three projections, for a total of 12\,003 {\it virtual-ETG} systems.} The general properties and range of the main structural parameters of the sample are summarized in Table \ref{range_parameters_before_and_after_dataclean} (left column).

\begin{table}
    \centering
    \caption{Interval where the fitted parameters are defined. In general, the parameters on the clean dataset have a slightly small range with respect to the ones in the full dataset.}
    \begin{tabular}{cc}
    \hline
    \multicolumn{2}{c}{Range of the parameters} \\   
    \hline
    Full dataset (\# of gal $= 12003$) & Clean dataset (\# of gal $= 10121$) \\
    \hline
     $2.00 \leq n \leq 14.00$    & $2.00 \leq n \leq 14.00$ \\
      $0.34 \leq \log R_{\rm e} \leq 2.38$   & $0.34 \leq \log R_{\rm e} \leq 1.40$ \\
       $2.02 \leq \log I_e \leq 6.42$ &  $4.80 \leq \log I_e \leq  6.34$\\
      $0.00 \leq K \leq 1.00$ & $0.27 \leq K \leq 1.00$\\
      $2.60 \leq \gamma_{\star} \leq 4.29$ & $2.64 \leq \gamma_{\star} \leq 3.44$\\
      $1.35 \leq \gamma_{\rm DM} \leq 2.15$ & $1.41 \leq \gamma_{\rm DM} \leq 2.1$\\
      $1.46 \leq \gamma_{\rm tot} \leq 2.68$ & $1.61 \leq \gamma_{\rm tot} \leq 2.42$ \\
    \hline
    \end{tabular}
\label{range_parameters_before_and_after_dataclean}
\end{table}

\subsection{Mass density profiles}\label{density_profiles}

In this section, we consider the 3D distribution of matter (dark matter, gas, and stars) in each galaxy to construct the radial density profiles   {for the sample of ETGs selected in \S\ref{classification}}. % In summary, each galaxy's 3D particle distribution is decomposed into three independent 2D projections and each of these 2D projections are associated to the same value of   {3D mass and density}. 
The method used here to construct these 3D radial density profiles is similar to \citet{Y.Wang}, and the procedure is briefly summarized below. 

We adopt a simple power-law density profile expressed by
\begin{equation}
\rho(r) =  \rho_0 r^{-\gamma},
\label{gen_SIS_model}
\end{equation} 
where $\rho_0$ is a proportionality constant. Theoretically, assuming a flat circular velocity in the galaxy centres, one expects the central total density profiles of ETGs to have a $r^{-2}$ dependence, 
%which characterizes the density distribution of 
e.g., of the form of a singular isothermal sphere (SIS) model (see, e.g., \citealt{Mo}),   {but observations show that this radial dependence varies around the canonical value of $\gamma = 2$ (e.g., \citealt{Tortora+2014, Cappellari_2015, Derkenne})}.   {For this reason, a generalization of the SIS model -- i.e., as in Eq. (\ref{gen_SIS_model}) -- is more suitable for describing the central matter distribution of galaxies}.
%This power-law model has been shown to be effective in describing the central total matter (dark matter $+$ stars $+$ gas) density characteristics of ETGs. 
Adopting this generalization,  {for all mass components (i.e., dark matter, gas, stars),} 
%and using the available data of dark matter, gas, stellar masses, and their respective positions, 
we can model the total radial density distribution and break this into the three main components, for all TNG ETGs. As this mass density modelling is confined in the very central regions, it is reasonable to use a simple power-law for all of them and so quantify the different slopes as a proxy of the different concentrations of each one of them, keeping in mind that these models are far from being an accurate mass model at all distances from the centre. 
 {Coming to the details of this mass density analysis, similarly to the procedure applied to the construction of the surface brightness profiles, we start by} 
%Our approach involves 
partitioning the radial 3D matter distributions into $100$ evenly spaced logarithmic bins. Each bin representing a spherical shell, and each shell being concentric with the ``galaxy'', whose 3D centre is defined as the position of the particle with minimum gravitational potential,  {(differently from the centres of the 2D stellar distribution defined as the centroid of the light distribution)}.  {The reason to choose a spherical symmetry for the total mass is that the total potential of galaxies is dominated by the dark component, which is eventually more spherical than the baryonic component \citep{Price+2022}.} Again, similarly to what is done in \cite{Y.Wang}, we adopt bins spanning an interval ranging from $0.4r_{1/2}$ to $4r_{1/2}$, where $r_{1/2}$ is defined as the stellar half-mass radius, i.e., the 3D radial distance from the galactic centre at which half of the total stellar mass is enclosed.  {We note here that for the density profiles, there is no need to impose the softening length as a lower limit, unlike the Sérsic profile fitting, since in our sample the minimum value of $0.4r_{1/2}$ is $0.9$ kpc.}

\begin{figure}
%    \centering
    \hspace{-0.5cm}
    \includegraphics[width=0.53\columnwidth]{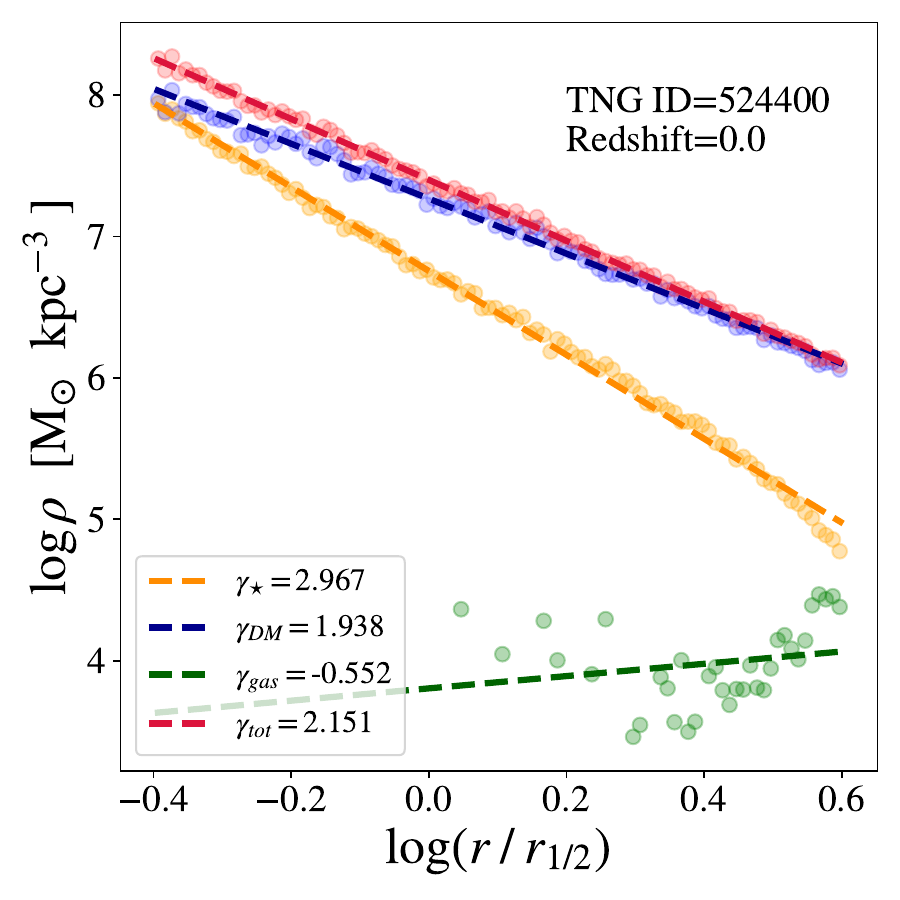}
    \includegraphics[width=0.53\columnwidth]{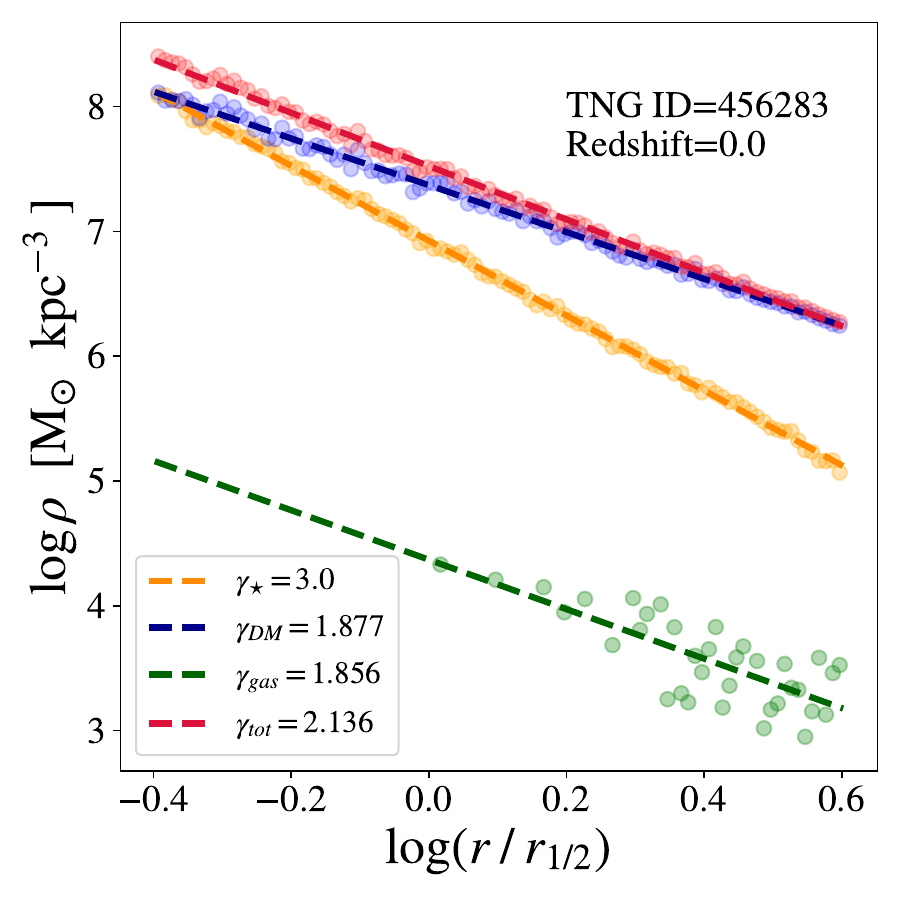}

    \hspace{-0.5cm}
    \includegraphics[width=0.53\columnwidth]{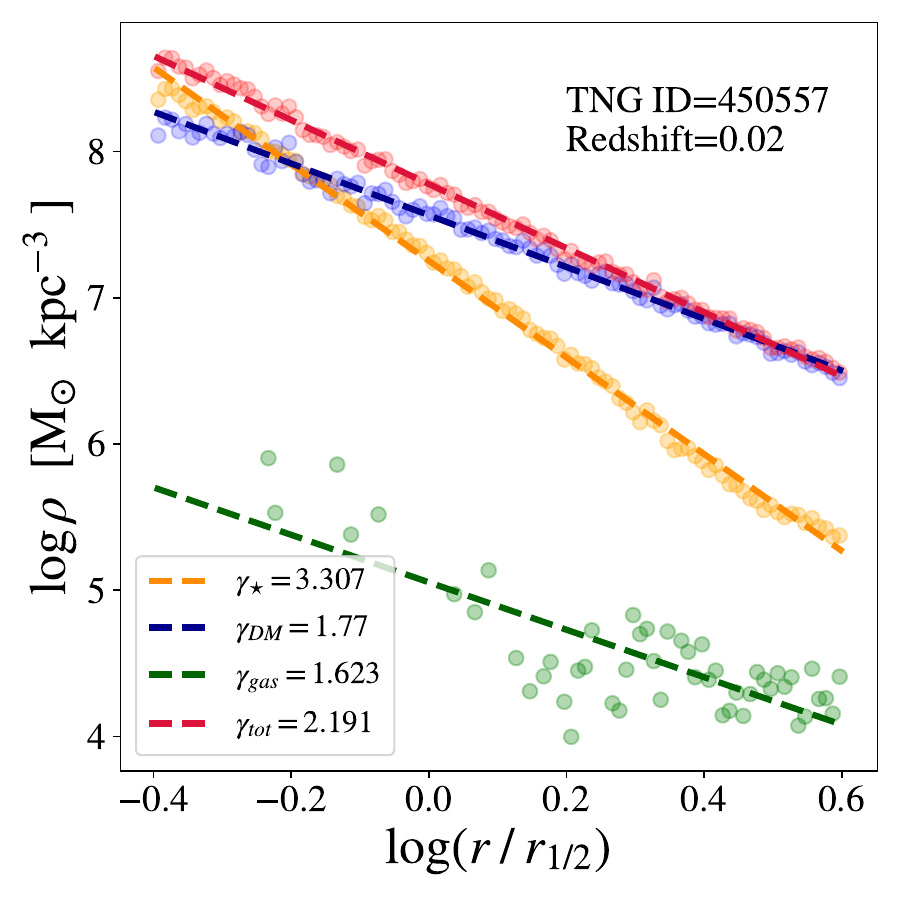}
    \includegraphics[width=0.53\columnwidth]{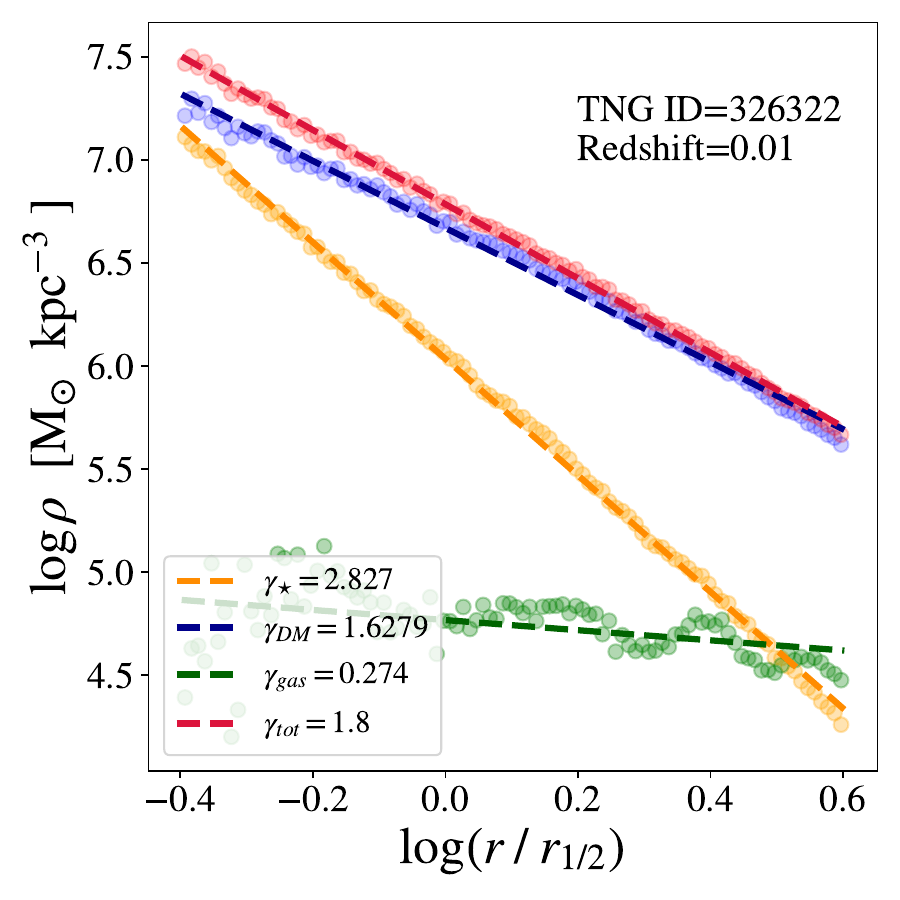}

    \caption{Radial density distributions for the various components of a pair of TNG ETGs, along with their best-fit lines (dashed lines) and slope values.   {The quantities $\gamma_{\star}$, $\gamma_{\rm DM}$, $\gamma_{\rm gas}$ and $\gamma_{\rm tot}$ are, respectively, the slope of the stellar component, dark matter, gas, and total matter}. In general, the gas component does not have a clear linear behaviour as the other components. The TNG ID and redshift of these particular objects are also shown in each panel.}
    \label{fig_density_profiles}
\end{figure}

We then perform a least-squares linear regression with equal bin weighting to the log-scaled density profile:
\begin{equation}
  \log \rho = -\gamma \log r + \log \rho_0, \quad r \in [0.4r_{1/2} , 4r_{1/2}]. \label{power_slope_diff_form}
\end{equation}
This approach yields the best-fit slope $\gamma$ for each galaxy. As an illustration, in Fig. \ref{fig_density_profiles} we present the radial density profiles of the three components for two particular galaxies, along with their best-fit lines and slope values. 
 {However, in order to more closely match the $\gamma$ estimates from observations that are usually performed in a radial range shorter than $4r_{1/2}$ (see \S\ref{density_slopes_results}), we also fit Eq. (\ref{power_slope_diff_form}) in the range $[0.4r_{1/2} , 2r_{1/2}]$. We will use for this latter definition the symbol $\gamma_2$ to distinguish from the one in Eq. (\ref{power_slope_diff_form}).}
We observe that, in general, the gas component is not nicely described by a linear relation between $\log\rho_{\rm gas}$ and $\log(r/r_{1/2})$ as the other components. This may be explained by the noisy radial gas distribution due to the low gas fraction in ETGs. 
%Therefore, the gas component should be better characterized by its total mass content, rather than its density profile slope. 
%
  {Hence, the density slopes of the gas will not be used as a proxy of the 
%do not reliably indicate the 
radial gas  distribution of the ETGs sample}. 

%For the density profiles we do not have to worry about profiles reaching radial values below the softening length, as in our sample, the minimum value for $0.4r_{1/2}$ is $0.9$ kpc.
% {For LTGs....}

\subsection{Systematics, outliers, and data cleaning}
%\subsection{Systematics}
%\label{sec: systematics}
%In this section we discuss some possible source of systematics introduced by the main assumptions in our procedure. This is particularly important before we collect the final catalogue of virtual-ETGs. 
%\subsubsection{Outliers and data cleaning}
\label{data_cleaning}
 {As introduced in \S\ref{sbp}, in  this section we quantify the effect of the assumptions made to model the virtual-ETG sample,
%Inevitably, the procedures used in this work, as the assumption that the galaxies' radial brightness distribution are well described by a 
in particular, the single-Sérsic profile and the 3D power-law density profiles.
%(\S\ref{sbp}), 
Other assumptions that might also produce systematic effects are }the methods adopted to measure the luminosity and shape parameters (\S\ref{geometric_params} and \ref{sbp}), 
and even the criteria used to classify the galaxies (\S\ref{classification}).
%are all susceptible to shortfalls. 

 {One physically motivated way to pinpoint the presence of systematic effects is provided by the scaling relations (see \S\ref{sec:intro}), as anomalous estimates might be spotted as outliers in the smooth correlations among the parameters.}
 {Since the definition of outliers is a statistical operation, which depends on the methodology one decides to adopt, we have decided to apply this ``data cleaning'' step to the full released catalogue produced in \S\ref{catalogue_section}, leaving it up to the generic user to apply further selection criteria if needed.} 
%With that in mind, some correlations naturally expected from observational data may be mitigated due to the impact of outliers added during the construction of the parameters. 

 {In our analysis, to classify outliers 
%based on the galaxy properties' 
deviating from some standard scaling relation,} 
%the influence of the outliers and hence, preserve as much as possible the expected physical correlations between the parameters included in our catalogue,  
we apply a statistical method based on Mahalanobis distance \citep{Mahalanobis} and on the Minimum Covariance Determinant approach (MCD for short; e.g., \citealt{Rousseeuw1984871, Rousseeuw1985}). 

Let $\mathbf{X}$ be a vector belonging to a multivariate distribution with mean $\mathbf{\mu}$ and covariance matrix $\mathbf{\Sigma}$. The Mahalanobis distance $D_M(\mathbf{X})$ of $\mathbf{X}$ from the distribution is defined as:
\begin{equation}
    D_M(\mathbf{X}) = \left[ \left(\mathbf{X}-\mu \right)^{T} \mathbf{\Sigma}^{-1} \left(\mathbf{X}-\mu \right)\right]^{1/2}, \quad \mathbf{X}\in \mathbb{R}^{d}, d \in \mathbb{Z}^+,
\end{equation}
where $d$ is the dimension of the distribution. In other words, $D_M$ can be interpreted as the distance of each point from the mean $\mu$ of the distribution. However, this criterion uses the multivariate sample mean and covariance matrix that are particularly sensitive to outliers \citep{Ghorbani}. Thus, we adopt a robust way of computing the sample mean and covariance using the MCD approach, which consists in finding a fraction $h$ of multivariate data points that are not considered to be outliers and then
compute the sample mean and covariance from this subsample. This robust distance  ($D_R$) indicator is defined as (see, e.g., \citealt{Hubert,Rousseeuw&Zomeren}):
\begin{equation}
    D_R(\mathbf{X}) = \left[ \left(\mathbf{X}-\mu_{\text{MCD}} \right)^{T} \mathbf{\Sigma}_{\text{MCD}}^{-1} \left(\mathbf{X}-\mu_{\text{MCD}} \right)\right]^{1/2},
\end{equation}
where $\mu_{\text{MCD}}$ and $\mathbf{\Sigma}_{\text{MCD}}$ are the mean and covariance estimated with the MCD approach. With this definition, for a point in the multivariate distribution to be considered as an outlier, it must satisfy $D_R > c_k$, where $c_k = \sqrt{\chi^2_{k, 1-\alpha}}$. Here, $\chi^2_{k, 1-\alpha}$ represents the critical value of the chi-squared distribution with $k$ degrees of freedom at the $(1-\alpha)$ significance level. This criterion is used because the squared Mahalanobis distance, when calculated using the Minimum Covariance Determinant (MCD) method, approximates a chi-squared distribution \citep{Li2019OutlierDetection}. For our work, we chose $1-\alpha = 99\%$.

To optimize the estimation of the covariance and sample mean, we chose to remove the outliers in relation to a lower-dimensional space spanned by the set $\{n, \log R_{\rm e}, \log I_e, K, \gamma_{\rm star}, \gamma_{\rm DM}, \gamma_{\rm tot}\}$, where the Sérsic parameters are those extracted using the $r_{30\rm kpc}$ cut and the density slopes are the ones computed over the radial range $[0.4, 4]r_{1/2}$. We chose this set of parameters because: \textit{i)} the Sérsic parameters are  {the ones the we concern the most to be a source of bias, for the reasons discussed in \S\ref{sbp}},
%probably the greatest source of errors, as 
and because they come from a non-linear regression task, 
\textit{ii)} the slopes are still prone to failures in the whole measurement procedure, even if they might not suffer from other clear biases as the softening length cut;
%, and they are obtained via a linear regression task, they
hence, 
%the errors from the procedure, and as they are 
being the most physically significant parameters newly derived in this analysis,
%in what concerns the density profiles, we only include them and not the offsets for the outlier removal task and, finally, 
we include them in the definition of the outliers and their removal;
\textit{iii)} 
%this choice also assumes that 
the parameters extracted directly from the TNG simulations and the stellar population synthesis (e.g.,$M_{\rm tot}, M_{\rm tot}(r_{1/2}), \Upsilon_{\star}$, etc.) 
%are the \textit{ground truth} and do not carry the same level of errors as the derived ones.
  {correspond to a \textit{ground truth} and do not carry the same level of error/bias as the  {parameters derived by the fitting procedures, i.e.} 
%ones in the set 
$\{n, \log R_{\rm e}, \log I_e, K, \gamma_{\rm star}, \gamma_{\rm DM}, \gamma_{\rm tot}\}$.} It should be clear that the choice of the Sérsic parameters derived using the $30$ kpc cut and the slopes within $4r_{1/2}$ to remove the outliers is driven by the fact that these are regarded as the canonical parameters in this work, serving as the primary metrics used for galaxy classification, being also the parameters that best describe the structure of the galaxies in our sample. % and for comparison to observations.

The application of the Robust Mahalanobis distance to our sample of galaxies provides a reliable method of detecting data points that significantly deviate from the norm, but we point out that the choice of threshold $c_k$ is mostly determined by eye and there is no ideal value for $D_R$, but rather a trade-off between data quality and quantity: if $D_R$ is too small, the range of each parameter is also constrained, meaning that \textit{regular} data points will be interpreted as out-of-trend points and excluded from the catalogue, and on the other hand, if $D_R$ is large, very few outliers will be removed, and the final catalogue will have a larger amount of samples but with a higher amount of outliers. Thus, the choice $D_R> \sqrt{\chi^2_{k, 99\%}}$ for the detection of outliers in the data is essentially made with the purpose of having a catalogue as clean as possible and with an appreciable number of samples. 

%\begin{figure}
%    \centering
%    \includegraphics[width=0.75\columnwidth]{figures/Re_SB_Re_30kpc.pdf}\\
%    \hspace{-0.2cm}
%    \includegraphics[width=0.76\columnwidth]{figures/Lr_SB_Lr_30kpc.pdf}\\
%    \includegraphics[width=0.75\columnwidth]{figures/n_SB_n_30kpc.pdf}
%    \caption{$R_e$ (upper row), $L_{\rm tot}$ (middle row) and $n$-index (lower row) of the two upper radius cut ($r_{\rm 30kpc}$ and $R_{26.5g}$), face to face.}
%    \label{fig:rmax_cut_comparison}
%\end{figure}

After the outlier removal task, $1882$ galaxies were selected out and the range of the parameters became more constrained. This is clearly shown in Appendix \ref{Appendix_scatter_matrix}, where we demonstrate the robustness of the data cleaning method by comparing the scatter matrix and distributions for the set $\{n, \log R_{\rm e}, \log I_e, K, \gamma_{\rm star}, \gamma_{\rm DM}, \gamma_{\rm tot}\}$ before and after the process.   {The general statistical properties for all the parameters included in the catalogue can be found in Appendix \ref{appendix_catalogue_stat_properties}.} In Table \ref{range_parameters_before_and_after_dataclean}, we compare the changes in the range of the quantities used for the outlier removal, before (left column) and after (right column) the data cleaning: the selected sample of $10121$ virtual-ETGs will be included in the final catalogue.

%Moreover, beyond the simulation issues discussed previously, additional systematics may stem from the incomplete representation of galaxy sizes in our sample for $\log R_{\rm e} \lesssim 0.4$ and $\log R_{\rm e}\gtrsim 1.4$ (see, e.g., Table~\ref{range_parameters_before_and_after_dataclean} and Fig. \ref{fig:rmax_cut}). This deficit of smaller galaxies is likely a consequence of our imposed stellar mass cut (at $\log M/M_\odot<10.5$) as well as Sérsic fittings.

%Particular values of $n$ in Eq.~(\ref{sersic_profile}) include $n=1$ that is the so-called exponential profile, and characterizes the surface brightness distribution of disky or other light-diffuse systems. For $n=4$ we have the De Vaucouleurs profile \citep{DeVaucouleurs}, that is better suited for the description of spheroidal systems, such as bulges and elliptical galaxies. Although these two models are particularly good at describing such systems, it is observed that, in general, the surface brightness of a particular galaxy is better described by Eq.~(\ref{sersic_profile}). However, it is common to adopt these two particular cases as a further step in galaxy morphological classification (e.g., \citealt{Y.Wang, D.Xu}), combined with the Sérsic index filtering. In this work, we do not use this step as a classification criterion, as we are confident that $sSFR$ and Sérsic index combined are enough to give a reliable distinction between LTGs and ETGs.

\subsection{The catalogue}
\label{catalogue_section}
\begin{figure}
    \centering    \includegraphics[width=\columnwidth]{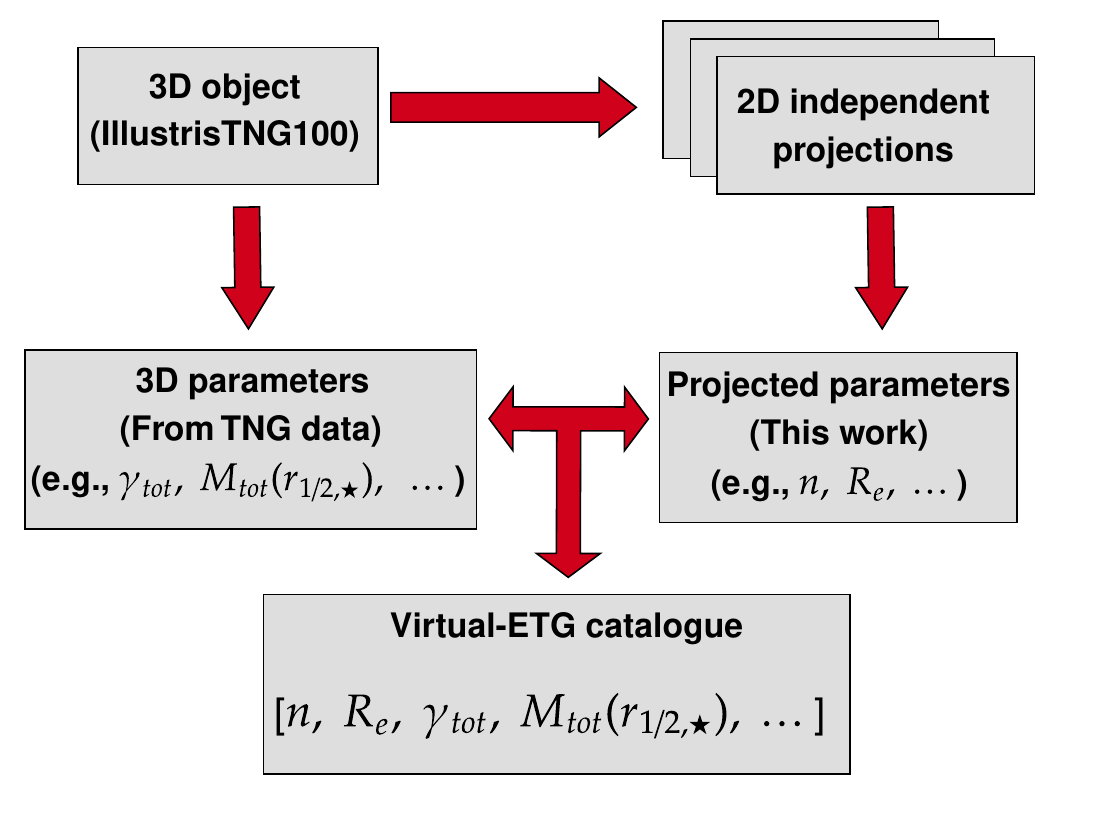}
    \caption{Workflow for the catalogue construction.   {Except for $\gamma_{\rm tot}$, luminosity information and masses within 2D radii}, the 3D quantities are derived or exported from the information contained on the TNG \texttt{SUBFIND} catalogue, while the 2D parameters are the ones obtained in this paper.}
    \label{fluxogram}
\end{figure}

In this section, we describe the quantities listed in the catalogue of selected ETGs, which is attached to this paper. Fig. \ref{fluxogram} provides a summary of the workflow used to estimate the various parameters and illustrates how they are assembled into the final catalogue. Our TNG catalogue includes both projected and 3D parameters. One of the objectives of this work is to offer a sample of TNG ETGs with properties closely resembling those found in observational datasets, enabling a meaningful comparison with existing literature. To achieve this, we incorporate commonly used observational parameters from both photometric and spectroscopic analyses, such as luminosity and velocity dispersion, alongside 3D quantities, such as central masses and density profiles. Below, we present the main parameters included in the catalogue.

%NRN: from here
% {In this section, we describe the quantities listed in the catalogue of selected ETGs, attached to this paper. }
%In Fig. \ref{fluxogram}, we summarize the workflow to estimate the different and how these are assembled in the final catalogue. Our TNG catalogue is composed of both projected and 3D parameters. As one of the goals of this work is to provide a sample of TNG ETGs with properties similar to observational datasets,   {in order to compare their properties to the literature}, we include common observational parameters from both photometric and spectroscopic analysis, as luminosity and velocity dispersion, along   {with 3D quantities,} as central masses and density profiles. Below, we introduce the main parameters included in the catalogue.
%discuss the main steps involved in the construction of our catalogue, as well as the description of its main parameters. The flowchart in Fig. \ref{fluxogram} summarizes the steps taken to data gathering and assembly in the final catalogue.

%We selected only central TNG galaxies whose stellar masses span in the log-scaled interval $[10.3-11.9]~\log(M_{\odot})$. With the 3D matter distribution, as described in section \ref{density_profiles}, we found the slopes of the density profiles for the gas, dark matter and stellar components. We then stored the values along with the galaxy's total mass (dark matter + stars + gas), which is a quantity provided by the simulation.

%\subsubsection{Parameters} 

%\paragraph
\subsubsection{Three-dimensional quantities}
\label{sec:3D_quantities}
\begin{enumerate}
    \item Stellar half-mass radius ($r_{1/2}$): defined as the 3D radius of the sphere that encloses half of the simulated total stellar mass, where the latter is defined as the total contribution of mass from all the stars assigned to a subhalo by \texttt{SUBFIND}. It is stored in the TNG \texttt{SUBFIND} catalogue in \texttt{SubhaloHalfmassRadType} for \texttt{PartType4}.
    
    \item Aperture $r$-band luminosities from \texttt{FSPS} ($L_{\rm 3D}(< r)$): the total 3D luminosity used here is, by definition, the contribution of the luminosity from all the stellar particles enclosed within a sphere of radius equals to $30$ kpc concentric with the galaxy. We also include two central luminosity definitions for this case, namely the 3D light inside one effective radius $L_{\rm 3D}(1R_{\rm e})$ and the 3D light inside two effective radii $L_{\rm 3D}(2R_{\rm e})$,   {where the effective radius $R_{\rm e}$ is defined in \S\ref{sbp}.} 

    \item Aperture stellar masses ($M_{\rm 3D,\star}(< r)$): similar to the luminosities defined above, these parameters represent the 3D stellar masses in three different apertures. According to the 3D radial cut performed to avoid a possible contribution of intracluster light (see, e.g., \S\ref{geometric_params}), the total stellar mass is defined here as the sum of the mass of all stellar particles within a sphere of $30$ kpc concentric with the galaxy. {We also include central stellar masses within $xR_{\rm e}$ ($x=1,2$). These 2D radii apertures were defined in order to compare them with their projected counterparts (see \S\ref{pararaph:projected_quantities} and \S\ref{projected_versus_3d_stellar_mass}).} 
    
    \item Total mass ($M_{\rm tot}$): the total (DM + stars + gas) mass of each galaxy. It is defined as the total contribution of mass from all the types of particles assign to this subhalo. This parameter comes straight from the simulations, where it is referred to as \texttt{SubhaloMass}.
    
    \item Total dark matter mass ($M_{\rm DM}$): the total mass of the dark matter content of each galaxy. It is extracted from the simulations and its definition is similar to $M_{\rm tot}$, except by the obvious fact that it only considers dark matter particles. It is stored in the TNG100 {\tt SUBFIND} catalogue in \texttt{SubhaloMassType} for \texttt{PartType1}.

    \item Total 3D stellar mass, ($M_{\rm 3D,\star}^{\rm tot}$): this is simply derived as  $M_{\rm 3D,\star}^{\rm tot} = M_{\rm tot}- M_{\rm DM} - M_{\rm gas}$. All these quantities are derived from the TNG100 {\tt SUBFIND} catalogue\footnote{  {For a projected total stellar mass definition, a straightforward approach is given by Eq. (\ref{luminosity}), that gives a logical way of estimating the modelled luminosity of a galaxy at any radius \citep{Trujillo}. Formally, the total luminosity of a galaxy is defined at an infinite extent, $r \to \infty$ in Eq. (\ref{luminosity}). Based on this total luminosity definition, the total stellar mass of a galaxy can be approximated as $M_{\star}(r \to \infty) = \Upsilon_{\star} L(r \to \infty)$. One should not expect this stellar mass definition to equalize the total stellar mass $M_{\rm 3D,\star}(<30~\text{kpc})$, as the latter is truncated at a finite radius while the former allows for an analytical extrapolation up to infinity. In fact, based on this extrapolation, $M_{\rm 2D,\star}^{\rm tot}(<\infty
   % r_{max}
    )$ should be greater than $M_{\rm 3D,\star}(<30~\text{kpc})$, specially at the bright end, where we would expect a greater impact of the intracluster light.
    %, but as part of the light contribution was ruled out by performing the spherical $30$ kpc cut, then the projected stellar mass will overcome its 3D counterpart by a higher amount than for fainter galaxies. 
    %For this reason, we do not compare the total luminosity and stellar mass computed from the 3D distribution of particles with their 2D counterparts. Once this \textit{outskirt} problem is assumed to have negligible effect on the central regions of galaxies, 
    With that in mind, we only discuss the comparison between the aperture luminosity and stellar mass definitions and their projected counterparts 
    %is studied here and the main findings are discussed 
    in \S\ref{projected_vs_3d_light} and \S\ref{projected_versus_3d_stellar_mass}.}}. 

    \item Total gas mass ($M_{\rm gas}$): the total gas content. It is defined as the mass contribution of all the gas particles assigned to the galaxy. It is a quantity extracted directly from the simulations and is included for sanity check. It is found in the TNG database in \texttt{SubhaloMassType} as \texttt{PartType0}.

    \item  Mass at stellar half-mass radius ($M_{\rm tot}(r_{1/2})$): this parameter represents the total mass of the galaxy enclosed in a sphere of radius $r_{1/2}$, defined as the radius at which half of the total stellar mass is confined. It is stored in the TNG catalogue as \texttt{SubhaloMassInHalfRad}.

    \item  Dark matter mass at stellar half-mass radius ($M_{\rm DM}(r_{1/2})$): similar to $M_{\rm tot}(r_{1/2})$, but to DM particles only. We chose to include these central masses definitions because they are more common in the \textit{real} galaxy catalogues than the total quantities (e.g., \citealt{Zhu}). This is found in the TNG database in \texttt{SubhaloMassInHalfRadType} for \texttt{PartType1}.

    \item Stellar mass at the stellar half-mass radius ($M_{3D,\star}(r_{1/2})$): similarly to the two last items, this parameter is defined as the total mass of stars inside a $r_{1/2}$ spherical aperture. It is stored in the TNG catalogue in \texttt{SubhaloMassInHalfRadType} for \texttt{PartType4}.
    
    \item Gas mass at the stellar half-mass radius ($M_{\rm gas}(r_{1/2})$: total gas mass inside $r_{1/2}$. Stored in the simulations in \texttt{SubhaloMassInHalfRadType} for \texttt{PartType0}.
    
    \item Total density slope ($\gamma_{\rm tot}$) and central density ($\rho_{\rm 0, tot}$): the trend of the inner total (DM+stars+gas) density profile of the ETGs. It is a quantity estimated in this work, assuming a density profile that scales as $\log(\rho) = -\gamma \log(r) + \log \rho_0$ (see Eq. (\ref{power_slope_diff_form})).  {In the catalogue, we report the slope obtained for two different radial intervals: $\gamma\rightarrow r \in [0.4,4]r_{1/2}$ and  $\gamma_2\rightarrow r \in[0.4,2]r_{1/2}$} (see \S\ref{density_profiles}). 
    
      {Another useful measure is the average central density slope, estimated within a spherical shell of inner radius $r_1$ and outer radius $r_2$ as (see \S\ref{sec:slope_definitions}): 
    \begin{equation}
        \gamma^{\rm AV}(r_1, r_2)  = \frac{\log[\rho(r_2)/\rho(r_1)]}{\log(r_2/r_1)}.
    \end{equation}
    We also include this slope definition in the catalogue computed over the radial range $r \in[0.4,2]r_{1/2}$. These slopes, calculated over smaller radial ranges, enable more consistent comparisons with observational data, as they are more closely aligned with the definitions used for real systems. Further details are provided in \S\ref{density_slopes_results}.} For the other components (see below), the same $\gamma$ definitions (i.e., $\gamma_{2}^{\rm}$ and $\gamma^{\rm AV}$) are also computed.
    
    \item Dark mater density slope ($\gamma_{\rm DM}$) and central density ($\rho_{\rm 0,DM}$): the slope of the dark matter component, defined similarly to $\gamma_{\rm tot}$, but considering only the inner density profile of the dark component in the galaxies. 

    \item Stellar density slope ($\gamma_{\star}$) and central density ($\rho_{\rm 0,\star}$):  the slope of the stellar component, defined in the same manner as $\gamma_{\rm tot}$ and $\gamma_{\rm DM}$, but describing the trend of the inner density profile of the stellar component of the galaxies. As shown in \S\ref{density_slopes_results}, this is steeper than the other slopes. 
    
    \item Gas density slope ($\gamma_{\rm gas}$): the slope of the gas component $\gamma_{\rm gas}$. Although this is not a robust estimate (see \S\ref{density_profiles}) we keep them in the catalogue for completeness. It is defined similarly to the other components.
    %, but only focus on stellar, dark matter and total slopes. 
    
    %\item Density profile offsets: for completeness, we also include the offsets of the linear fits to the density profiles for each component, so if one wants to retrieve the density slope $\rho = \rho_0 r^{-\gamma}$, with the correct normalization $\rho_0$, one should use:
    %\begin{equation}
     %   \rho(r) = 10^{\beta}\left(\frac{r}{r_{1/2}} \right)^{-\gamma} h^{-\gamma},
    %\end{equation}
    %where $\gamma$ is the slope of a particular component, $\beta$ is its offset and $h$ the reduced Hubble constant. The factor $ h^{-\gamma}$ comes from the fact that the linear fittings were done with $r_{1/2}$ in code units, i.e., kpc/$h$. Thus, as it is stored in the catalogue in physical units (kpc), this term appears as the conversion factor for the units.
    \item Star Formation Rate ($SFR$): star formation rate of the galaxy, stored in the TNG database as \texttt{SubhaloSFR}.
    
    \item Stellar mass-to-light ratio ($\Upsilon_{\star}(r)$): this quantity is defined as being the ratio of the three-dimensional stellar mass and $r$-band luminosity within a 3D distance $r$ from the centre, i.e., $\Upsilon_{\star}(r) = M_{\rm 3D,\star}(r)/L_{\rm 3D}(r)$. In particular, we've defined $\Upsilon_{\star}(r)$ inside a $r = 30 \rm kpc$ aperture, however, given the 3D central stellar masses and luminosities, the generic user can also define this quantity inside $r = R_{\rm e}$ and $r= 2R_{\rm e}$. These various apertures are particularly useful to access eventual variations of the stellar mass-to-light ratio as a function of the radial distance from the galactic centre. 
    %$30$ kpc from the galactic centre $M_{\rm 3D,\star}(<30~\rm kpc)$ and the three-dimensional $r$-band luminosity within $30$ kpc, $L_{\rm 3D}(<30~\rm kpc)$, derived using \texttt{FSPS} (see \S\ref{fsps}), i.e., $\Upsilon_{\star}(r = \rm 30 kpc) = M_{3D, \star}(<30~\rm kpc)/ L_{3D}(<30~\rm kpc)$. We 
\end{enumerate}

\subsubsection{Projected quantities}
\label{pararaph:projected_quantities}
\begin{enumerate}
    \item Axis ratio ($K=b/a$): this parameter is defined in terms of the second moments of luminosity (see \S\ref{geometric_params}).
    %, and represents how much the galaxies' projected light distribution deviates from a circular shape. If it is close to one, then the light distribution is closer to a circular shape. On the other hand, if it is close to zero, then the light distribution tends to have a more flattened elliptical shape. It is clearly a dimensionless quantity.
    
    \item Sérsic index ($n$): one of the parameters of the Sérsic profile model of the 1D surface brightness profile, as described in \S\ref{geometric_params}. It describes the inner slope of the 2D light density profile.
    %the Sérsic index gives information about how concentrated or diffuse the light profile of a galaxy can be. 
   % Galaxies with higher $n$ have more concentrated luminosity distributions, and galaxies with smaller $n$ have more diffuse luminosity distributions. The Sérsic index is also a dimensionless parameter.
    
    \item Effective Radius ($R_{\rm e}$): it represents the half light radius and it is also obtained via Sérsic profile fitting as described in \S\ref{geometric_params}. 
    %As shown in \S\ref{sec:max_rad}, the effective radius is almost equivalent regardless the outermost radius selection ($r_{\rm 30kpc}$ or $R_{26.5g}$). Hereafter, we use the radii obtained from the 30 kpc cut for all the projected quantities inside the $R_e$, unless otherwise specified.
    The $R_{\rm e}$ parameter should not be confused with the half-mass radius. According to \S\ref{sbp}, the former represents the radius from the galactic centre at which half of the total luminosity is contained, while the latter is the radius at which half of the mass (for a particular component, their sum, etc.) is contained.

    \item Sérsic $r$-band Luminosities ($L_{\rm 2D}(<R_{\rm e})$, $L_{\rm 2D}(<2R_{\rm e})$ and $L_{\rm 2D}^{\rm tot}$): the integral of the surface brightness profile (up to $xR_{\rm e}$, for $x=1,2$ and $x \to \infty$) measured with respect to the SDSS $r$-band (see \S\ref{sbp}). %Note that the three Sérsic parameters above ($n$, $R_{\rm e}$, and total luminosity), will be derived using both a 30 kpc cut and the 26.5 mag/arcsec$^2$ surface brightness limit (see \S\ref{sbp}).  
    %As it comes from the projected  Sérsic profile, it is clearly a projected quantity. It assumes that the light distribution is well described by a one-component Sérsic profile, i.e., it does not take into account any possible features that may be present on the true projected light profiles.
    
    \item Projected stellar masses ($M_{\rm 2D,\star}(R_{\rm e})$, $M_{\rm 2D,\star}(2R_{\rm e})$ and $M_{\rm 2D, \star}^{\rm tot}$): these parameters are estimates of the  stellar content of the galaxies within different apertures, assuming that for each, there is a constant mass-to-light ratio $\Upsilon_{\star}$ (see its definition above). For such, we use $\Upsilon_{\star}$ inside a 3D aperture of $30$ kpc\footnote{We also used $\Upsilon_{\star}$ within apertures of $R_{\rm e}$ and $2R_{\rm e}$ to define the projected stellar mass. However, given that the average variation on the stellar mass estimates is only about $0.02$ dex, we chose to retain $30$ kpc as our baseline.}. %Using the mass information of the individual stellar particles belonging to each TNG-galaxy,
    %We define the quantity $\Upsilon_{\star}$ as the ratio of the three-dimensional stellar mass  within $30$ kpc from the galactic centre $M_{3D, \star}$ and the three-dimensional $r$-band luminosity within $30$ kpc, $L_{3D}$, derived using \texttt{FSPS} (see \S\ref{fsps}), i.e., $\Upsilon_{\star} = M_{3D, \star}/ L_{3D}$. 
    Then, the projected stellar mass is estimated as $M_{\rm 2D,\star}(xR_{\rm e}) = \Upsilon_{\star} \times L_{\rm 2D}(xR_{\rm e})$, for $x=1,2$ and $M_{\rm 2D,\star}^{\rm tot} = \Upsilon_{\star} \times L_{\rm 2D}^{\rm tot}$ for $x \to \infty$. %This stellar mass definition applies to both $r_{30\rm kpc}$ and $R_{26.5g}$ radial cuts.

    \item Line-of-sight (LOS) velocity dispersion  {within the effective radius aperture} ($\sigma_{\rm e}$): this parameter is computed using the stellar \texttt{Velocities} field from the IllustrisTNG simulations. For each projection,  {we use the line-of-sight velocities (i.e., the ones along the projection direction) of the selected stars within a circular aperture of $R_{\rm e}$ and excluding the stars within the softening length, i.e., $ v_i(<R'):\epsilon_\star<R'<R_{\rm e}$. The reason we exclude the regions within the softening length is that, by definition, the gravitational field is artificially smoothed and the velocity do not trace the true potential. In Appendix \ref{app:veldisp_profiles}, we show the velocity dispersion profile of a sample of galaxies with mass ranging from $10.5< \log M_\star/M_{\sun} < 11.5$, almost all exhibiting a drop of velocity dispersion values within $\epsilon_\star$. Including these particle would unphysically dilute the aperture velocity dispersion (we will refer to this as ``softened velocity dispersion'' for short, see also a discussion in Appendix \ref{app:veldisp_profiles}). We then proceed to define two types of $\sigma_{\rm e}$:}
    
    The one defined as the velocity dispersion for 
    %velocity field 
    selected star particles within $\epsilon_\star\leq R'\leq R_{\rm e}$, with respect to a given projection.  {This is simply defined as,}
    %  {To do so, we construct the velocity dispersion profile $\sigma(r)$ of each galaxy with $N = 40$ bins (similar to the procedure done to the light profiles) and then weight $\sigma^2(r)$ in each bin by its corresponding luminosity, which can written in the discrete form as:}
%    \begin{equation}
%    \sigma_{[\epsilon_\star,R_e]} \equiv \sqrt{\sum_{i=1}^{N}\frac{\sigma_i^2 L_i}{L(<R_{\rm e})}}, 
%    \end{equation}
%    where $L(<R_{\rm e})$ is the luminosity inside the effective radius and the $\sigma_i$'s are defined as the standard deviation of the velocity of the stars within the ring $R'$, i.e.,
    \begin{equation}
    \sigma_{[\epsilon_\star,R_{\rm e}]} \equiv \sqrt{\sum_{i=1}^{N}\frac{\left(v_i - \bar{v}\right)^2}{N}},
        \label{eq:sig_e_stand}
    \end{equation}
    {where $\bar{v}$ is the mean velocity of the same $v_i$ sample defined above \footnote{We have also computed a light-weighted velocity dispersion, using the ratio of the particle luminosity over the total luminosity as weight, obtaining the same distribution of $\sigma_{[\epsilon_\star,R_e]}$, consistently with the recent findings from \citet{Sohn+2024_TNG_veldis}, hence we decided to keep the simple Eq. (\ref{eq:sig_e_stand}).}.
   %  {To make its definition closer to the velocity dispersion obtained from spectroscopic measurements, we break the profile of each galaxy into $40$ bins (similar to the procedure done to the light profiles) and then weight $\sigma_{[\epsilon_\star, R_{\rm e}]}$ in each bin by its corresponding luminosity. Doing so, for each galaxy we obtain a luminosity-weighted (LW) velocity dispersion defined in a ring $\epsilon_\star\leq R'\leq R_{\rm e}$, namely $\sigma_{\rm LW}(R')$.}
   This is the most unbiased LOS velocity dispersion we can ``measure'' in simulations that fully traces the internal dynamics of the simulated galaxies and that is closely connected to the ``virial'' velocity dispersion that one should include in the galaxy fundamental plane (\citealt{deGraaff_eagle}),  {although in Eq. (\ref{eq:sig_e_stand}) we include both random and rotational motions making this definition intrinsically larger than the rotation subtracted velocity dispersions derived from galaxy spectroscopy, especially toward the low mass end ($\log M_\star/M_\odot <11$) of our sample that is usually dominated by fast-rotators (see, e.g., \citealt{2020MNRAS.491..773G}).}
   %However this is, by definition, different from the LOS velocity dispersion of observed galaxies that, instead, 
   Furthermore, real spectroscopy, yet differently from the $\sigma_{[\epsilon_\star,R_{\rm e}]}$ above, considers all stars down to the very galaxy centres, where the velocity dispersion profile of ETGs, usually show their maximum value\footnote{ {For observations, the main ``dilution'' factor is the atmospheric seeing, which is generally ignored on fiber/long-slit observations, but it is accounted for in integral field spectroscopy (\citealt{2021_Chung_PSF_spec} and references therein).}}. To take this central component, missed by Eq. (\ref{eq:sig_e_stand}), into account, we introduce a ``corrected'' $\sigma_{\rm e}$, which is fully detailed in Appendix \ref{app:sigma_corr} but
   %, to account for the central component, missed by Eq. \ref{eq:sig_e_stand} and 
   that we also report here below for sake of completeness:
    \begin{equation}
        \sigma_{\rm e,corr}^2 = \sigma_{[\epsilon_{\star}, R_{\rm e}]}^2\frac{L(\epsilon_\star < r < R_{\rm e})}{L(<R_{\rm e})}\left( 1 - \frac{L(<\epsilon_\star)}{L(<R_{\rm e})} \left(\frac{\epsilon_\star}{R_{\rm e}}\right)^{-0.132} \right)^{-1}
        \label{eq:sig_e_corr}
    \end{equation}  
   % where $\sigma^2(R_{\rm Ap}= \epsilon_\star)$ is the aperture corrected velocity dispersion obtained using the results by \citet{Cappellari+2006} starting from the inferred $\sigma_{[\epsilon_\star,R_e]}$, as in Eq. \ref{eq:sig_e_stand} %and the $L(<X)$ are the total luminosities in different radial ranges used as weights 
    %weighted by the total luminosities $L(<X)$ in the different radial ranges 
{This $\sigma_{\rm e, corr}$ is the one we will use in the analysis and comparisons with literature, and will be included in the catalogue. Thus, from now on we will drop the subscript $\rm corr$ and refer to it just as $\sigma_{\rm e}$, unless it is needed to be specified.}
   %In Fig. \ref{fig:sig_comp} we compare the distribution of the $\sigma_{[\epsilon_\star,R_e]}$ defined above and two galaxy samples, introduced in \S\ref{sec:intro}, that we will use further in this paper: the SPIDER (\citalt{tortora+14}) and the DynPop (\citaalt{dynpop}) samples. We see that the virtual-ETG is systematically lower than the two observed samples, eventually because the simulated galaxies miss the typical central peaks seen in the observed velocity dispersion profiles. To try to account for this missing ``central component'' we introduce here an heuristic correction based on standard aperture corrections applied to observed galaxies (see, e.g.,Jorgensen+, \citealt{Ziegler-Bender_apcorr}, Cappellari+2006)
 }

 \item Dynamical LOS velocity dispersion ($\sigma_{\rm e, dyn}$): as a sanity check, we include in the catalogue a dynamically derived $\sigma_{\rm e, dyn}$ of each virtual-ETGs based on the stellar luminosity and mass distribution properties reported in the catalogue and computed via the radial Jeans equation (assuming isotropic orbits), projected along the LOS and integrated in an effective radii aperture, as described in \citet[][Eqs. 2--6]{Tortora+2022}. Precisely, for the stellar component, we have used the S\'ersic profile and related quantities as described in \S\ref{sbp} with
%with Eq.~(\ref{eq:inferred_sigma}) and using 
%the photometric parameters 
$n, ~R_{\rm e},~ L_{\rm tot}$ as derived from this section's items (ii), (iii), and (iv), respectively, and, for the total mass density, the $\rho_{\rm tot}$ as in Eq. (\ref{power_slope_diff_form}) with the $\rho_0$ and $\gamma$ defined within $r \in [0.4, 4]r_{1/2}$ as in item (x) in \S\ref{sec:3D_quantities}.
A full discussion on the comparison between the $\sigma_{\rm e, corr}$ and the $\sigma_{\rm e, dyn}$ is provided in Appendix \ref{app:sigma_corr}. Here, we anticipate that the correction works reasonably well for virtual-ETGs with total luminosity $\log L_{3D,r}/L_{\odot}>10$, as shown in Fig. \ref{fig: sig_corr-dyn}. It is evident, though, a trend for the low $\sigma_e$ galaxies to show lower $\sigma_{\rm e, corr}$ than the one predicted by the Jeans analysis, which is exhacerbated in the lower luminosity systems (see Appendix \ref{app:sigma_corr} for a more detailed discussion).

\end{enumerate}

{\it Units and remarks}: light and mass measurements are in solar units and logarithm scale, i.e., $\log L_{\odot}$ and $\log M_{\odot}$ respectively. The log-scaled sizes and velocities have physical (not-comoving) units of 
%kpc and km/s, 
$\log \text{kpc}$ and $\log \text{km/s}$, respectively.
%, and are also log-scaled, i.e., $\log \text{kpc}$ and $\log \text{km/s}$. 
Note that, in particular, the central gas mass $M_{\rm gas}(r_{1/2})$ is  stored in our catalogue as $\log(1+M_{\rm gas}(r_{1/2}))$ to avoid infinite values, given that there is a relevant number of ETGs with $M_{\rm gas}(r_{1/2}) =0$.
One may check a summary of the statistical properties of the main parameters in the final catalogue in Appendix \ref{appendix_catalogue_stat_properties} (Table~\ref{catalogue_stat_properties}), where we provide the full list of the parameters included in the catalogue together with their range and some statistics about their distributions.

\begin{figure}
    \centering
    \includegraphics[width=\columnwidth]{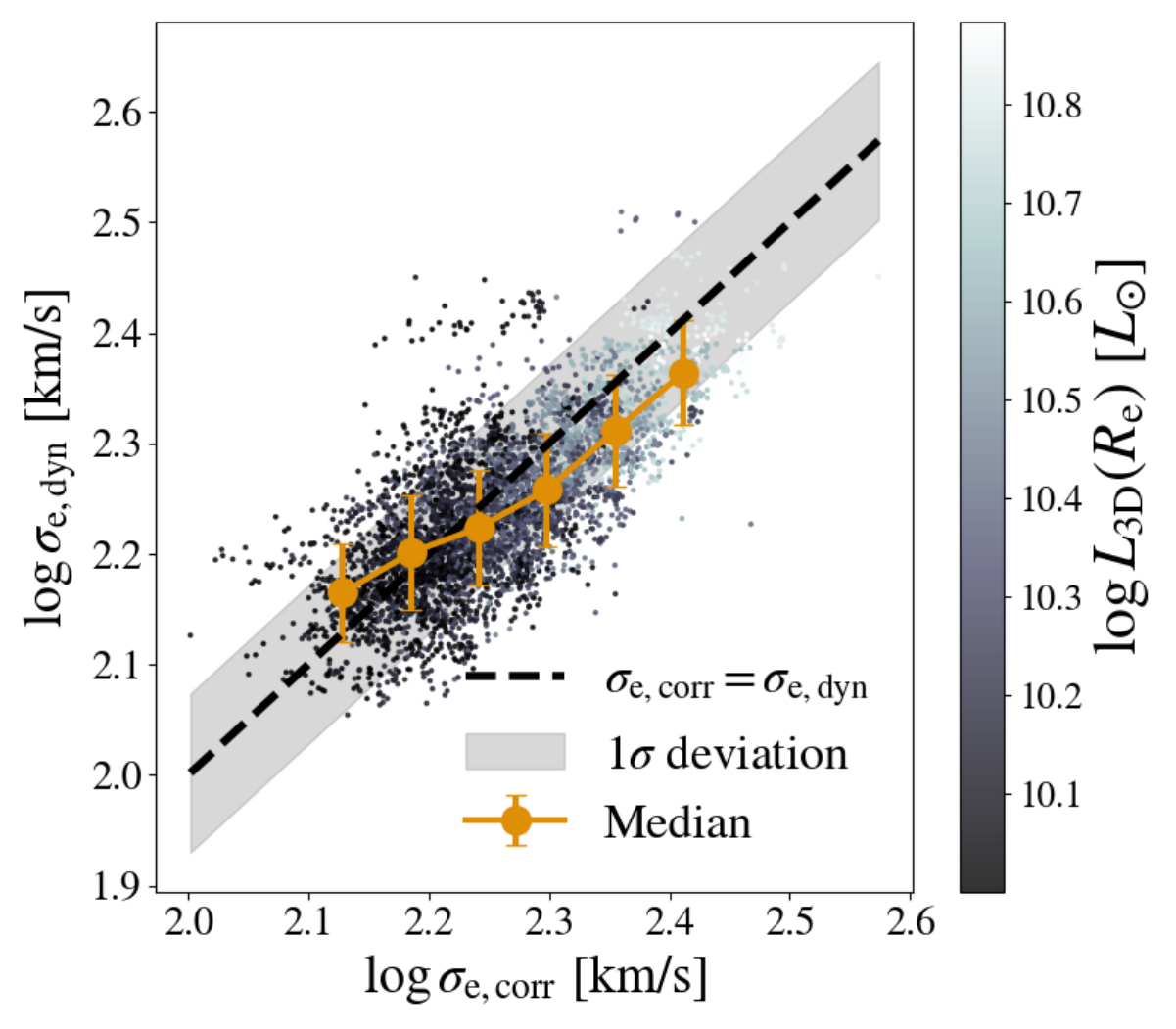}
    \caption{Comparison of the corrected virtual-ETG $\sigma_{\rm e, corr}$ (Eq. (\ref{eq:sig_e_corr})) and the dynamical inferred $\sigma_{\rm e, dyn}$ from the circular-aperture projected radial Jeans equation assuming isotropic orbits. Here we plot only galaxies with $\log L_{3D}(R_{\rm e})/L_{\odot}>10$, for which the corrected formula shows a rather accurate approximation (see Appendix \ref{app:sigma_corr} for a more detailed discussion).}
    \label{fig: sig_corr-dyn}
\end{figure}

\section{Results and discussion}\label{results_discussion}
 {In this section, we intend to science validate the virtual-ETG catalogue.} 
%We now turn to the discussion of the catalogue properties and the results we extracted from it. 
 {In particular, based on the observed-like quantitites derived in \S\ref{catalogue_section} and ``cleaned'', as discussed in \S\ref{data_cleaning} and Appendix \ref{Appendix_scatter_matrix}, we compare 1) some virtual-TNG scaling relations, including the size luminosity, $R_e-L$, size stellar mass, $R_e-M_\star$, and the size-effective surface brightness relation, $R_e-I_e$, and 2) the density slopes of the three components of the virtual-ETG sample, i.e., stars, gas and dark matter, as well as the total mass density slopes, with the  with the ones from the observed samples and some other literature datasets. We concentrate, in particular, 
%Then, we investigate 
on the correlations of the total slopes
%the derive: 
%1)} the virtual-ETG fundamental plane, and 2) 
%the correlations %statistical properties 
%of the density slopes 
with 1) the stellar mass, 2) the effective radius, 3) the S\'ersic index and 4) velocity dispersion and 5) compare them with literature data of both simulations and observations.
%, correlations between common galaxy structural parameters and the total density slope, and finally, we finish the section by discussing how galactic 
 {Finally, we discuss the relation between the derived 2D} sizes, luminosities and stellar masses, %are affected when projected onto the plane of the sky.
 {and the corresponding 3D quantities, as a function of the galaxy parameters.}
}

\begin{figure*}
    \centering
    \includegraphics[width=0.92\columnwidth]{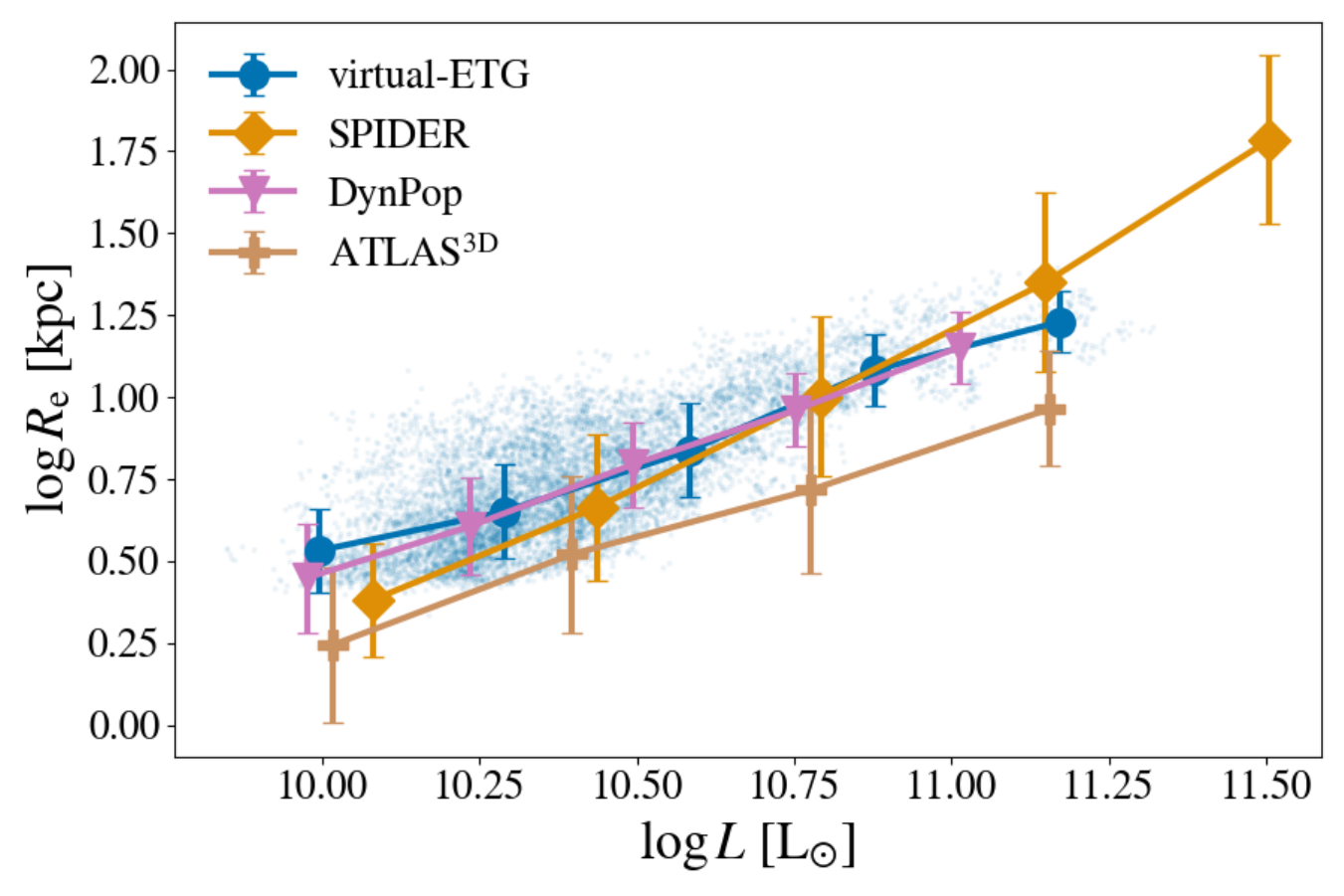}
    %\centering
    \hspace{-0.2cm}
    \includegraphics[width=0.95\columnwidth]{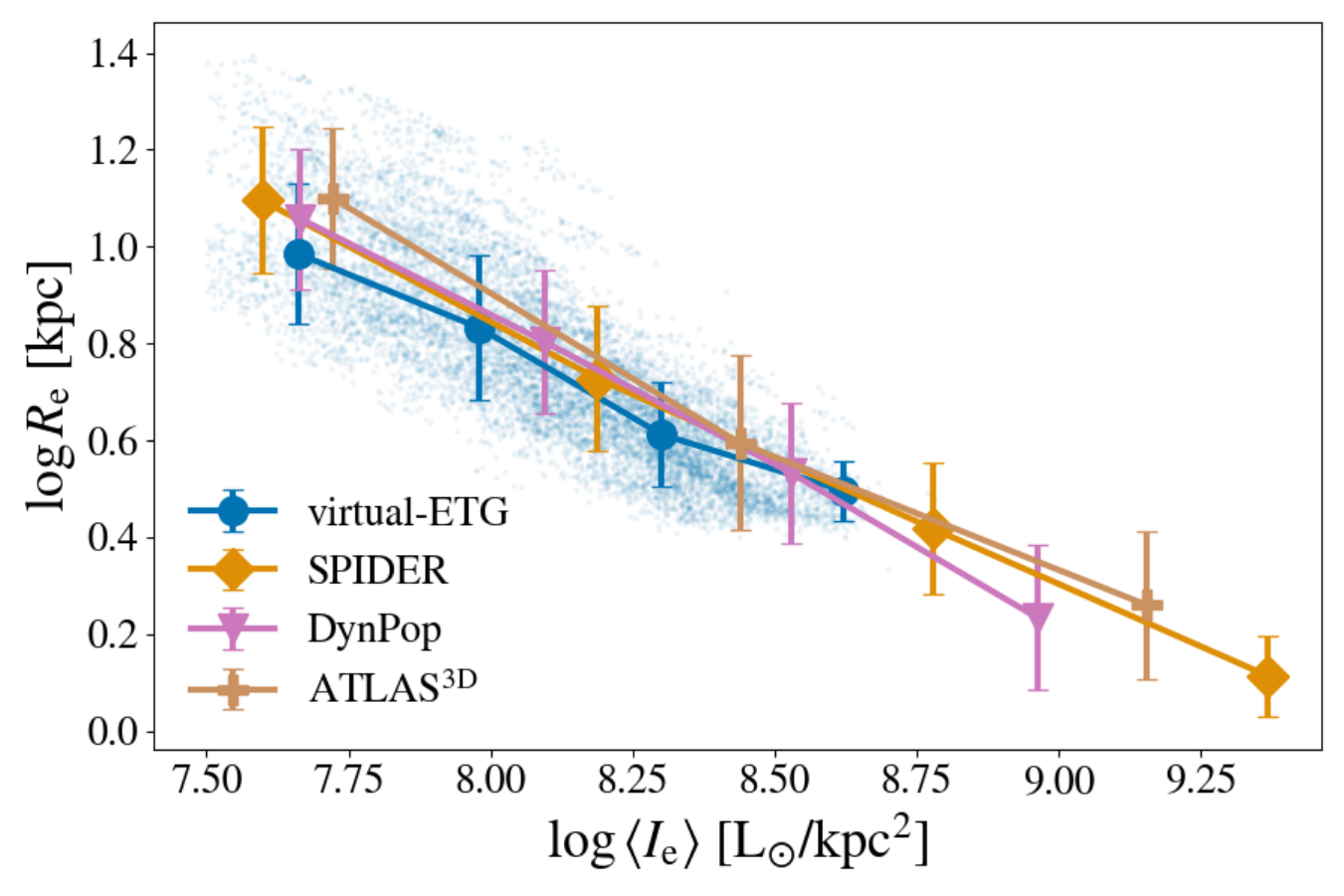}
    \centering
    \includegraphics[width=0.92\columnwidth]{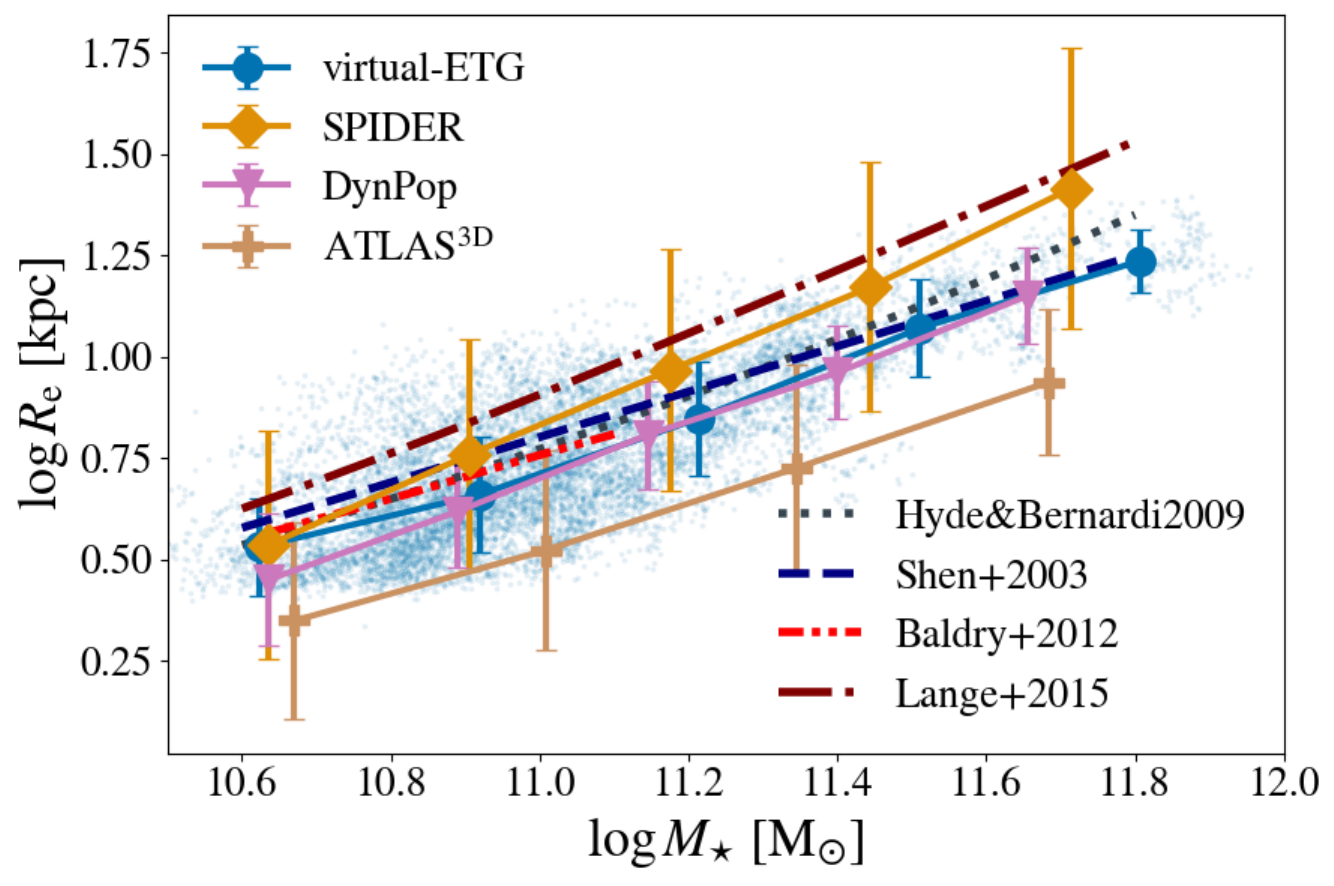}
    %\centering
    \hspace{-0.2cm}
    \includegraphics[width=0.95\columnwidth]{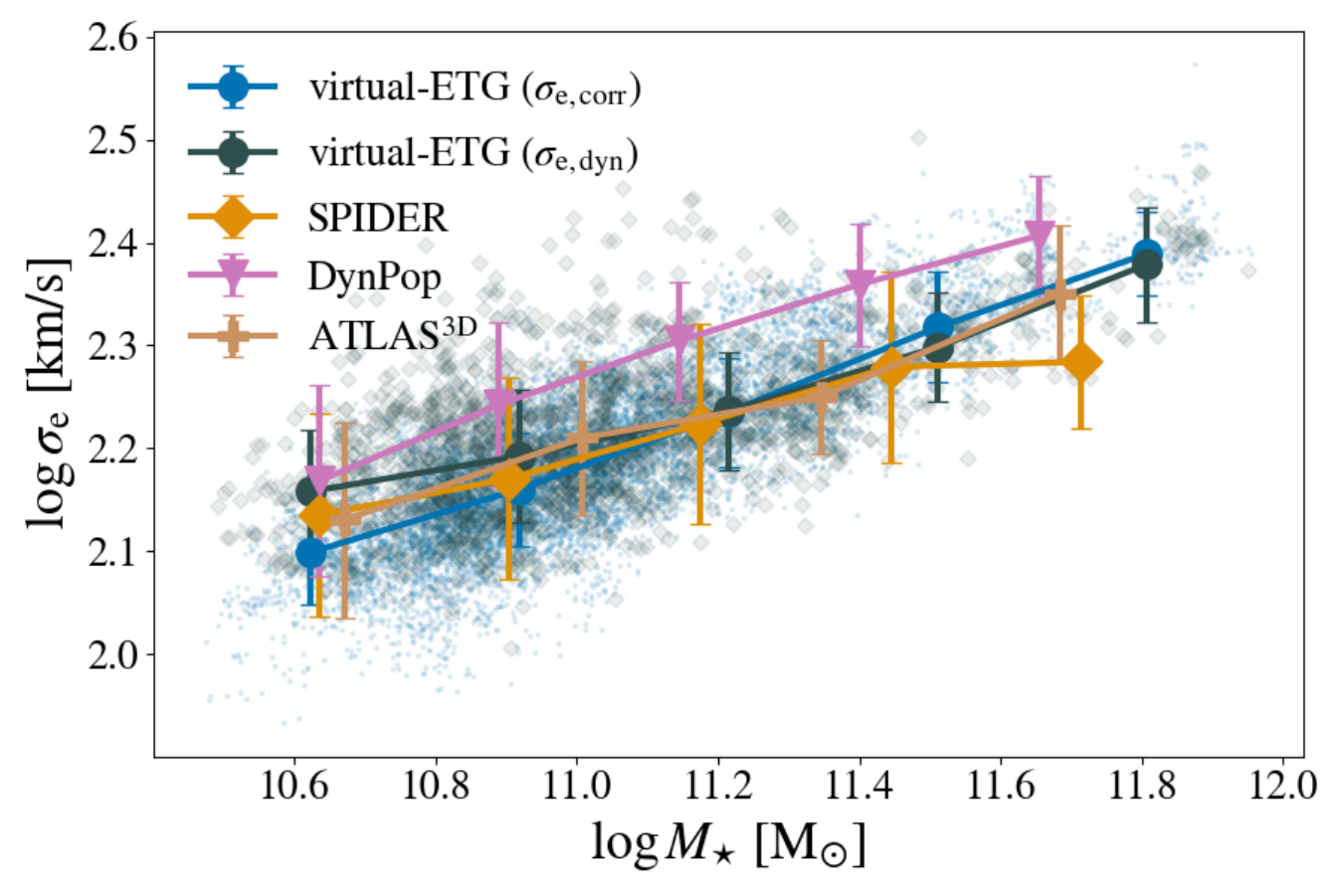}
    \caption{$R_{\rm e}$-$L_{\rm 2D}^{\rm tot}$ (upper left), $R_{\rm e}$-$\langle I_{\rm e} \rangle$ (upper right), $M_{\rm 2D,\star}^{\rm tot}$-$R_{\rm e}$ (bottom left) and the $M_{\rm 2D,\star}^{\rm tot}$-$\sigma_{\rm e}$ (bottom right) relations of the virtual-ETG and observed samples. The blue dots represent the scatter plot of each relation and, in particular, we represent the scatter plot of the $M_{\rm 2D,\star}^{\rm tot}-\sigma_{\rm e,dyn}$ relation by dark slate grey tetragons.}
    \label{fig:Mstar_relations}
\end{figure*}

\subsection{Scaling relations of the stellar component}
\label{sec:size-mass}
ETGs  are known to follow well-defined scaling relations, i.e., empirical correlations between pairs of parameters. These relations are thought to encapsulate the physics of the assembly of systems; hence, it is crucial to check whether hydro-dynamical simulations can reproduce these relations. Sticking to the photometric scaling relations, based on the quantities involving the photometric structural parameters, relevant scaling relations are the size vs. luminosity or equivalently the size vs. stellar mass relations (e.g., \citealt{Shen+2003},
\citealt{Hyde+2009}) and the $\langle I_{\rm e}\rangle-R_{\rm e}$
relation (\citealt{Kormendy+77}, \citealt{Capaccioli+92a}), where the mean surface brightness is defined as $\langle I_{\rm e}\rangle = L_{\rm 2D}^{\rm tot}/(2 \pi R_{\rm e}^2)$. Including the central velocity dispersion, $\sigma_{\rm e}$, we can also mention the 
Faber-Jackson (FJ; \citealt{FaberJackson1976}) and the fundamental plane (\citealt{Dressler+1987};
\citealt{DOnofrio+97_distance}). {For consistency, we strictly use the projected parameters from our sample and the equivalent definitions for the observational ones. The photometric structural parameters are computed with respect to the $r$-band for all samples, and central velocity dispersions are computed within an $R_{\rm e}$ aperture}. %For the sake of simplicity, we refer to total projected stellar masses and $r$-band luminosities from the virtual-ETG and the observational samples as $M_{2D,\star}^{\rm tot}$ and $L_{\rm tot, r}$, respectively.}

In this section, we do not intend to perform a full comparison of scaling relations from the virtual-ETG catalogue quantities with literature data. We want to only show a representative (albeit incomplete) check against the observational samples used as reference in this paper, and add some other literature results where needed, as a science validation of the virtual-ETG catalogue.

%we start by  {Finally, we also remark that the adoption of different radial cuts can potentially affect the S\'ersic fit and the consequent parameters. In
%In Fig. \ref{fig:rmax_cut_comparison} we show the $R_e$, $L_{\rm tot}$ and $n$-index of the two methods, face to face.
%As we can see, the $R_{\rm e}$ is the least affected parameter, while the $L_{\rm r, tot}$ shows a systematic overestimate for the $r_{\rm 30kpc}$ cut sample (as expected), except for the massive galaxies ($\log M_\star/M_\odot\gtrsim 10.9$) where the two outer radius definitions produce almost equivalent results (meaning that the SB$\sim26.5\rm mag/arcsec^2$ is reached at $\sim$30kpc). Finally, the $n$-index has a rather strong variation, that is quite worrisome because it enters into the classification discussed in the previous section. For this latter step, indeed, we decided to keep the $n$-index from the 30 kpc selection, but we have checked that the use of the 26.5 mag/arcsec$^2$ cut would minimally impact the number of galaxies with $n\geq2$ and thus the classification of the ETG sample (of the order of a few systems moving from ETGs to LTGs).
%We also further evaluate the impact of these systematic effects on the two outermost radius selection on the scaling relations of the virtual-ETG.
%, in the perspective of the discussion we will have in \S\ref{sec:FP}. 
In Fig. \ref{fig:Mstar_relations}, we collect four of the main scaling ETG relations, namely 
%we can collect using the content of the virtua-ETG sample. These are 
the $L_{\rm 2D}^{\rm tot}-R_{\rm e}$, the $M_{\rm 2D,\star}^{\rm tot}-R_{\rm e}$, the $\langle I_{\rm e} \rangle -R_{\rm e}$, and the $M_{\rm 2D,\star}^{\rm tot}-\sigma_{\rm e}$ relations. For comparison, we also plot the same relations from the SPIDER, DynPop, and ATLAS$^{3\rm D}$ samples (see \S\ref{sec:external_data}). {The relevant data for SPIDER comes from \citet{Tortora+2014}, and for the DynPop sample, it comes from \citet{Zhu}. For the ATLAS$^{\rm 3D}$ sample, effective radii and $r$-band luminosities from Sérsic fits to the light distribution were extracted from \citet{Krajnovic-2013}, central velocity dispersions from \citet{Cappellari+2013}  and, lastly, stellar masses were estimated as $\Upsilon_\star \times L_{2D}^{\rm tot}$, where $\Upsilon_\star$ is the 3D stellar mass-to-light ratio extracted from \citet{Cappellari+2013b}, assuming a \citet{Salpeter+1955} IMF, which we then converted to a Chabrier IMF by using $M_{\star,\rm Chab} =0.61M_{\star, \rm Salp}$ \citep{Madau+2014}. For the case of the $M_{2D,\star}^{\rm tot}-R_{\rm e}$, we will further add some results from \citet{Shen+2003}, \citet{Hyde+2009}, \citet{Baldry+12_gama_mass}, and
%\citet{Mosleh+13_size_evol}, 
\citet{Lange+2015}, all referred to a similar redshift range as the virtual-ETG sample.  {Furthermore, we selected only ETGs with stellar mass compatible with ours (i.e., $\log M_\star \gtrsim 10.5$) and, when possible, with Sérsic index $n\geq 2$. For the DynPop sample specifically, we restricted the Sérsic index to the range $2\leq n \leq 5.9$, to mitigate potential biases from the high-$n$ end, where a disproportionate number of objects have $n\approx 6$. This excess may be attributed to truncation effects during the Sérsic fitting process for these objects.}
Starting from the luminosity- and mass-size relations, overall, the TNG100 galaxies are nicely consistent with the observed samples within the errors, despite they tend to be over-luminous with respect to the observed analogue systems for $\log L_{\rm 2D}^{\rm tot}/L_\odot\gtrsim 10.8$, and under-luminous below $\log L_{\rm 2D}^{\rm tot}/L_\odot \approx 10.6$ (but just with respect to the SPIDER). 
%  {and ATLAS$^{\rm 3D}$} samples). 
However, in the former case, the luminosities seem to be more consistent within the scatter, with the real galaxies. We finally stress that the ATLAS$^{\rm 3D}$ sample looks overall systematically discrepant from the other virtual and real galaxies, eventually because of its smaller sizes. %, derived from multi-Gaussian expansion instead of the standard S\'ersic fit.
For the size mass relation, we also see a very good match with a large variety of real samples (e.g., \citealt{Shen+2003}, \citealt{Hyde+2009}, \citealt{Lange+2015}, as well as the SPIDER and DynPop), with the ATLAS$^{\rm 3D}$ being again the only sample systematically more compact.  

Similarly, the virtual-ETG sample shows a substantial agreement with the observed sample in the $R_{\rm e}$-$\langle I_{\rm e}\rangle$ relation, although we remark systematically lower mean $\langle I_{\rm e}\rangle$ in the centre of larger galaxies ($\log R_e/{\rm kpc}>0.6$), which would not be really significant if not coupled with the tilt seen in the $R_e$-$L_{\rm 2D}^{\rm tot}$ relation.
%, suggesting the presence of more compact cores. 
%%We will return to this properties in \S\ref{sec:FP}. Here we just notice that one explanation of the tilt of the $R_e$-$L_{\rm tot,r}$ and the $R_{\rm e}$-$I_{\rm e}$ relations
%\footnote{Notice that, in order to keep a more homogeneous comparison with the observational samples, we have defined a mean surface brightness: $\langle I_{\rm e} \rangle = L_{\rm tot}/(2\pi R_{\rm e}^2)$.} 
%%can come from a weak AGN feedback that can act on small scales 
%in high-luminosity galaxies 
%%(e.g., facilitating dense $I_{\rm e}$) and 
%a weak SN feedback that is expected to impact more the low-luminosity galaxies 
%%and large scales (e.g., causing larger star-formation activities and so larger stellar masses) of ETG systems (see, e.g., \citealt{2021_horizon}).
{Finally, the $\sigma-M_\star$ relation, shows a striking consistency with the real data samples, especially in the high mass range ($\log{M_\star/M_\odot}>11$) regardless of the correction adopted; while at the lower mass range the $\sigma_{\rm corr}$ and the $\sigma_{\rm dyn}$ diverge with the latter being more aligned to the SPIDER, DynPop and ATLAS$^{\rm 3D}$. 
%if we consider the corrected or the dynamically inferred $\sigma$....}

The main conclusion of this section is that the virtual-ETG sample catalogue of structural parameters, derived using our ``observation-like'' procedure, is consistent with all the main scaling relations involving the luminosity and mass of stars, as well as their kinematics. In this latter case, we stress that this is especially true if we account for the effect of the softened dynamics affecting the central $\sim 1$ kpc, via the $\sigma_{\rm e, dyn}$. With this result in mind, we can move to the properties of the total mass of galaxies.

\subsection{Total mass density slope distributions}
\label{density_slopes_results}
 
%Having established the fundamental plane of TNG ETGs, we now turn our focus to the analysis of 
The central density slopes are key parameters that offer information into the internal structure and matter distribution of ETGs. The total density slopes in our catalogue are calculated assuming three components, namely gas, dark matter and stars, hence providing an insight of the interplay between these different ingredients. We assume that the mass density of each of these components follows a power-law distribution, as written in Eq. (\ref{gen_SIS_model}).
%As discussed before, the gas component is not to be trusted, as in general it is not smoothly radially distributed, but we keep it in the catalogue for completeness. 

%It has been verified   {by observations }
%verified 
 {Measurements of the total central density slope of ETGs, mainly from strong gravitational lensing (e.g., \citealt{Koopmans_2006}), have shown}
  {that the total power-law density slopes of ETGs stay closer to isothermal 
%up to the outermost parts of the galaxies 
up to a few times the effective radius
%and as demonstrated 
%  {In fact, using the SLACS survey 
(see also \citealt{SLACS}, \citealt{Gavazzi})}.  {There is still a debate about whether this isothermal slope is universal and/or evolves with redshift, considering both simulations and observations (see, e.g., \citealt{Remus+2017}). Part of the discrepancies between observations and simulations might yet come from the different definitions of the density slopes in the two environments. We will comment more later on that, based on our results.}  
%  {However, we also need to remind that, although the total density slopes seems to stay closer to isothermal,} 
 {It goes without saying that} neither the baryons nor the dark matter follow such a density trend individually.
%(see, e.g.,Table \ref{slopes_stats}). 
Rather, they \textit{conspire} together to generate the total density profile following the simple isothermal sphere (SIS) model. This phenomenon is also referred to as the \textit{bulge-halo conspiracy} (e.g., \citealt{Keeton_2001, Gavazzi, Jiang_2007}). Depending on the method adopted to measure the total slope, a source of systematics can come from the relative weight one gives to the baryonic vs. dark matter component. This is well known in dynamical methods, where one has to constrain the mix of stars and dark matter (e.g., by constraining the stellar mass-to-light ratio, $M_\star/L$) and one can overestimate or underestimate the density slope by over/underestimating the $M_\star/L$ (see, e.g., \citealt{Tortora+2014} and \citealt{Tortora+2022}
for a discussion). These systematics cannot be totally alleviated, also including strong lensing (see, e.g., \citealt{Sonnenfeld_2013}).

\subsubsection{Definitions}
\label{sec:slope_definitions}
%found} that the approximate isothermal behaviour of the total density profile extends up to one hundred times the effective radius. 
%, given that the  density slope of baryons is typically steeper and for the dark matter it is  shallower, their joint behaviour is consistent with a SIS density profile over a radial range of several effective radii. 
%

\begin{table*}
\centering
\caption{Summary of statistical properties for the slopes within the catalogue. The slopes $\gamma_{i}^{\rm PL}$ and $\gamma_{i,2}^{\rm PL}$ are computed according to Eq. (\ref{power_slope_diff_form}) in the radial ranges $[0.4,4]r_{1/2}$ and $[0.4, 2]r_{1/2}$ respectively. The slopes $\gamma_{i}^{\rm AV}$ are computed according to Eq. (\ref{average_slope_def}) in the radial range $[0.4, 2]r_{1/2}$. }
\begin{tabular}[hbt!]{ccccccccccccc} % eight columns, alignment for each
\hline
{Statistics} & {$\gamma_{\star}^{\rm PL}$} & {$\gamma_{\rm gas}^{\rm PL}$} & {$\gamma_{\rm DM}^{\rm PL}$} & {$\gamma_{\rm tot}^{\rm PL}$} & {$\gamma_{\star,2}^{\rm PL}$} & {$\gamma_{\rm gas,2}^{\rm PL}$} & {$\gamma_{\rm DM,2}^{\rm PL}$} & {$\gamma_{\rm tot,2}^{\rm PL}$} & {$\gamma_{\star}^{\rm AV}$} & {$\gamma_{\rm gas}^{\rm AV}$} & {$\gamma_{\rm DM}^{\rm AV}$} & {$\gamma_{\rm tot}^{\rm AV}$}\\
\hline
{Mean} & 3.038 & 0.668 & 1.742 & 1.998 & 2.960& 1.113&1.717 & 2.055 & 2.959 & 1.266 & 1.714 & 2.057\\
{SMW mean} & 3.036 & 0.706 & 1.709 & 1.939 & 2.960& 0.911& 1.669& 1.972 & 2.927 & 1.030 & 1.669 & 1.976\\
{Median} & 3.035 & 0.745 & 1.742 & 2.003 &2.923 & 1.303& 1.721& 2.049 & 2.934 & 1.380 & 1.716 & 2.056 \\
{Std. Err.} & 0.001 & 0.010 & 0.001 & 0.001 &0.002 &  0.017& 0.001&0.002 & 0.002 & 0.016 & 0.002 & 0.002\\
{Scatter} & 0.113 & 0.983 & 0.103 & 0.148 & 0.186& 1.720&0.135 & 0.203 & 0.167 & 1.648 & 0.163 & 0.195\\
\hline
\end{tabular}
\label{slopes_stats}
\end{table*}

Before proceeding to the results, it is important to remind the reader of the following. 
%note that 
different slope definitions that are used in the literature and adopted consistently here for comparison. For instance, according to the procedure adopted in \S\ref{density_profiles} (Eq. (\ref{gen_SIS_model})), 
%the one used in this work is the 
we use the power-law  ($\rm PL$) slope $\gamma^{\rm PL}$, defined as the best-fit slope of a line for the log-scaled radial density distribution between two radii $r_1$ and $r_2$. We recall that, besides $\gamma^{\rm PL}$ defined over the radial range $r \in [0.4,4]r_{1/2}$, we also defined a $\gamma_2^{\rm PL}$ over a smaller radial range, i.e., $r \in [0.4, 2]r_{1/2}$. We will refer in the following to these two definitions directly as $\gamma_i^{\rm PL}$ and $\gamma_{i,2}^{\rm PL}$ for the $i$-th component (i.e., gas, stars, DM, or total).

We can also define the average slope $\gamma^{\rm AV}$ in a spherical shell as: 
\begin{equation}
    \gamma^{\rm AV}(r_1, r_2) \equiv \frac{\log \rho(r_2) - \log \rho(r_1)}{\log r_2 - \log r_1}  = \frac{\log[\rho(r_2)/\rho(r_1)]}{\log(r_2/r_1)}.\label{average_slope_def}
\end{equation}
  {As anticipated in \S\ref{catalogue_section} (see item (x)), for the ETGs in our sample, we defined an average slope over the radial range $[0.4, 2]r_{1/2}$ (see Table~\ref{slopes_stats}) in order to access possible discrepancies due to slope definitions between simulations and real observations. We chose to define an average slope as it is one of the most commonly used definition among the observational samples we compare to.}

Another common definition used in literature is the mass-weighted density slope $\gamma^{\rm MW}$ 
%is used in the literature 
(see, e.g., \citealt{Dutton}) which describes the trend of the density profiles in the central parts of the galaxies. It is defined at a radius $r$ as:
{\begin{equation}
    \gamma^{\rm MW}(r) \equiv \frac{4\pi}{M(r)}\int_{0}^{r} \gamma(r') \rho(r')r'^2 dr', \label{MW_slope_def}
\end{equation}
where $\gamma(r)$ is the local slope, defined as 
\begin{equation}
\gamma(r) = -\frac{\rm d\log \rho(r)}{\rm d\log r},  
\label{eq:slope_r}
\end{equation}
while $M(r)$  and $\rho(r)$ are, respectively, the 3D mass  and density enclosed within a radius $r$, related by:
\begin{equation}
M(r) = 4\pi\int_0^r\rho(r')r'^2dr'.
\end{equation}
It can be easily demonstrated that:
\begin{equation}
    \gamma^{\rm MW}(r) = 3 - \frac{\rm d \log M(r)}{\rm d \log r}. 
\end{equation}}
The equations above are valid for any component. % {However, it is worth to mention that there is a substantial difference }
For the sake of a better understanding, in the next section, we will refer to each slope definition using its respective notation.

\subsubsection{Results and comparison with literature}
\label{sec:slope_res_vs_lit}
  {In this subsection, we discuss the results for each of the mass component of the virtual-ETG sample and how their distributions compare to previous studies using both simulated and observed data}.

In Table \ref{slopes_stats} we report some statistical properties of the slopes {for the total mass density distributions and also the three components, i.e., star, gas and dark matter, separately. We also report the $\gamma_{\rm tot, 2}^{\rm PL}$ values discussed in \S\ref{density_profiles} and the average slope defined in Eq. (\ref{average_slope_def}): these latter estimates are, by construction, better references for the observed values as they are defined on a radial sample closer to the typical values adopted for real systems.} In the same table, we include the mean value, the stellar mass-weighted (SMW for short) mean -- where the weights assigned for each virtual-ETG system are their 2D total stellar mass $M_{2D,\star}^{\rm tot}$ --, median value, standard error (Std. Err.) of the mean and the observed scatter. %The distributions for the stellar, dark matter and total slopes are shown in Fig. \ref{fig_slopes_distro}. 
{These are the most relevant quantities we will discuss in the following, having already mentioned that the $\gamma_{\rm gas}$ is little robust and poorly physically meaningful for ETG samples}. {Looking at the statistics in Table \ref{slopes_stats}, we find that the $\gamma_{\rm tot}^{\rm PL}$ is nicely ``isothermal'', according to the different statistical indicators, 
%(see Table \ref{slopes_stats}), 
while $\gamma_{\rm 2, tot}^{\rm PL}$ and $\gamma_{\rm tot}^{\rm AV}$ are slightly steeper than 2, although in all the cases ``isothermality'' is included in the large scatter (0.148, 0.203 and 0.195 for $\gamma_{\rm tot }$, $\gamma_{\rm tot, 2}$ and $\gamma_{\rm tot}^{\rm AV}$, respectively). The reason for $\gamma_{\rm tot}^{\rm PL}$ to be shallower than $\gamma_{\rm tot,2}^{\rm PL}$ and $\gamma_{\rm tot}^{\rm AV}$ comes from the fact that the last two slopes are defined over a radial range where the steeper stellar density profile ($\gamma_\star\sim3$, see Table \ref{slopes_stats}) dominates over the DM density profile which has a shallower slope ($\gamma_{\rm DM}\sim1.7$, see also Table \ref{slopes_stats} ).} 

\begin{figure*}
    \centering
    \begin{subfigure}{0.33\textwidth}
        \includegraphics[width=\textwidth]{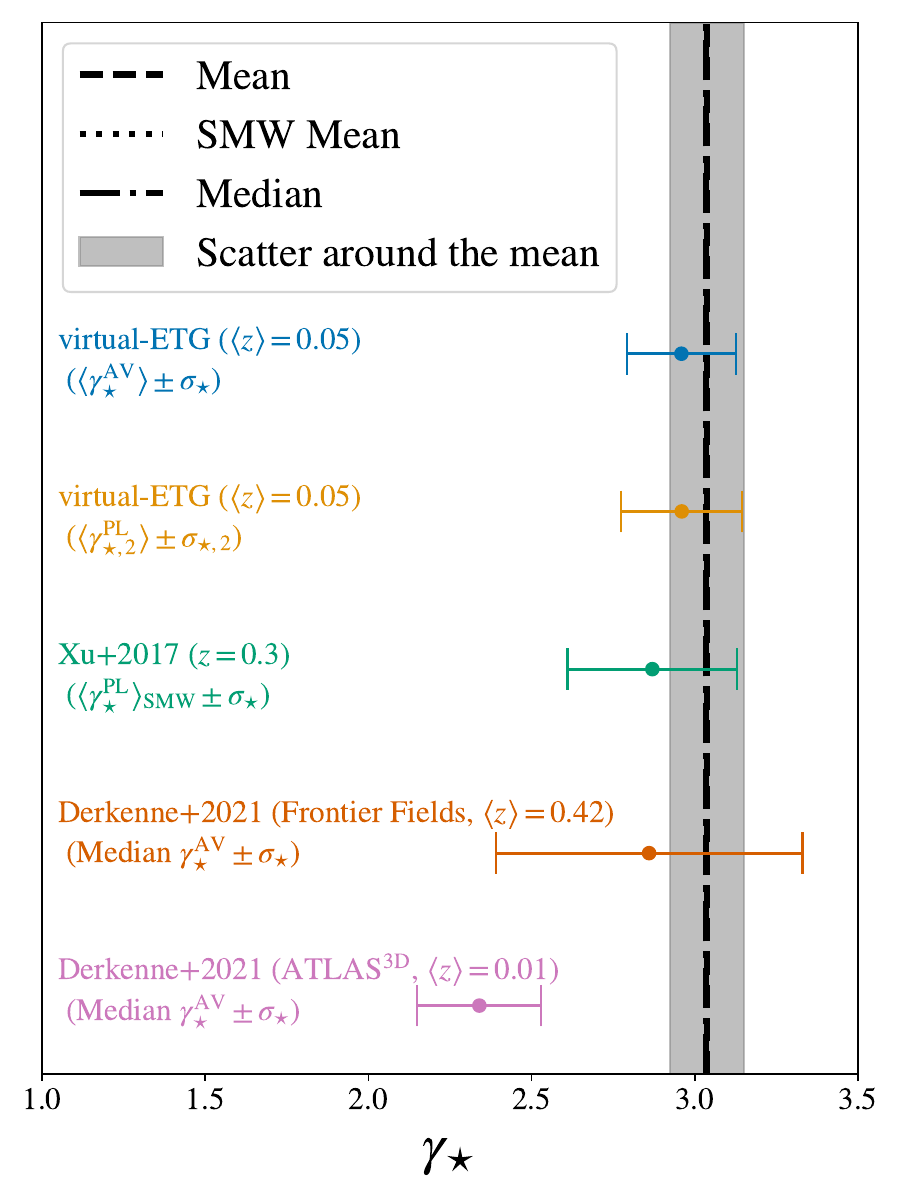}
    \end{subfigure}
    \begin{subfigure}{0.33\textwidth}
        \includegraphics[width=\textwidth]{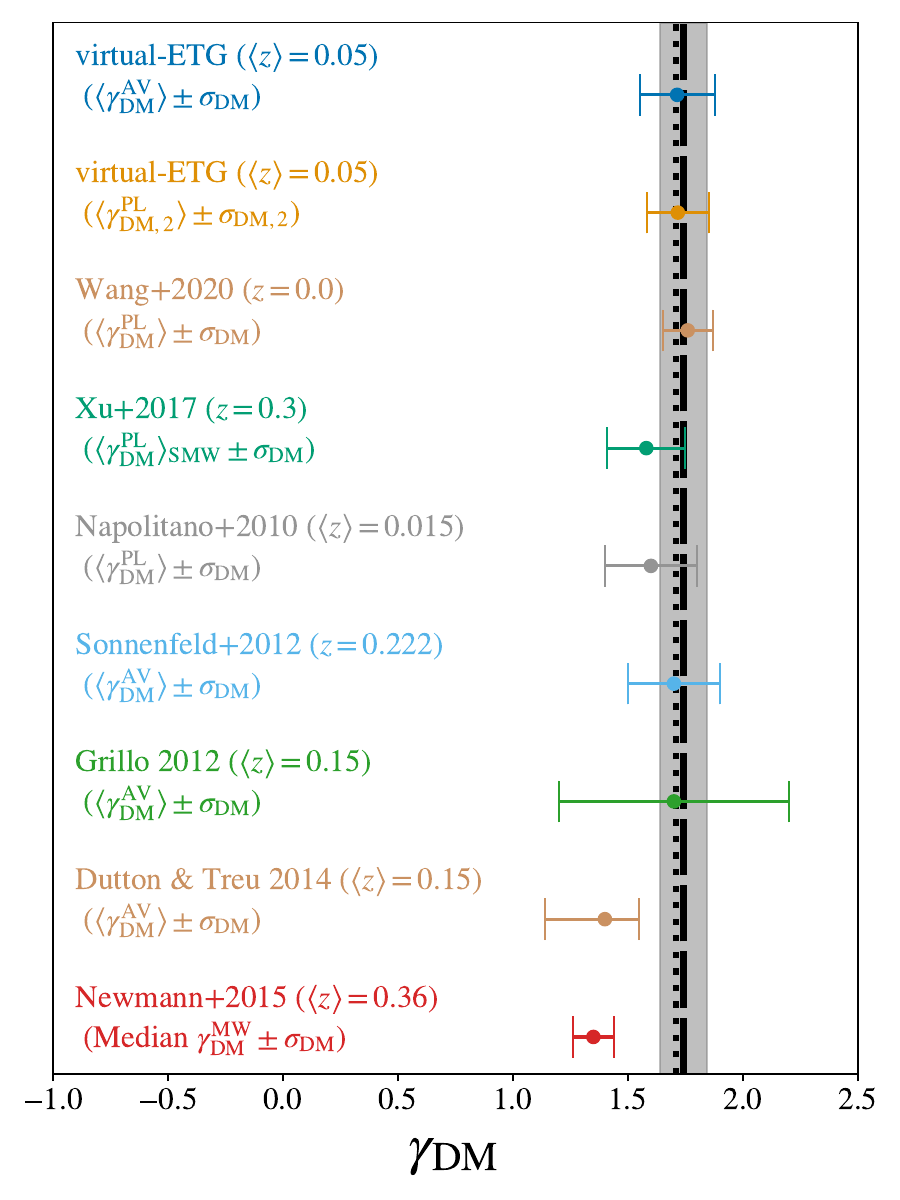}
    \end{subfigure}
    \begin{subfigure}{0.33\textwidth}
        \includegraphics[width=\textwidth]{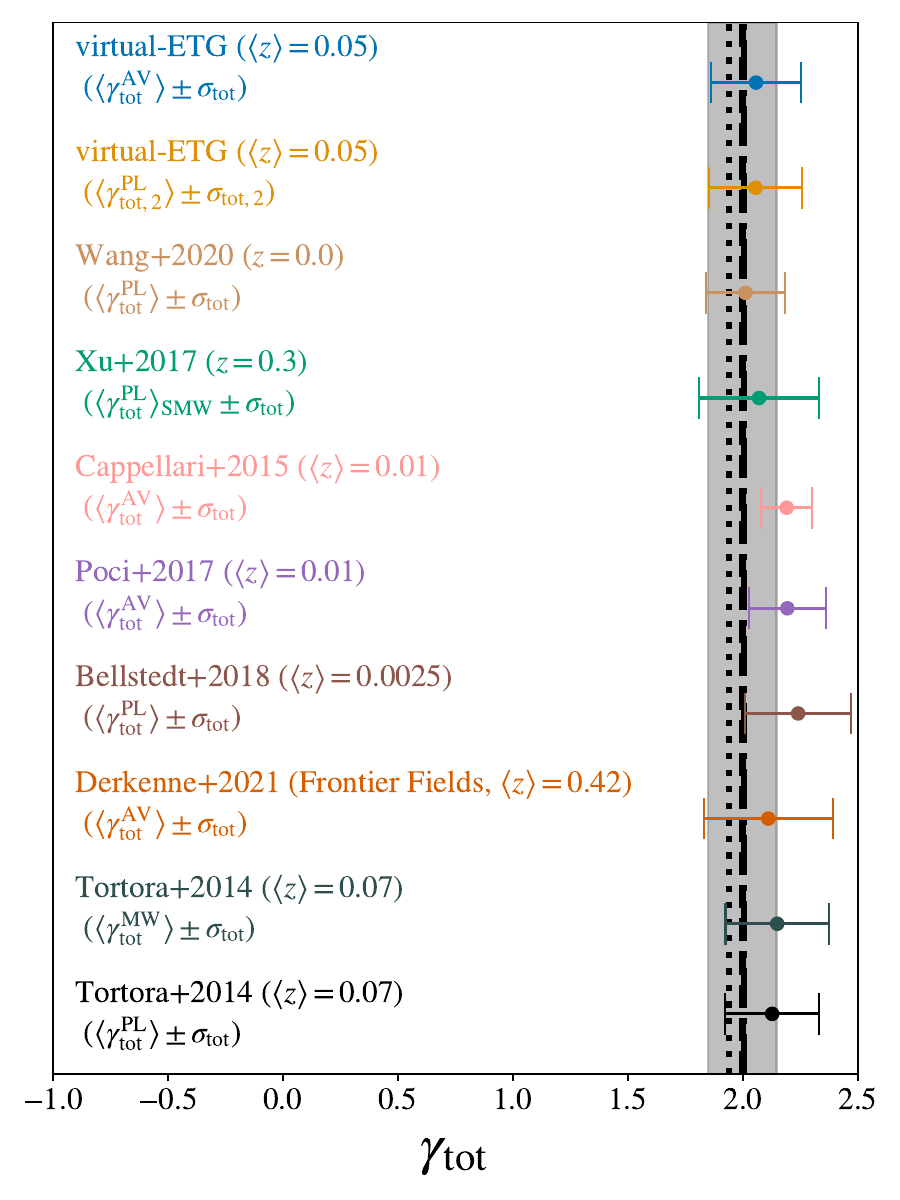}
    \end{subfigure}
\caption{Comparison of the density slopes obtained by the various methods and works. We use as reference the $\gamma_{\rm tot}^{\rm PL}$ defined over the radial range $[0.4, 4]r_{1/2}$. For all the samples, we show their mean redshift values. For the stellar density slopes (left panel), we see that all the results are in agreement within $1\sigma$, except for the result obtained by \citet{Derkenne} for the ATLAS$^{\rm 3D}$ sample. In the middle panel, We have the dark matter density slopes, which are generally all consistent within $1\sigma$, except for \citet{Dutton} and \citet{Newman_2015} (see text). For the total matter density slope (right panel), we notice that all the slopes are statistically consistent within $1\sigma$. For \citet{Derkenne} and \citet{Bellstedt+2018}, we estimated their respective standard deviations as $\sigma = \mathrm{Std. Err.} \times \sqrt{\mathrm{sample~size}}$.}
\label{summary_stat_slopes}
\end{figure*}

To put the results of the slopes of the virtual-ETG sample, reported in Table~\ref{slopes_stats}, more in the context of the current estimates from observations and simulations, 
%We can now compare the results of the slopes of the virtual-ETG sample, reported in Table~\ref{slopes_stats}, with other literature results, both from simulations and observations. % {As anticipated in \S\ref{density_profiles}, we also report the $\gamma_2$ values in Table \ref{slopes_stats}: this latter estimates are, by construction, a better reference for the observed values as they are defined on a radial sample closer to the typical values adopted for real systems.}\\ 
{in Fig. \ref{summary_stat_slopes}, we present a comparison of the slopes obtained in this work vs. literature values, making use of different methods as reported in the legend.
%by each method and galaxy sample (from both real and simulated systems). 
We take as reference the power-law slope defined over the radial range $[0.4, 4]r_{1/2}$, as it is the quantity obtained with the lowest scatter among the others we derived.}
In the following, we discuss the density slope results of the three components in detail. 
~\\

{\it Stars}.   {Starting from the stellar component, we have found (mean $\pm$ scatter) $\langle \gamma_{\star}^{\rm PL} \rangle \pm \sigma_{\star}= 3.038 \pm 0.113$, $\langle \gamma_{2,\star}^{\rm PL} \rangle \pm \sigma_{\star,2} = 2.960 \pm 0.186$ and $\gamma_{\star}^{\rm AV} \pm \sigma_{\star} = 2.959 \pm 0.167$ for galaxies with 2D (3D) stellar masses in the range $10^{10.47} \leq M_{2D,\star}^{\rm tot} \leq 10^{11.95}$ ($10.39 \leq M_{\rm 3D, \star}(< 30~\mathrm{kpc})\leq 11.70$) (Table~\ref{catalogue_stat_properties}). Further statistical properties for the slopes can be found in Table~\ref{slopes_stats}. } Compared with other datasets based on simulations, using the data from Illustris-1 simulations\footnote{\url{https://www.illustris-project.org/data/downloads/Illustris-1/}}, \cite{D.Xu} found a stellar mass-weighted mean value $\langle \gamma_{\star}^{\rm PL} \rangle_{\rm SMW} = 2.87$ with $\sigma_{\star} = 0.26$ by considering the radial interval $0.5R_{\rm e}<r<2R_{\rm e}$ at $z=0.3$ for ETGs with stellar masses $M_{\star} \geq 10^{10} M_{\odot}$. This looks smaller but consistent with our $\gamma_{2,\star}$ estimate (specially the median value), covering a similar radial range
%, although systematically shallower 
(see also Fig. \ref{summary_stat_slopes}). Surely, part of the differences can be ascribed to the different mass range, as they include lower mass systems than ours. 

Moving to observed galaxies, considering 90 early-type galaxies in the redshift window $0.29 < z < 0.55$ from the Frontier Fields project \citep{FrontierFields}, \citet{Derkenne} found a median $\gamma_{\star}^{\rm AV}$  equals to $2.86 \pm 0.05$ (standard error) computed over the  radial range $0.16$ arcsec to the maximum kinematic extent of the data (which is $3R_{\rm e}$ on average).  {This is also  consistent with us but yet shallower than our estimate, which might be possibly not related to their fixed lower radial limit, i.e., 0.16 arcsec, that at the centre of their redshift range ($z\sim0.42$) corresponds to 0.9 kpc, i.e., in line with our softening length.}  {Instead, a plausible reason of the systematic shallower slopes with both the simulated sample from Illustris and IllustrisTNG simulations and the observed sample from Frontier Fields can be due to the different radial range that for both samples is shorter than the $4r_{1/2}$ we adopt, which, for a S\'ersic profile, can produce slightly shallower profiles as the $d\log I(R)/d\log R$ becomes increasingly steeper at $R>R_e$ (see, e.g., \citealt{Vitral_Mamon_2020}). However, there might be other effects for systematic discrepancies, like the redshift or the overall star formation efficiency \citep{Wang+2019}, that can also impact stellar profile slopes.}

Going to lower redshift, using the Multi Gaussian Expansions (MGE) from \cite{Scott} for the ATLAS$^{\text{3D}}$ sample of nearby galaxies \citep{Atlas3D}, \citet{Derkenne} found that the median $\gamma_{\star}^{\rm AV}$ is $2.34 \pm 0.02$, computed over the radial range of $2$ arcsec to the maximum kinematic range ($1R_{\rm e}$ on average), making the ATLAS$^{\text{3D}}$ sample the one with the shallowest central stellar density profiles. {This result is inconsistent with ours within $1\sigma$ (see Fig. \ref{summary_stat_slopes})}.  {In this case, the discrepancy might come from both the extremely shorter radial range, which is in line with our comment above, but also to the MGE technique which is different from the standard parameterization used in literature and ourselves, i.e., the S\'ersic profile.

%However, using the SPIDER sample of ETGs from \citet{Tortora+2014}, we estimated a shallower mean stellar density slope with respect to our result: $\langle \gamma_\star \rangle = 2.463$ with a scatter $\sigma_\star = 0.120$ computed over the radial range $[0.3, 4]r_{1.2}$. These are more consistent with the results of \citet{Derkenne}. This result from \citet{Tortora_2014} comes by assuming a deprojected Sérsic profile to model the 3D stellar distribution and then fitting

Overall, it is clear that {\it in order to reach  a consensus about the alignment between observations and simulations, we should start by uniforming the slope definitions, in order to leave the remaining systematics caused only by physical discrepancies.}}\\
%Considering Illustris-1 and Frontier Fields samples, the results for the stellar slopes at intermediate redshift values are in agreement with our work, in which computations were made to galaxies at low redshift values, i.e., $0<z<0.1$. We argue that the stellar slopes within our catalogue are steeper in comparison to Illustris-1, Frontier Fields and ATLAS$^{\text{3D}}$ because our choice for the radial interval goes up to the outermost regions of the galaxies, where the dark matter dominates over baryons, and the stellar component only dominates over DM at the innermost regions, typically $\lesssim R_{\rm e}$ \citep{Totani}. Thus, it suggests that the stellar slope becomes steeper for greater radial intervals, where dark matter influences the most \citep{Poci}.  In addition, as pointed by \cite{Derkenne} the stellar density distribution is not generally described by a simple power-law model, but is more curved, making the stellar slopes steepen more for larger radial intervals. 
~\\

{\it Dark Matter.} {For the DM component, we have found (mean $\pm$ scatter) $\langle \gamma_{\rm DM}^{\rm PL} \rangle \pm \sigma_{\rm DM} = 1.742 \pm 0.103$, $\langle \gamma_{\rm DM,2}^{\rm PL} \rangle \pm \sigma_{\rm DM,2} = 1.717 \pm 0.135$ and $\langle \gamma_{\rm DM}^{\rm AV}\rangle \pm \sigma_{\rm DM} = 1.714 \pm 0.163$. This component is the one with the minimum slope scatter in each respective definition}. The DM also has a shallower radial distribution compared to the stellar and total slopes (see Table~\ref{slopes_stats}). Using TNG100 ETGs at $z=0$, and considering the radial range $0.4r_{1/2}\leq r\leq4r_{1/2}$, \cite{Y.Wang} found $\langle \gamma_{\rm DM}^{\rm PL} \rangle = 1.760 \pm 0.005$ and $\sigma_{\rm DM} = 0.108$ for early-type galaxies with stellar masses in the range $10^{10.7} \leq M_{\star}/M_{\odot} \leq 10^{11.9}$, which is unsurprisingly consistent with the one found in our work {, as in our analysis we are closely following their procedure, and just differing for the range of stellar masses adopted that in our case includes ETGs with masses as small as $10^{10.3} M_{\odot}$ and up to $10^{11.9} M_{\odot}$.} With Illustris-1, using different radial and redshift intervals (again, $0.5R_{\rm e} < r < 2R_{\rm e}$ and $z=0.3$), \cite{D.Xu} found $\langle \gamma_{\rm DM}^{\rm PL} \rangle = 1.58$ with a scatter $\sigma_{\rm DM} = 0.17$, i.e., shallower than the results found in TNG100, but yet in agreement within $1\sigma$, as we illustrate in Fig. \ref{summary_stat_slopes}. The shallow value obtained from Illustris-1 is likely because $\gamma_{\rm DM}$ steepens by increasing the radial interval where the density profile is defined, as suggested by the results in \citet{Y.Wang}. Thus, for a dark matter density slope defined within $R_{\rm e}$, one should expect a shallower $\gamma_{\rm DM}$. 
%Overall, despite the differences discussed above, all simulations seem to indicate a $\gamma_{\rm DM}$ in the range $1.6-1.7$. Interestingly, this can be interpreted as being consistent with a contracted halo (due to baryon physics) from a  ``collisionless'' NFW-like DM slope $\sim-1$ (see, e.g., \S 3.2 in \citealt{Napolitano+2010}).  
The results of TNG100 above suggest {\it a contraction of the dark matter halo from a ``collisionless'' NFW-like DM slope $\sim-1$}, in agreement with the results of \citet[][see discussion in their \S 3.2]{Napolitano+2010} (see also \citealt{Tortora_2010} for further discussion) which, assuming a Chabrier IMF, found an average $\langle \gamma_{\rm DM}\rangle \sim 1.6 \pm 0.2$ from the Jeans dynamical analysis of a local sample of ETGs ($z\lesssim 0.03$). 
A consistent result has also been found by \citet{Sonnenfeld_2012} 
%(please cite also Napolitano et al. 2010, where we find gamma~-1.6, using a Chabrier IMF), 
who measured a dark matter slope $\gamma_{\rm DM}^{\rm AV} = 1.7 \pm 0.2$
in a strong lensed ETG 
%of an ETG at redshift 
at $z = 0.222$. %which is the strong gravitational lens of the system SDSSJ0946+1006 \citep{Bolton+2004} and has a stellar mass estimated from lensing and dynamics $M_{\star}^{\rm LD} = 5.5_{-1.3}^{+0.4}\times 10^{11} M_{\odot}$. It was found a value $\gamma_{\rm DM}^{\rm AV} = 1.7 \pm 0.2$ within the radius of the outer ring. 
From strong lensing again, using the SLACS %strong lenses from 
sample \citep{SLACS} 
%for galaxies with 
at $z<0.3$, and assuming density distribution inside the effective radius, %it has been found by  
\citet{Grillo} have found a mean dark matter slope of $\langle \gamma_{\rm DM}^{\rm AV} \rangle = 1.7 \pm 0.5$, in agreement with our estimate and \citet{Sonnenfeld_2012}. However,  \citet{Dutton}  found a contradicting result using the same 
%corrected the result in 
\citet{Grillo} sample and claimed a mean $\langle \gamma_{\rm DM}^{\rm MW} \rangle = 1.40^{+0.15}_{-0.26}$ interior to $R_{\rm e}$. The main difference between \citet{Dutton} analysis and the one mentioned above resides in the different IMF adopted, with the latter assuming a Salpeter IMF, which favours mild contraction.
This is also found by 
%Combining data from 10 group lenses, and assuming a Salpeter IMF, 
\citet{Newman_2015} which, using a Salpeter IMF and combining data from 10 group lenses, found $\langle \gamma_{\rm DM}^{\rm MW}\rangle = 1.35 \pm 0.09$ inside $R_{\rm e}$ for ETGs within $10^{13}-10^{15} M_{\odot}$ halos, mean redshift $\langle z \rangle = 0.36$ , also favouring mild contraction. 
This IMF vs. DM degeneracy is a well known issue in both dynamical (e.g., \citealt{Napolitano+2010}) and strong lensing analysis (e.g., \citealt{Treu+2010}), which also requires uniformity of approaches/assumptions when it comes to comparison among different analyses. In this respect, our result from TNG100 allows us to set a reference for the inferences based on Chabrier IMF. \\
%Apart from \citet{Dutton} and \citet{Newman_2015}, all the other results are in agreement with our work. In both works, the dark matter density slope is defined within $R_{\rm e}$, similarly to \citet{D.Xu}, and the three results nicely agree. Therefore, we believe that the divergence between these results and ours is, again, likely due to the radial interval definition.
~\\

{\it Total mass.} Moving to the total mass density slope,
%the mean value derived for our sample is $\langle \gamma_{\rm tot}^{\rm PL} \rangle = 1.998$ with a scatter $\sigma_{\rm tot} = 0.148$, 
we have found that (mean $\pm$ scatter) $\langle \gamma_{\rm tot}^{\rm PL}\rangle = 1.998 \pm 0.148$, $\langle \gamma_{\rm tot, 2}^{\rm PL}\rangle = 2.055 \pm 0.203$ and $\langle \gamma_{\rm tot}^{\rm AV}\rangle = 2.057 \pm 0.195$. These results show a nice agreement with a singular isothermal sphere model ($\gamma =2$ in Eq.~\ref{gen_SIS_model}).  In each definition, the slightly larger scatter of $\gamma_{\rm tot}$ with respect to $\gamma_{\rm DM}$ and $\gamma_{\star}$ is likely inherited from the gas distribution, which is the greatest source of stochasticity among the three components. Using the same radial range ($0.4r_{1/2}<r< 4 r_{1/2}$), \citet{Y.Wang} found $\langle \gamma_{\rm tot}^{\rm PL}\rangle = 2.011$ and $\sigma_{\rm tot} = 0.171$, which (again) is in perfect agreement with our result. 
%We indeed expected their result to  closely resemble the ones we report, since the same definitions and approach to compute the density slopes were applied here, and the only divergence is the mass interval, for which we consider galaxies with stellar masses as low as $10.3 \log M_{\odot}$ up to $11.9 \log M_{\odot}$. 
Using Illustris-1, \citet{D.Xu} found a stellar mass-weighted mean total slope $\langle \gamma_{\rm tot}^{\rm PL}\rangle_{\rm SMW} = 2.07$ and scatter $\sigma_{\rm tot} = 0.26$, again, for galaxies at $z=0.3$ and over the radial range $0.5R_{\rm e} < r < 2R_{\rm e}$, which is steeper compared to ours ($\langle \gamma_{\rm tot} \rangle_{\rm SMW} = 1.939$, see Table~\ref{slopes_stats}). 
  {
%As reported in the beginning of this paragraph, 
As mentioned, we have also estimated the mean total slopes by using $2r_{1/2}$ as the outer radial limit, namely $\gamma_{\rm tot,2}^{\rm PL}$ and $\gamma_{\rm tot}^{\rm AV}$, and we found that they are slightly steeper than the result considering an upper bound of $4r_{1/2}$, as also seen in Fig. \ref{summary_stat_slopes}. This is expected as the shorter radial range makes the total slope more aligned to the steeper stellar density slope, dominating in the very centre of galaxies.}

  {Looking at the observed galaxies, for a small sample of local fast-rotator ETGs with $10.2 < \log (M_{\star}/M_{\odot})<11.7$ from ATLAS$^{\rm 3D}$ \citep{Atlas3D} and SLUGGS \citep{Brodie+2014, Forbes+2016} surveys based on 2D stellar kinematics, \citet{Cappellari_2015} derived a mean total slope (defined over the radial range $0.1R_{\rm e} < r < 4R_{\rm e}$)
%, and found a mean 
of $\langle \gamma_{\rm tot}^{\rm AV} \rangle = 2.19 \pm 0.03$ with a rms scatter $\sigma_{\rm tot} = 0.11$. On the other hand, again for a small sample of fast-rotators from the SLUGGS survey of ETGs, using the JAM approach (see \S\ref{sec:intro}),  \citet{Bellstedt+2018} found $\langle \gamma_{\rm tot}^{\rm PL}\rangle = 2.24 \pm 0.05$ over the radial range $0.1 R_{\rm e} \leq r \leq 4 R_{\rm e}$. For the ATLAS$^\text{3D}$ sample, \citet{Poci} found $\langle \gamma^{\rm AV}_{\rm tot}\rangle = 2.193 \pm 0.016$ with $\sigma_{\rm tot} = 0.168\pm0.015$ over the radial range $0.1 R_{\rm e} \leq r \leq R_{\rm e}$. For the Frontier Fields data,
%(in which the radial profiles extend to $3R_{\rm e}$ on average) for galaxies at $0.29<z<0.55$, 
\citet{Derkenne} found a median $\gamma_{\rm tot}^{\rm AV}=2.11\pm 0.03$ (standard error), i.e., steeper than ours. } Lastly, we report the density slope results for the SPIDER sample of local ETGs from \citep{Tortora+2014}: for the mean mass-weighted density slope within $R_{\rm e}$, we found $\langle \gamma_{\rm tot}^{\rm MW} \rangle  = 2.149$ with a scatter $\sigma_{\rm tot} = 0.225$, which is shallower than the ones we found for TNG100-1, for the same reasons discussed before. On the other hand, using a larger radial interval, i.e., $r \in [0.3, 4] r_{1/2}$ (which is pretty similar to the one we used to define $\gamma^{\rm PL}$), we found the following power-law density slope to the SPIDER sample: $\langle \gamma_{\rm tot}^{\rm PL}\rangle = 2.127$ with a scatter $\sigma_{\rm tot} = 0.205$. Using this definition, the result is shallower than the former and with a slightly larger scatter, which is in line with our results for the virtual-ETG.

%  {Lastly,  for the DynPop sample of low-redshift ETGs \citep{Zhu+2023}, the mass-weighted slope, defined within the 3D half-light radius, has an average value $\langle \gamma^{\rm MW}_{\rm tot }\rangle = 2.218$ with $\sigma_{\rm tot} = 0.159$, over the same stellar mass range we used.}

Although all results above are statistically consistent within $1\sigma$ (see Fig. \ref{summary_stat_slopes}), %with the ones we report, 
we remark that our total slopes tend to be systematically shallower than the central values of all other samples, either if we consider the standard $4 r_{1/2}$ or the $2 r_{1/2}$ upper limit, and regardless the redshift. Once again, part of this discrepancy comes from the radial range adopted for the fit: 
%in comparison to the ones mentioned above. Using the same arguments as before, we emphasize again that our slopes are computed over a radial interval that goes up to the outermost parts of the galaxy, where the dark matter dominates, while the slopes in 
Illutris-1 is limited to 
%go a little beyond the innermost parts of the galaxies 
$2R_{\rm e}$, while Frontier Fields reaches $3R_{\rm e}$ (while as discussed previously, the lower bound of 0.16 arcsec is in line with our softening length). 
%, where the stellar component has a significant contribution. Thus, because the stellar component tend to have steeper slopes than the dark matter component,  it is expected to have a steeper $\gamma_{\rm tot}$ when considering smaller radial intervals, while at larger radii, the total slope becomes shallower as a consequence of the dark matter dominance. A similar line of thought can be applied to explain the divergence in relation to the Frontier Fields sample and the while 
 {On the other hand, both ATLAS$^\text{3D}$ \citep{Cappellari_2015} and SLUGGS  \citep{Bellstedt+2018} ETGs
%: because the effective radius is, in general, more compact than the stellar half-mass radius (see \S\ref{re_versus_hmr_sec} below), the Frontier Fields radial intervals likely get further into the cores of the galaxies (as close as $0.16$ arcsec), where the stellar matter dominates, and as it has a slope steeper than dark matter, its contribution steepens the total slope. The same applies for the ATLAS$^\text{3D}$ galaxies, where the 
have a lower bound of the radial interval, of $0.1 R_{\rm e}$. A typical approximation for $r_{1/2}$ in terms of $R_{\rm e}$ is given by $r_{1/2} \approx 1.33 R_{\rm e}$ \citep{Wolf}, so we estimate $0.1R_{\rm e}$ to be about $0.075r_{1/2}$, or roughly more than five times smaller than our lower-bound choice of $0.4r_{1/2}$. As we cannot check this short-range limit due to the poorer resolution of TNG100, we can only postpone this check to an analysis with higher-resolution simulations, where we expect that this systematics will go away.}

 {The bottom line of this section is that, looking at the slopes of the stars and dark matter component and the total ones, we find a global agreement if the underlying assumptions in terms of IMF and the radial range adopted for the slope derivation are consistent between simulations and observations. However, to make this comparison more robust, we will look now into the correlations of the total mass density slope with other galaxy properties.}

\subsection{Total density slope correlations}
%In this section, we turn our attention to the validation of typical correlations between the total density slope $\gamma_{\rm tot}$ and galactic parameters. 
In this section, we quantitatively compare the correlations of the total slopes and the other galaxy parameters derived for the virtual-ETG sample with the same relations derived 
%to the ones obtained 
from observational data.  {\it This is a very stringent validation test for our catalogue, 
%and, to our knowledge, the firt attempt of this kind, where
as we use a full set of observational-like parameters to derive scaling relations from simulated galaxies.} 
In particular, we compute the correlation of the total density slope with some Sérsic parameters (i.e., $n, R_{\rm e}$ and the derived stellar mass $M_{2D,\star}^{\rm tot}$) and kinematics, i.e., $\sigma_{\rm e}$,   {
%For our sample, we compare how these scaling relations vary with 
using the different slope definitions adopted as in \S\ref{sec:slope_definitions}.} In the analysis, we denote the slopes of the best-fit line as $\partial \gamma_{\rm tot}/\partial{X}$, for a particular parameter $X$ and the strength of the correlation by the Pearson's correlation coefficient $r$\footnote{\url{https://docs.scipy.org/doc/scipy/reference/generated/scipy.stats.pearsonr.html}}. We regard a correlation as being significant %if the corresponding slope deviates from zero for less than $3\sigma$, being $\sigma$ the error on the best fit parameters and also 
if the p-value of the best-fit linear regression is less than or equal to $0.01$ (see, e.g., \citealt{moore2009introduction} for a statistical framework). This ensures that we can reject the null hypothesis, which, in our case, states that there is no correlation in the data.

%  {Although we have confirmed that the slopes defined within smaller radial ranges (i.e., $[0.4, 2]r_{1/2}$) align more closely with those obtained from real galaxies, in this section, we use the power-law total mass density slopes defined within the radial range $[0.4, 4]r_{1/2}$. While not explicitly shown, all the results in this section remain statistically consistent with the other slope definitions we previously introduced (i.e., $\gamma_{\rm tot,2}^{\rm PL}$ and $\gamma_{\rm tot}^{\rm AV}$)}. 

For comparison, we use data products from the Sloan Lens ACS (SLACS) Survey \citep{SLACS}, the SL2S Galaxy-Scale Gravitational Lens Sample \citep{Gavazzi_2012}, the Dynamics and stellar Population (DynPop) from the MaNGA survey \citep{Zhu} and the SPIDER sample \citep{LaBarbera+2010}. %and ATLAS$^\text{3D}$ \citep{Atlas3D}. 
  {In particular, for DynPop we consider only the objects with the dynamical fits quality greater than zero, \texttt{QUAL} $> 0$ (i.e., acceptable fit for the velocity dispersion map). We use the mass-weighted total density slope within a sphere defined by the 3D half-light radius, derived from spherical Jeans Anisotropic Modeling (JAM) combined with three distinct dark matter models: a generalized Navarro-Frenk-White (gNFW) model, a free NFW model, and a fixed NFW model. In the fixed NFW model, parameters like the break radius are determined using scaling relations, which reduces the number of free parameters in the model.
%plus a generalized Navarro-Frenk-White (gNFW) profile to describe the dark matter distribution
} We also use the effective velocity dispersion within elliptical half-light isophote, circularized effective radius, calculated as the product between the root of the axial ratio $b/a$ from 2D Sérsic fit in SDSS $r$-band and the Sérsic 50\% light radius along major axis, the stellar mass from K-correction fit for Sérsic fluxes and, lastly, Sérsic index from 2D Sérsic fit in SDSS $r$-band. %For ATLAS$^{\text{3D}}$, the central velocity dispersion values  and logarithmic density slopes for ETGs with stellar masses between $10^{10}$ and $10^{11}$ $M_{\odot}$ are taken from \citet{Serra+2016}. 

For the SPIDER sample we use the Sérsic index, half-light radii, stellar mass (assuming a Chabrier IMF), stellar velocity dispersion within $R_{\rm e}$ and mass-weighted density slopes within $R_{\rm e}$ 
%for SPIDER galaxies were taken 
from \citet{Tortora+2014}, in which ETGS have stellar masses $\gtrsim 10^{10}~M_{\odot}$ . For the SL2S sample, we use the logarithmic density slopes, stellar masses assuming a Salpeter IMF and effective radius from \citet{Sonnenfeld_2013} for objects with stellar masses $\gtrsim 10^{10.8}~M_{\odot}$ for $0.2<z<0.8$. For consistency, we converted these stellar masses to a Chabrier  using $M_{\star,\text{Chab}} = 0.61 M_{\star, \text{Salp}}$ \citep{Madau+2014}, where $M_{\star,\text{Chab}}$ is the stellar mass derived assuming a Chabrier IMF and $M_{\star,\text{Salp}}$ is the one derived assuming a Salpeter IMF. For SLACS sample, we extract the power-law slope, effective radius, and stellar mass with Chabrier IMF from \citet{Auger+2010}, where the ETGs have stellar masses $\gtrsim 10^{10} M_{\odot}$ and the redshift range is $0.24< z < 0.78$. Although these ETGs samples have (projected) total stellar masses approximately equivalent to ours, we selected only those with $M_{\rm 2D,\star}^{\rm tot} \geq 10.5$ to ensure a more consistent comparison. %, i.e., they are in the range $10^{10.3}\leq M/M_{\odot}\leq 10^{11.9}$.  
Moreover, for \citet{Tortora+2014}, we selected only those galaxies with $n \geq 2$ and for \citet{Zhu} those with $2\leq n\leq5.9$ (see \S\ref{sec:size-mass}). After filtering the relevant data from the samples, we are left with $504$ objects for \citet{Zhu}, $25$ for \citet{Sonnenfeld_2013}, $59$ for \citet{Auger+2010}, and the greatest dataset is from \citet{Tortora+2014}, with $4112$ galaxies.

\begin{figure*}
    \centering
    \includegraphics[width=0.95\columnwidth]{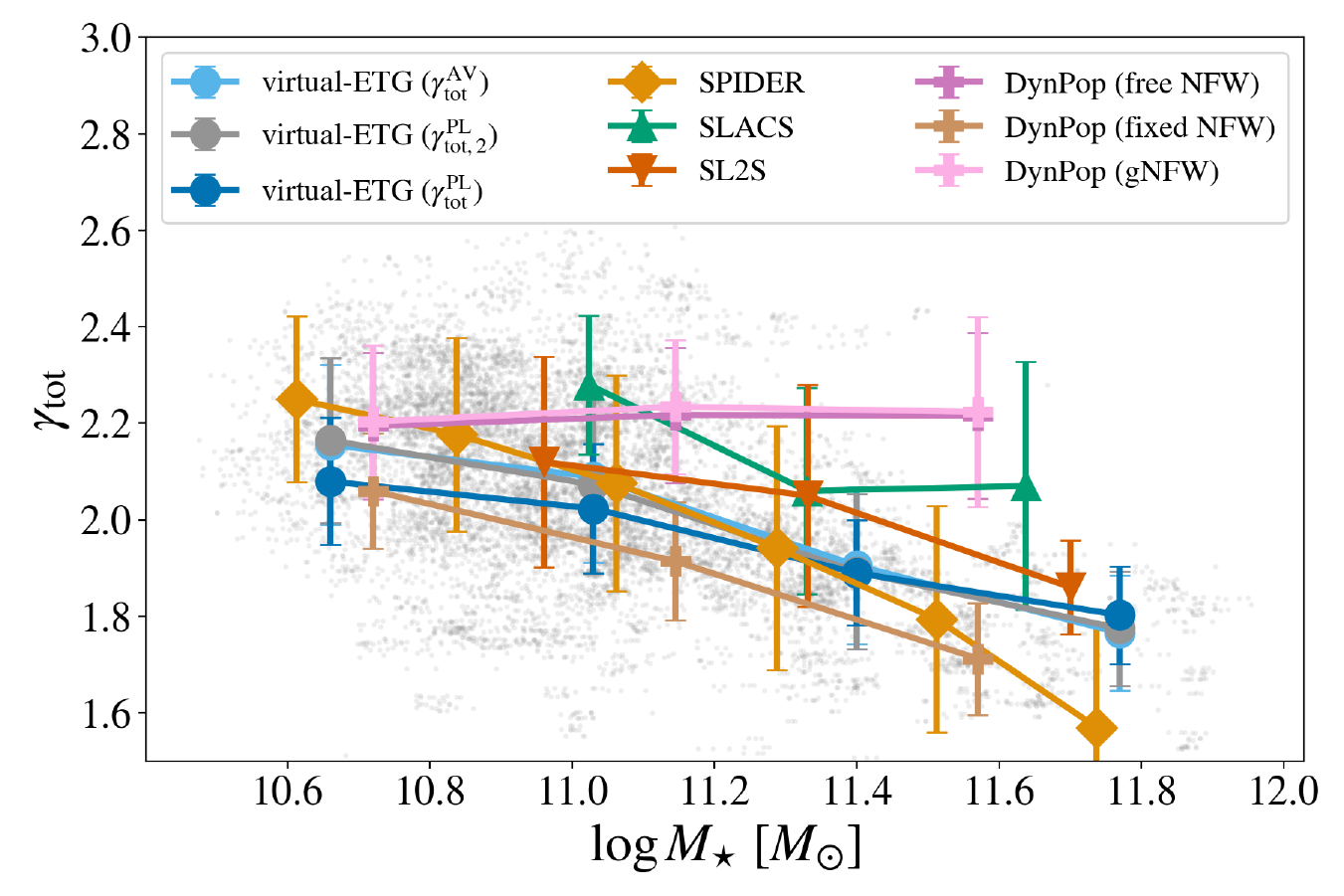}
    %\centering
    \hspace{-0.2cm}
    \includegraphics[width=0.95\columnwidth]{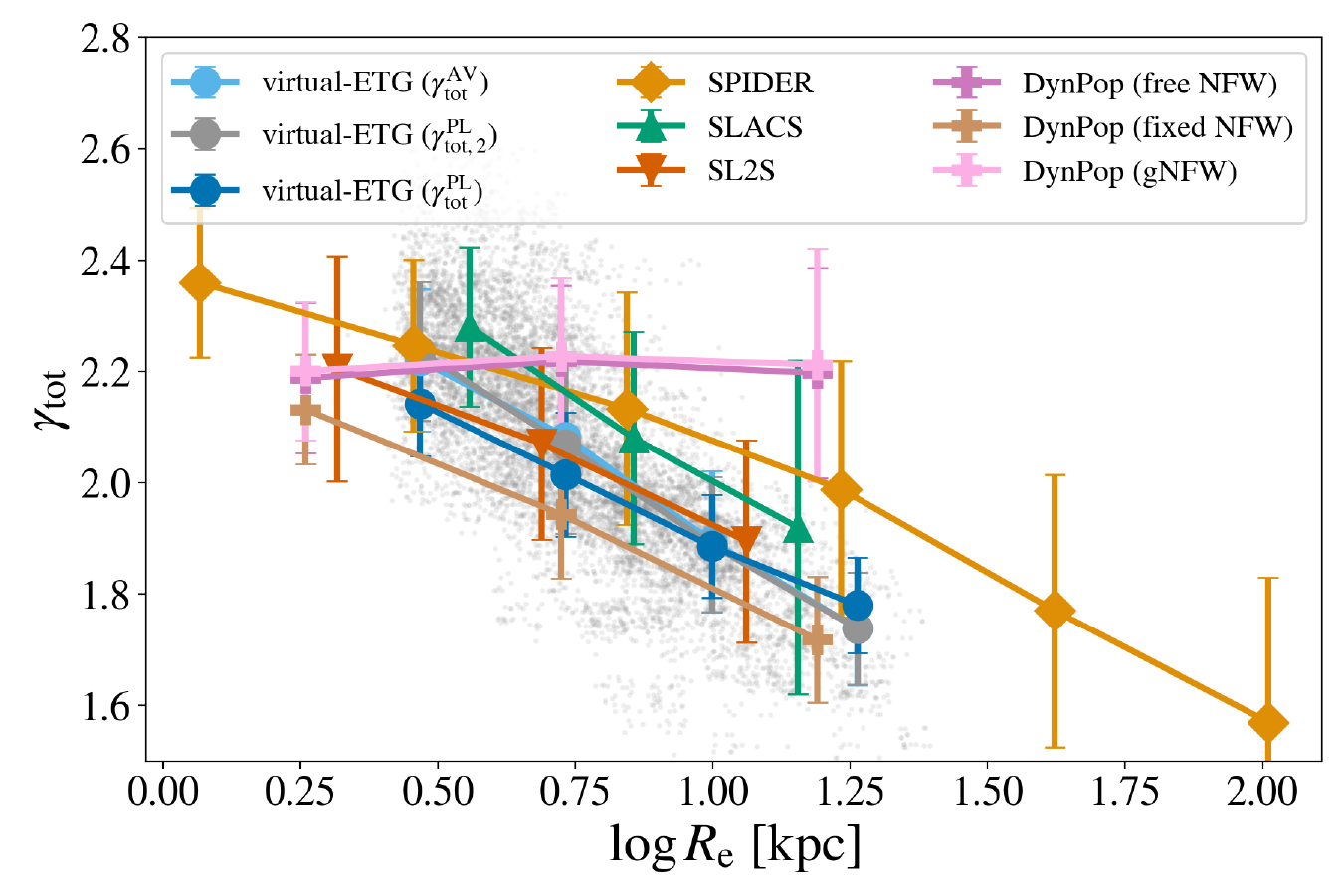}

    \centering
    \includegraphics[width=0.95\columnwidth]{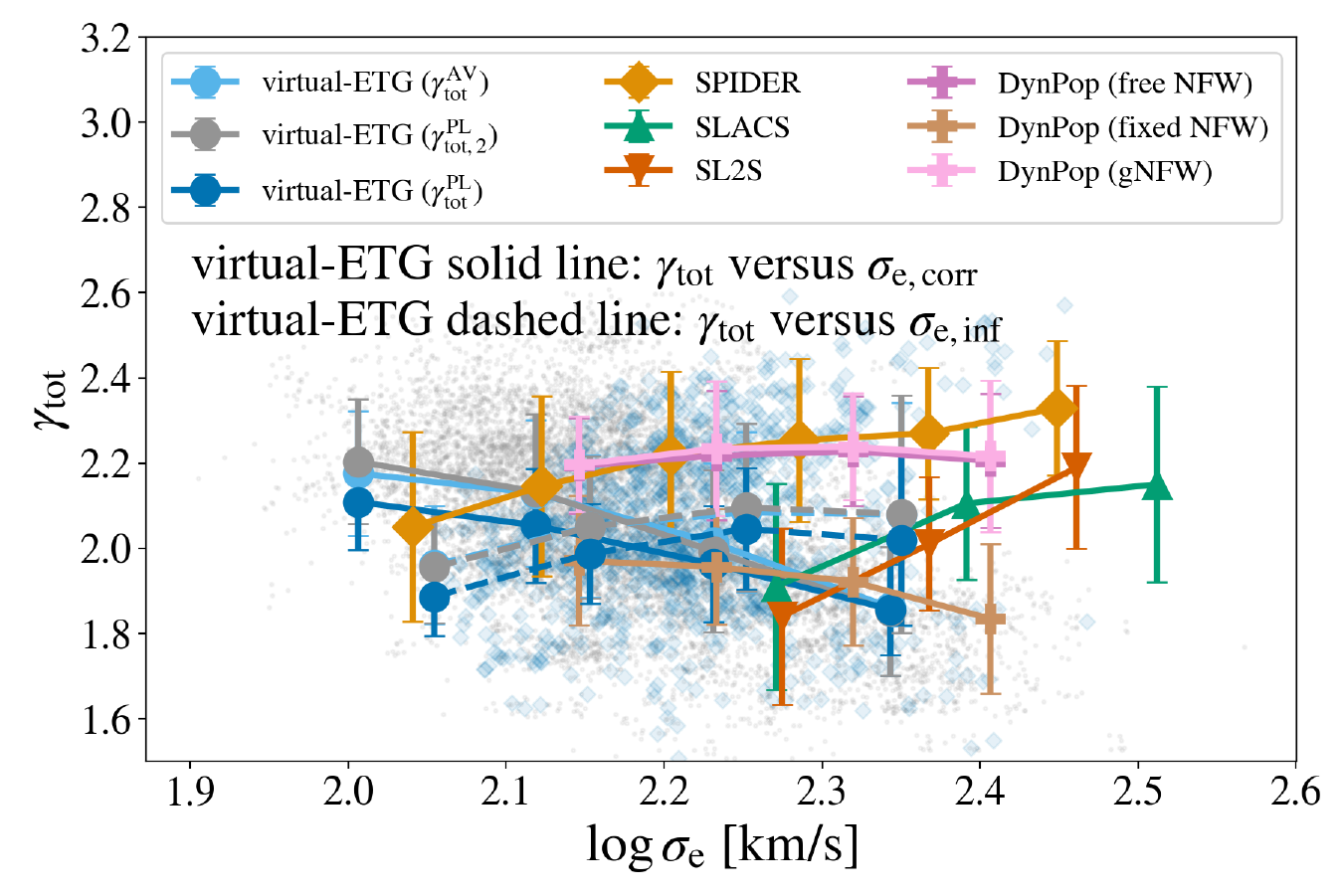}
    %\centering
    \hspace{-0.2cm}
    \includegraphics[width=0.95\columnwidth]{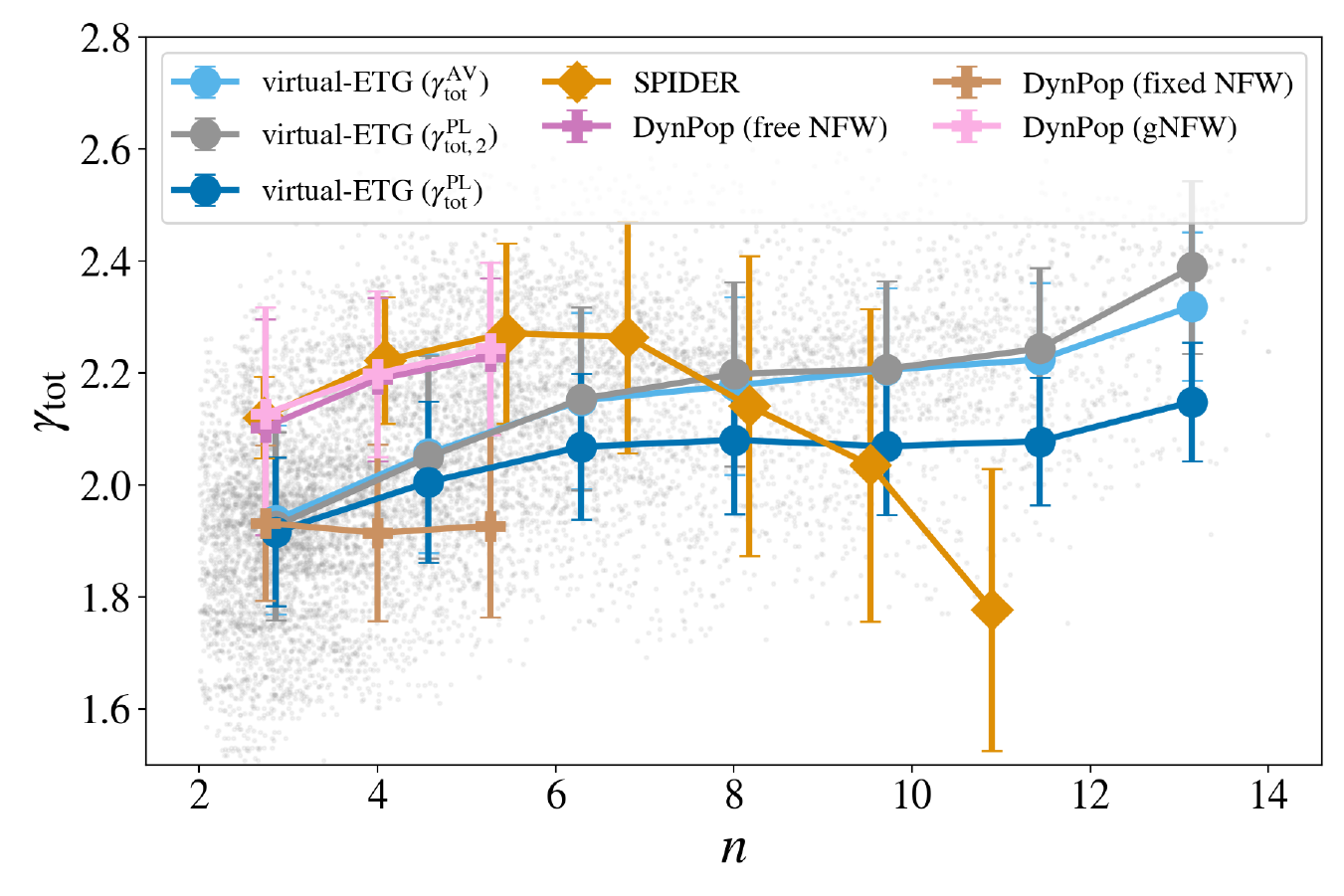}
    \caption{Relation between the total density slope $\gamma_{\rm tot}$ and the stellar mass $M_\star$ (upper left), effective radius $R_{\rm e}$ (upper right), central velocity dispersion $\sigma_{\rm e}$ (lower left) and Sérsic index $n$ (lower right). In each figure, we show the median trend ($\pm$ the standard deviation) of our sample for the three slope definitions we introduced before. The grey points are the scatter of each correlation with respect to the $\gamma_{2}$ slope. In particular, for the $\gamma-\sigma$ correlation, we show the scatter plot of the $\gamma_2$-$\sigma_{\rm e, corr}$ relation in grey and the scatter plot of the $\gamma_2$-$\sigma_{\rm e, dyn}$ in blue.}
    \label{fig:gamma_correlations}
\end{figure*}

%FROM HERE
\subsubsection{Correlation with stellar mass}\label{corr_stellar_mass}

%We now turn our attention to the correlation between the total density slope and the total stellar mass of the ETGs. 
We start the inspection of the $\gamma_{\rm tot}$ correlations with galaxy parameters from the stellar mass, which is possibly the less biased quantity one can obtain from an observed sample, provided that the IMF assumption is clearly stated (see, e.g., \citealt{Xie+2023SCPMA}).
%But before proceeding, we may add a discussion about the derivation of a total stellar mass based on the Sérsic parameters.
  {Typically, the projected ``total'' stellar mass should closely approximate its 3D counterpart, making them equivalent (see \S\ref{projected_versus_3d_stellar_mass} for non total quantities). However, to account for the effects of projection, we opted to use a projected definition, derived from Sérsic photometry, instead of a 3D definition, obtained directly from the simulation. For consistency, we also keep this 2D stellar mass definition for the observational samples.} 

For each slope definition, the best fit line to the $\gamma_{\rm tot}-\log M_{\star}$ relation gives us $\partial \gamma_{\rm tot}^{\rm PL}/\partial \log M_{\star} = -0.281 \pm0.005$, $\partial\gamma_{\rm tot,2}^{\rm PL}/\partial \log M_{\star}  = -0.388 \pm 0.007$ and $\partial\gamma_{\rm tot}^{\rm AV}/\partial \log M_{\star}  = -0.369 \pm 0.006$. Their Pearson's correlation coefficient are, respectively, equal to $-0.505$, $-0.507$ and $-0.5$. In Fig. \ref{fig:gamma_correlations} (upper left corner) we present the median trend for these various definitions and compare them with observational samples. For comparison with other literature results using the IllustrisTNG, \citet{Y.Wang} performed a similar study but using the 3D stellar mass defined as the sum of all the stellar particles assigned to its host
subhalo by \texttt{SUBFIND}. With this definition, they obtained $\partial \gamma_{\rm tot}^{\rm PL}/\partial \log M_{\star, \rm 3D} = -0.41  \pm 0.03$ with a Pearson's correlation coefficient $r=-0.58$ at $z=0$. Knowing that their $\gamma_{\rm tot}$ and stellar mass range are consistent with ours, it is clear that the $\gamma_{\rm tot} -\log M_{\star}$ correlation is diluted when one considers a projected definition for the stellar mass, rather than the one directly extracted from the simulation.

Moving to the comparison with the observational datasets, among all the samples, DynPop \citep{Zhu} is the one that gives the shallowest $\gamma - M_\star$ correlation when considering the slopes extracted from the free NFW and generalized NFW models. In fact, these correlations are negligible, with a p-value $> 0.01$. However, for slopes obtained from the fixed NFW model, the $\gamma -M_\star$ correlation aligns with our results within the scatter, yielding a slope of $\partial \gamma_{\rm tot}^{\rm MW}/\partial \log M_{\star} = -0.408 \pm 0.019$ and a stronger Pearson's correlation coefficient of $ -0.693$. This slope is $1\sigma$ consistent with our result for the correlation between the stellar mass and the slope $\gamma_{\rm tot,2}^{\rm PL}$, with the main difference in the trend being an offset. We attribute this offset to the fact that the total density slopes from the fixed NFW model are systematically shallower than ours, with the measured average being $\langle \gamma_{\rm tot}^{\rm MW}\rangle = 1.915 \pm 0.160$. 

For the SLACS sample \citep{Auger+2010}, we have $\partial \gamma_{\rm tot}^{\rm PL}/\partial \log M_{\star} =-0.106 \pm 0.142$, but with a very weak anti-correlation, $r = -0.099$, which is negligible (p-value $>0.01$). This is the second most deviating trend compared to our results. The steepest   $\gamma_{\rm tot}-\log M_{\star}$  correlation is observed for the SPIDER sample \citep{Tortora+2014}, with $\partial \gamma_{\rm tot}^{\rm MW}/\partial \log M_{\star} = -0.477 \pm 0.014$, and $r = -0.502$, Overall, this correlation is in good agreement with ours within the stellar mass interval $10.6 \lesssim \log M_\star \lesssim 11.5$. Lastly, for the SL2S sample from \citet{Sonnenfeld_2013} we have: $\partial \gamma_{\rm tot}^{\rm PL}/\partial \log M_{\star} = -0.226 \pm 0.168$, and a weak anti-correlation $r = -0.270$ that although is consistent with our result within the scatter, is also negligible due to the high p-value. %Apart from the DynPop with free NFW and gNFW models, as well as the SLACS sample, these observational results are consistent with ours within the scatter (see Fig.~\ref{fig:gamma_correlations}, upper left corner). 

These results show consistency in what regards the direction of the $\gamma-M_\star$ correlation, which is always negative. An anti-correlation between $\gamma_{\rm tot}$ and the total stellar mass is expected, since the stellar component of less massive galaxies is generally more compact, implying steeper density slopes and vice versa (see, e.g., \citealt{Remus+2013,Tortora+2014} for more theoretical considerations). 

\subsubsection{Correlation with effective radius}
%For this case, we use $\log R_{\rm e}$ to study the trend of the total slope, because, by definition, $\gamma_{\rm tot} = - \partial \log\rho/\partial\log r$ (\ref{power_slope_diff_form}), so it is expected to have a better size-slope relation when comparing $\gamma_{\rm tot}$ with the log-scaled size, which in this case is $R_{\rm e}$. 
For each $\gamma_{\rm tot}$ definition, the slope of the $\gamma_{\rm tot}-\log R_{\rm e}$ relation is found to be $\partial \gamma_{\rm tot}^{\rm PL}/\partial \log R_{\rm e} = -0.553 \pm 0.005$, $\partial \gamma_{\rm tot,2}^{\rm PL}/\partial \log R_{\rm e} = -0.756 \pm 0.006$ and $\partial \gamma_{\rm tot}^{\rm AV}/\partial \log R_{\rm e} = -0.71 \pm 0.006$. Their Pearson's correlations are, respectively $-0.773$, $-0.769$ and $-0.749$. Fig. \ref{fig:gamma_correlations} (upper right corner) summarizes the relationship between them, along with data from observational samples. The correlation between $\gamma_{\rm tot}$ and $\log R_{\rm e}$ was also examined by \citet{Wang+2019}, where they used the projected 2D stellar half-mass radius—calculated from all stellar particles assigned to the galaxy identified by \texttt{SUBFIND}—along the X-axis of the simulation box as their proxy of the effective radius $R_{\rm e}$. They reported a slope of $\partial \gamma_{\rm tot}^{\rm PL}/ \partial \log R_{\rm e} = -0.64 \pm 0.02$ and a correlation coefficient of $r = -0.80$ at $z=0$. Their $\gamma_{\rm tot} - \log R_{\rm e}$ relation is notably tighter than ours. Although both studies rely on projected galaxy sizes, their measurement of $R_{\rm e}$ is model-independent and may introduce less stochasticity compared to ours. Furthermore, they only considered a single projection direction from the simulation box, whereas we accounted for variance by using three orthogonal projections. This difference in methodology may contribute to the discrepancy in the slope and strength of the correlation obtained by Wang et al. and us.

In general, the direction of the $\gamma-R_{\rm e}$ correlation is consistent across all samples. However, the DynPop \citep{Zhu} sample with a free NFW and a gNFW model for the dark matter distribution represents the weakest correlations, which are, in fact, negligible, having a p-value above the threshold we imposed. On the other hand, the ETGs from DynPop considering a fixed NFW model show consistency with the virtual-ETG, although shallower, with a slope of $\partial \gamma_{\rm tot}^{\rm MW}/\partial\log R_{\rm e} =-0.556 \pm 0.019$ and a Pearson's correlation coefficient of $-0.788$. This slope is consistent with our result for the correlation between the effective radius and the slope $\gamma_{\rm tot}^{\rm PL}$ at the $1\sigma$ confidence level.

For the SLACS sample \citep{Auger+2010}, the best linear fit gives $\partial \gamma_{\rm tot}^{\rm PL}/\partial \log R_{\rm e} = -0.253 \pm 0.161$, and $r = -0.206$, which is also negligible because of its high p-value. Although this slope is measured with a high uncertainty, (possibly due to the small sample size), this is consistent with the virtual-ETG for the correlation between $R_{\rm e}$  and our $\gamma_{\rm tot}^{\rm PL}$ at the $2\sigma$ confidence level. The SLACS trend for this correlation aligns well with our results within the scatter (see Fig.~\ref{fig:gamma_correlations}, upper right corner), although with a systematic offset in the $\gamma_{\rm tot}$ direction, which stems from the fact that the observational total density slopes are generally steeper than the simulated ones (see, e.g., the results reported in \S\ref{sec:slope_res_vs_lit})

The SPIDER sample of galaxies from \citet{Tortora+2014} have a slope of $\partial \gamma_{\rm tot}^{\rm MW}/\partial \log R_{\rm e} = -0.382 \pm 0.009 $ with $r = -0.575$, This slope is significantly shallower than all our estimates, differing by more than $10\sigma$ from our results, although it has a matching median trend with our virtual-ETG sample within the scatter. Lastly, for the SL2S sample from \citet{Sonnenfeld_2013}, the results from the best linear fit are statistically compatible with the ones reported here: $\partial \gamma_{\rm tot}^{\rm MW}/\partial \log R_{\rm e} = -0.501 \pm 0.135$ and $r=-0.611$. This result is statistically consistent with our estimates within $1\sigma$ for the $\gamma_{\rm tot}^{\rm PL}-R_{\rm e}$ relation and within $2\sigma$ for both $\gamma_{\rm tot,2}^{\rm PL}-R_{\rm e}$ and $\gamma_{\rm tot}^{\rm AV}-R_{\rm e}$ correlations.

Galaxies with smaller $R_{\rm e}$ tend to have lower overall stellar masses (see \S\ref{sec:size-mass}), consistent with the galaxy-halo connection, which also suggests a smaller dark matter mass \citep{Wechsler}.
This implies that smaller galaxies tend to have a more concentrated matter distribution relative to larger ones. Consequently, the slope of the total density profile is likely to be steeper in smaller galaxies, while more diffuse galaxies (those with larger $R_{\rm e}$) are expected to have shallower total density slopes \citep{Remus+2013}. Overall, apart from the DynPop \citep{Zhu} sample using the slopes from the mass distribution inferred via the free NFW and gNFW models, the observational results reported in this section are in agreement with the virtual-ETGs sample within the scatter. Moreover, the general result is that an anti-correlation between $\gamma_{\rm tot}$ and $\log R_{\rm e}$ is evident among  the results discussed above.

\subsubsection{Correlation with
line-of-sight velocity dispersion}
%We define this correlation with $\sigma_{\rm e}$ in log scale. 
The global trend of the $\log \sigma_{\rm e}-\gamma_{\rm tot}$ correlation is shown in Fig. \ref{fig:gamma_correlations} (lower left panel). Here we report the two $\sigma_{\rm e}$ definitions we introduced in \S\ref{catalogue_section} and, as we can see, there are substantial differences, which we describe below. 

For the $\sigma_{\rm e, corr}$  (solid lines in tones of blue and grey) we find negative correlations: $\partial \gamma_{\rm tot}^{\rm PL} / \partial \log \sigma_{\rm e} = -0.808 \pm 0.014$, $\partial \gamma_{\rm tot,2}^{\rm PL} / \partial \log \sigma_{\rm e} = -1.118 \pm 0.019$ and $\partial \gamma_{\rm tot}^{\rm AV} / \partial \log \sigma_{\rm e} = -1.037 \pm 0.018$. The respective correlation strengths are $r=-0.504$, $-0.509$ and $-0.489$. The trend of the $\sigma_{\rm e, corr}$, is not strongly dissimilar from \citet{Y.Wang} who also explored the relationship between $\gamma_{\rm tot}$ and central velocity dispersion and reported a weaker Pearson correlation coefficient of $r=-0.37$, using the velocity dispersion from all star particles
%There are two main reasons for the divergence between our findings. First, they measured $\sigma_{\rm e}$ 
within an aperture of $0.5 R_{\rm e}$. 

  {These trends suggest that galaxies with a higher velocity dispersion have shallower density profiles which is in contrast with the results from most of the observational samples, that generally show a positive correlation (see, e.g., Fig.~\ref{fig:gamma_correlations} and \citealt{Y.Wang}). Notably, the only observational sample that deviates from this pattern is DynPop \citep{Zhu}. This sample shows very weak $\gamma-\sigma$ correlations when using the free NFW and gNFW models ($r = 0.038$ and $0.045$, respectively, which are negligible, having p-value $>0.01$), that partially overlap with the SPIDER results. On the other hand, DynPop has a stronger anti-correlation if the fixed NFW model is adopted, which aligns nicely with our results for the corrected $\sigma_{\rm e}$ (see Fig.~\ref{fig:gamma_correlations}, lower left panel). For this DM model, the slope of the anti-correlation is $\partial \gamma_{\rm tot}^{\rm MW}/\partial \log \sigma_{\rm e} =-0.497 \pm 0.072$ and $r = -0.296$. A weak $\gamma-\sigma$ correlation (or even mildly negative) for MaNGA ETGs  has also been reported in other studies for $\log\sigma_{\rm e}~[\mathrm{km/s}]> 2.1$ \citep{Li+2019, Li+2024,Zhu+2024}. We possibly track this shallowness or even inversion in the correlation direction -- in contrast with findings from other observational samples -- to a strong incompleteness of objects below $\log\sigma_{\rm e}~[\mathrm{km/s}] \approx 2.2$ (see Fig.~\ref{fig:sig_comp}). As such, any correlation derived below this threshold is likely unreliable, possibly limiting the utility of the DynPop sample as a baseline for comparison.}

However, even forgetting the incomplete samples,
%letting aside possible sample incompleteness, as the one discussed above, 
the $\gamma-\sigma$ correlation is generally positive in all other observational samples, which is in contrast with the trend reported for
%. However, according to the results we reported for 
the virtual-ETG sample and also the results obtained by \citet{Y.Wang}.
%, regardless the regions of the galaxies used for measuring the kinematics, and the best correction one can apply to account for the missing kinematics from the central regions, the intrinsic kinematical properties of TNG100 galaxies cannot reproduce the observations. 
On the other hand, if using the dynamically motivated $\sigma_{\rm e, dyn}$, i.e., the one, we remind, emulating the realistic particle dynamics in absence of softened potential, the situation is enormously improved and in Fig. \ref{fig:gamma_correlations} (lower left panel), we see the virtual-ETGs nicely reproduce a positive trend, with $\partial \gamma_{\rm tot}^{\rm PL} / \partial \log \sigma_{\rm e} =0.477 \pm 0.019$, $\partial \gamma_{\rm tot,2}^{\rm PL} / \partial \log \sigma_{\rm e} = 0.423 \pm 0.027$, and $\partial \gamma_{\rm tot}^{\rm AV} / \partial \log \sigma_{\rm e} =0.32 \pm 0.026$, and the Pearson correlation coefficients are, respectively, $r = 0.243$, $0.157$ and $0.123$. 
%In terms of ``absolute value'', the virtual-ETG sample places itself in between the SPIDER sample and the strong lensing samples (SLACS ans SL2S), showing a rather large scatter, when using the $\sigma_{e, \rm dyn}$. This is due to the large shift toward larger dispersion values of the corrected $\sigma$ at the high $\gamma$ value on the top of the graphics, which is in line with our interpretation of the impact of the softening length at these high $\gamma$, discussed in \S\ref{app:sigma_corr}. 

%, as opposed to using the corrected $\sigma_e$ inside the full effective radius as we did
%, probing different regions of the galaxies. Second, and most importantly, they did not use a logarithmic scale for velocity dispersion but instead analyzed the linear velocity dispersion $\sigma_{\rm e}$. Because of this difference in scaling, we do not report their slope for comparison. The relationship between $\gamma_{\rm tot}$ and $\sigma_{\rm e}$ on a linear scale does not directly correspond to the one between $\gamma_{\rm tot}$ and $\log \sigma_{\rm e}$, as logarithmic scaling transforms the slope and tends to linearize power-law-like relationships. As a result, comparing the slopes would be misleading, since the nature of the correlations is fundamentally altered by the choice of scale.

%Thus, the combination of different apertures and the use of linear instead of logarithmic scaling in their analysis accounts for both the lower Pearson correlation of $-0.37$ found by Wang et al. and the lack of direct comparability in the reported slopes.

For a more quantitative comparison, 
the SPIDER sample of galaxies \citep{Tortora+2014} have $\partial\gamma_{\rm tot}^{\rm MW}/\partial \log \sigma_{\rm e} =1.007 \pm 0.033$, and $r = 0.461$.  We also show in Fig. \ref{fig:gamma_correlations} the strong lensing results from SL2S \citep{Sonnenfeld_2013} and SLACS \citep{Auger+2010}, that are the ones with steeper $\partial \gamma / \partial \log \sigma_{\rm e} $, possibly because of their very tight $\log \sigma_e$ range. For SL2S, $\partial\gamma_{\rm tot}^{\rm PL}/\partial \log \sigma_{\rm e} = 1.568 \pm 0.486$, $r =  0.558$ and, for SLACS, $\partial\gamma_{\rm tot}^{\rm PL}/\partial \log \sigma_{\rm e} = 1.57 \pm 0.359$, $r= 0.505$. Although both these results have slopes consistent with our results for the $\gamma_{\rm tot}^{\rm PL}- R_{\rm e}$ correlation within $2\sigma$ for SL2S and $3\sigma$ for SLACS, we stress that their reliability is questionable, since these samples' sizes are relatively small.

  {Lastly, referring again to Fig.~\ref{fig:gamma_correlations} (lower left panel), we stress that the trend of the $\gamma-\sigma$ correlation using the $\sigma_{\rm e,dyn}$ definition is consistent within the scatter with DynPop (free NFW and gNFW models) and SPIDER samples, with the main discrepancy being the offset, which is likely related to the slope definition  (see \S\ref{sec:slope_res_vs_lit}) and the choice of DM model (see also above). Although the $\gamma_{\rm tot}-\sigma_{\rm dyn}$ correlations observed in the virtual-ETG are weaker than in observations, our results show that the main reason for the discrepancy between observations and simulation for the $\gamma_{\rm tot}-\log\sigma_{\rm e}$, largely discussed in literature (see \citealt{Rodriguez+2019}, \citealt{Y.Wang}, \citealt{D.Xu}) is {\it mainly driven by the softened dynamics of the central regions}, although it is also partially depending on the radial range adopted for the $\gamma$ measurement, as well as the model choice to describe the distribution of the dark component. This leads us to reject the arguments that the mismatch between the $\gamma_{\rm tot}-\log\sigma_{\rm e}$ trend in simulations and observations is mainly linked to a limitation on the baryonic models of the TNG simulations (see, e.g., \citealt{Y.Wang}). This conclusion is corroborated from similar evidence from the fundamental plane (see de Araujo Ferreira et al., in preparation). }

\begin{figure}
    \centering
    \includegraphics[width=\columnwidth]{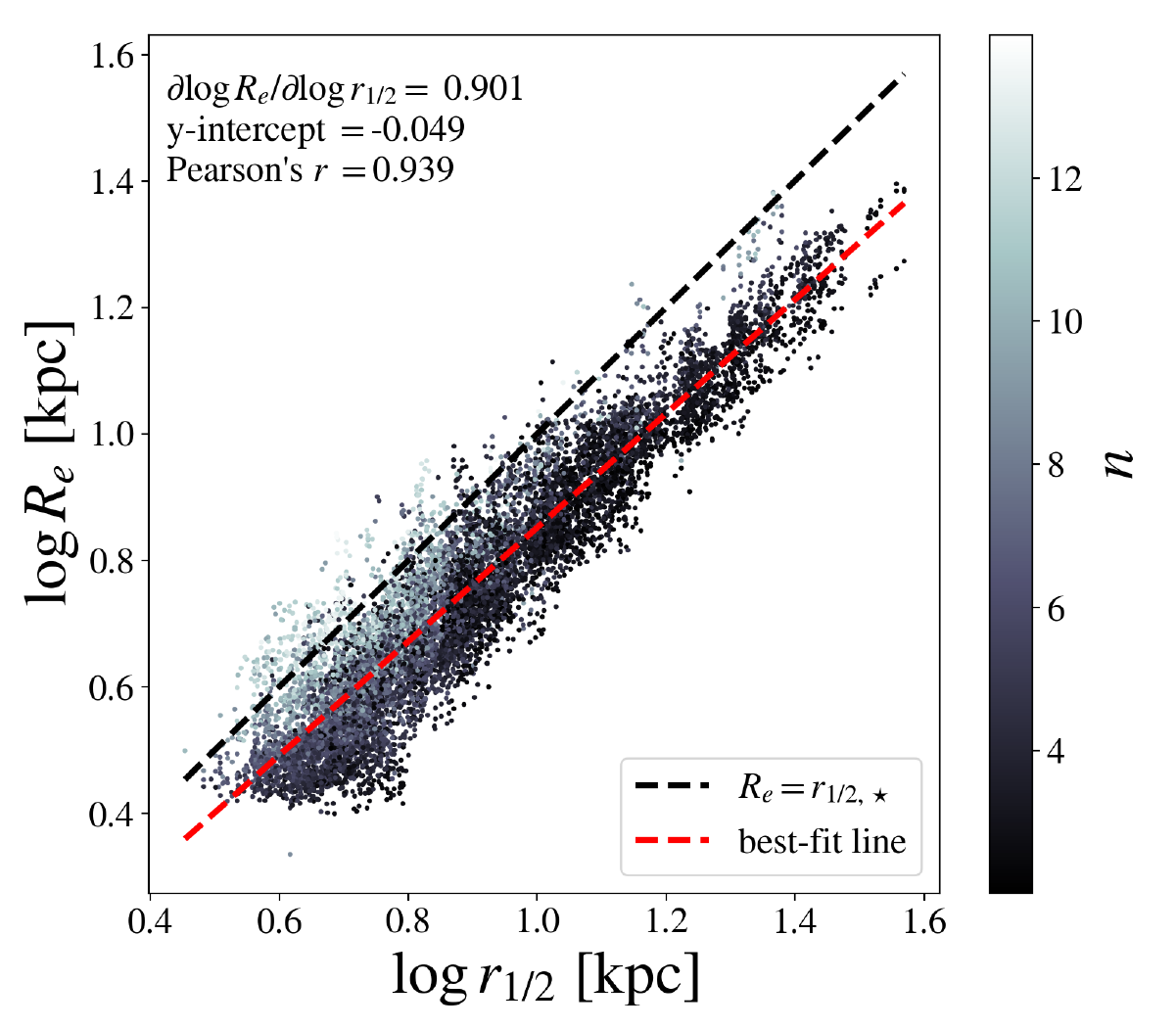}
    \caption{Relation between the stellar half-mass radius and the effective radius. In red is the identity line and in black is the best-fit line for the $r_{1/2\star}-R_{\rm e}$ relation. The parameters of the best-fit are included in the upper-left corner. The colours represent the Sérsic index value, where lighter dots have higher $n$ and darker ones represent lower $n$ values.}
    \label{Re_versus_hmr}
\end{figure}

\subsubsection{Correlation with Sérsic index}

Finally, we investigate the correlation between the total mass density slope $\gamma_{\rm tot}$ and the Sérsic index. This is shown in Fig. \ref{fig:gamma_correlations} (lower right panel) where we have measured the following slopes: $\partial \gamma_{\rm tot}^{\rm PL}/\partial n =0.021 \pm0.001$, $\partial \gamma_{\rm tot,2}^{\rm PL}/\partial n = 0.04 \pm 0.001$, $\partial \gamma_{\rm tot}^{\rm AV}/\partial n = 0.038 \pm 0.001$, and their Pearson's correlation coefficients are, respectively, $r = 0.386$, $0.535$ and $ 0.514$. %Fig. \ref{total_slope_versus_n} shows the trend between these two parameters. 
The overall correlations are weak, but given our criterion of correlation significance, they cannot be discarded. 

Looking at the full SPIDER samples of galaxies \citep{Tortora+2014}, we remark a descending trend for $n>6$, which is not found in any of the other datasets, either observed or simulated. Consequently, their slope is negative: $\gamma_{\rm tot}^{\rm MW}/\partial n = -0.026 \pm 0.002$ and $r= -0.246$. However, notice that the median suggests a non-linear relationship between $\gamma_{\rm tot}$ and $n$ over the full Sérsic index interval of the SPIDER sample. This non-linear relationship was explored in \citet{Tortora+2014} by adopting a quadratic dependence.

However, for low values of the Sérsic index, i.e. $n \lesssim 6$, the virtual-ETGs show trends that are similar to those of SPIDER and DynPop (free NFW and gNFW models). Conversely, the trend is reversed for DynPop models with a fixed NFW model, which has a negligible correlation, i.e., p-value $>0.01$. For a more quantitative comparison, we can compare the slopes by imposing this limit to the Sérsic index values on the virtual-ETG and SPIDER samples. With this limit, for virtual-ETGs the slopes are: $\partial\gamma_{\rm tot}^{\rm PL}/\partial n = 0.049 \pm 0.001$,
$\partial\gamma_{\rm tot,2}^{\rm PL}/\partial n = 0.077 \pm 0.002$, and $\partial\gamma_{\rm tot}^{\rm AV}/\partial n = 0.071 \pm 0.002$ with $r=0.366$, $0.443$ and $0.417$, respectively. For the SPIDER sample, its slope is $\partial \gamma_{\rm tot}^{\rm MW}/\partial n = 0.042 \pm 0.003$ with $r=0.295$. This SPIDER slope is now consistent with ours for the $\gamma_{\rm tot}^{\rm PL}-n$ correlation at the $2\sigma$ confidence level, but  $<7\sigma$ consistent with the correlations using $\gamma_{\rm tot,2}^{\rm PL}$  and $\gamma_{\rm tot}^{\rm AV}$. For the DynPop sample of galaxies, the slope (Pearson's r) of the correlation is $\partial \gamma_{\rm tot}^{\rm MW}/\partial n = 0.034 \pm 0.008$ ($r = 0.182$) for the free NFW model and $\gamma_{\rm tot}^{\rm MW}/\partial n =0.028 \pm 0.009$ ($r = 0.144$) for the gNFW model. The slope of the DynPop correlation with a free NFW model is consistent with ours within $2\sigma$ and within $3\sigma$ for the gFNW model.

\begin{figure*}
    \centering
    \includegraphics[width=0.95\columnwidth]{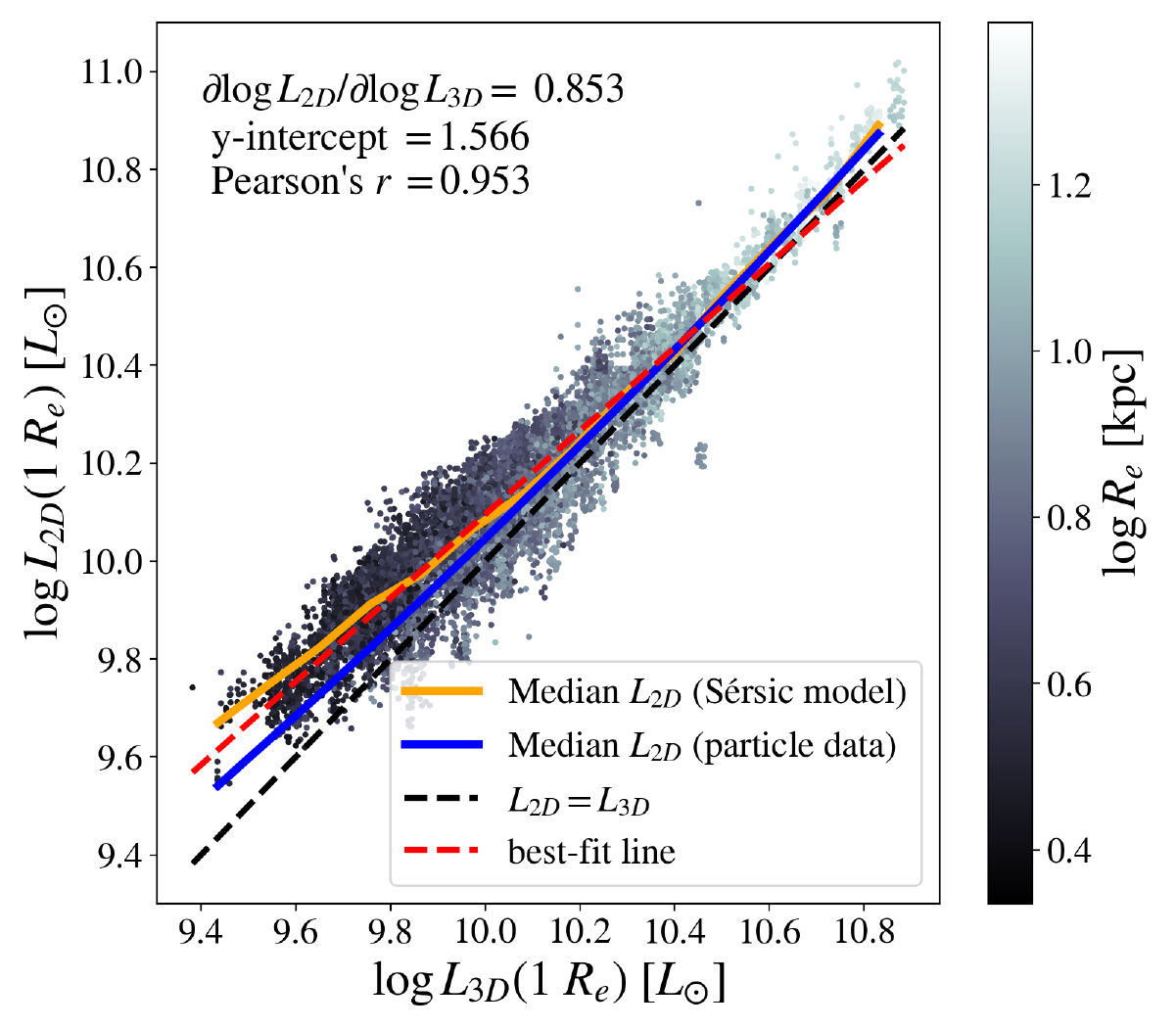}
    \includegraphics[width=0.95\columnwidth]{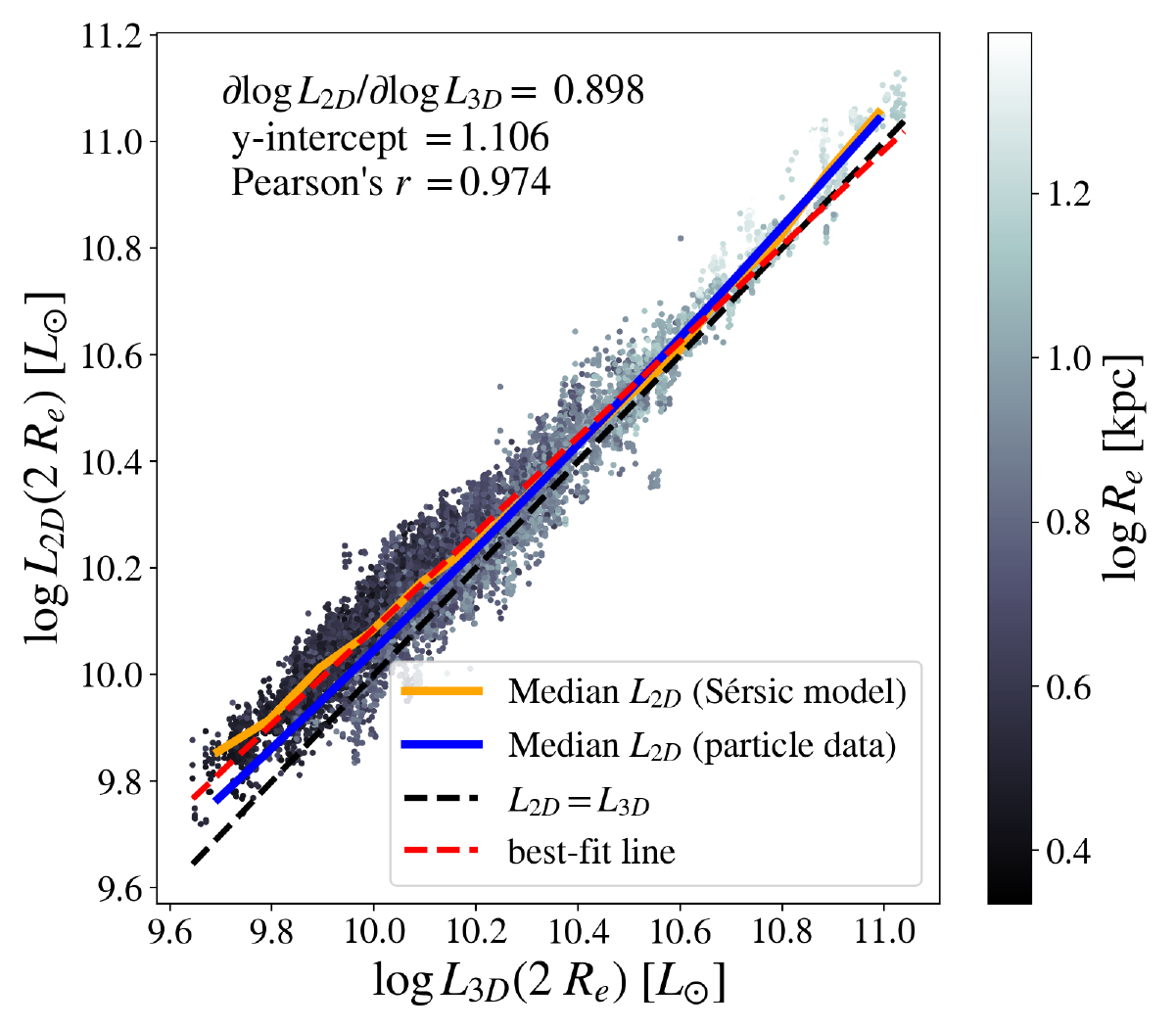}  
\caption{Projected luminosity inside one effective radius (left) and inside two effective radii (right) computed assuming a Sérsic model versus 3D luminosity inside the same aperture computed from the simulation's particle data. In red is the best-fit line to the relation, in orange is the median trend of the modelled $L_{2D}$ (in 20 bins of $L_{3D}$) computed from the points shown and, similarly, in blue we show the median trend of the projected light computed from the simulation's particle data.}
    \label{2D_versus_3D}
\end{figure*}

%For the DynPop galaxies, we found that $\partial \gamma_{\rm tot}/\partial n = 0.082\pm 0.005$, $\beta = 1.756 \pm 0.021$ and $r=0.453$, which is steeper than ours but on a tiner range of $n$ ($\sim 2-6$), where also vthe virtual-ETG has a locally steeper $\partial \gamma_{\rm tot}/\partial n$. Despite the larger offset, 
%the results   {are not fully consistent}, 

%  {the slope of the correlation is consistent with ours, so} the overall trend of $\gamma_{\rm tot}$ being directly proportional to $n$ is consistent between DynPop and our sample.
%, and agrees with the median trends. 

Overall, based on the samples discussed above, the results suggest that $\gamma_{\rm tot}$ is directly proportional to $n$ for $n\lesssim 6$. In other words, the mock galaxies from our catalogue well reproduce the behaviour of the $\gamma_{\rm tot}-n$ relation   {for low values of $n$}. In fact, one should expect a positive correlation between these parameters because the Sérsic index tends to be greater in objects with more concentrated light, i.e., typically systems dominated by a spheroidal component. The concentration of light is directly reflected on the concentration of stellar mass, and hence, on its density, that steepens the total density slope.

\subsection{Projected versus three-dimensional quantities}
In this section, we focus on the study of the correlations between the three-dimensional quantities and their projected counterparts or equivalent. In particular, we analyse the relation between the effective radius and the stellar half-light radius $r_{1/2}$, the relation between central Sérsic luminosity and \texttt{FSPS} luminosity and also between central projected and 3D stellar mass. For the luminosity and stellar mass, the central values are defined within apertures of $R_{\rm e}$ and $2R_{\rm e}$.

\subsubsection{Effective radius versus stellar half-mass radius}\label{re_versus_hmr_sec}
We start by studying the relation between $r_{1/2}$ and $R_{\rm e}$, summarized in Fig. \ref{Re_versus_hmr}. While $R_{\rm e}$ is the (projected) radius enclosing half of the total luminosity of a galaxy,  $r_{1/2}$ is the radius of the sphere enclosing half of its total stellar mass, meaning that is not the straight 3D definition of $R_{\rm e}$. To be fully consistent, one should compare $R_{\rm e}$ to the 3D half-light radius. However, following the arguments in \citet{wu2023total}, as the stellar mass-to-light ratio gradients in 3D are closer to unit than their 2D counterpart, we take the 3D stellar half-mass radius as being a reasonable approximation of the 3D half-light radius. We observe that galaxies associated with higher Sérsic index are typically the ones where $R_{\rm e} > r_{1/2}$. Besides these particular objects, the slope $\partial \log R_{\rm e}/\partial \log r_{1/2} = 0.87$ and offset value of $-0.049$ shows that $R_{\rm e}$ is systematically smaller than $r_{1/2}$ for the virtual-ETG sample, as one should expect. The linear relationship between these two quantities is, therefore, expressed as:
\begin{equation}
\log R_{\rm e}= 0.901\log r_{1/2} -0.049, \label{Re_HMR_relation}
\end{equation}
with a Pearson correlation coefficient of $0.939$.

{Using the fit provided in  \citet{LimaNeto} for the relation between $r_{1/2}/R_{\rm e}$ and the reciprocal of the Sérsic index $n$, \cite{Wolf}  verified that for most surface brightness profiles used to model galaxy light distributions, the ratio $r_{1/2}/R_{\rm e}\approx 4/3$ is an accurate approximation regardless the profiles' shapes}. By directly performing the ratio between $r_{1/2}$ and $R_{\rm e}$ for the galaxies in our catalogue, we have verified that the mean ratio is $\langle r_{1/2}/R_{\rm e}\rangle = 1.39$ with a scatter $\sigma = 0.23$, consistent with \cite{Wolf} statement. Typical values of $r_{1/2}/R_{\rm e}$ are also computed in \citet{Ciotti} (see their table 2) and they vary from $1.34$ to $1.35$. For ATLAS$^{\rm 3D}$ ETGs, \citet{Cappellari+2013} found a mean ratio $\langle r_{1/2}/R_{\rm e}\rangle \approx 1.42$ which is $1\sigma$ consistent with the ratio we found. Overall, these results show agreement between real world samples of galaxies and our virtual-ETG sample for the $r_{1/2} - R_{\rm e}$ relation. 

%  { We associate the systematically larger mean $r_{1/2}/R_{\rm e}$ ratio we observed to the definition of $r_{1/2}$, which, as previously mentioned, serves only as a proxy for the 3D half-light radius. Assuming a constant stellar mass-to-light ratio $\Upsilon_{\star}$, we find that, for the galaxies in our sample, $\Upsilon_{\star} > 1$ (see Table~\ref{catalogue_stat_properties}), indicating that 3D stellar mass-weighted sizes are generally larger than 3D luminosity-weighted sizes. Consequently, this results in an overestimation of the $r_{1/2}/R_{\rm e}$ ratio. Regardless, we believe this test sufficiently demonstrates the observational reliability of the effective radii derived in this work}

\subsubsection{Projected versus three-dimensional luminosity}\label{projected_vs_3d_light}

In \S\ref{sbp} we have introduced two luminosity definitions, one within $R_{\rm e}$ and the other within $2R_{\rm e}$. Here, we compare the two-dimensional definitions with the three-dimensional ones, the former being derived from an analytical approach and the latter defined as being the contribution of light from the stellar particles inside the corresponding apertures. In Fig. \ref{2D_versus_3D} we show the overall relationship between the projected and the three-dimensional counterpart. 

Inside $1R_{\rm e}$, The 2D luminosity of galaxies derived with Sérsic model is related  to the 3D luminosity by:
\begin{equation}
    \log L_{2D}(1R_{\rm e}) = 0.853\log L_{3D}(1R_{\rm e}) + 1.566.
\end{equation}
For the definition inside $2R_{\rm e}$, the relation is steeper:
\begin{equation}
    \log L_{2D}(2R_{\rm e}) = 0.898 \log L_{3D}(2R_{\rm e}) + 1.106.
\end{equation}
For the luminosity defined within $2R_{\rm e}$, both the scatter and tilt are apparently smaller than when considering an aperture of $1R_{\rm e}$ (see Fig. \ref{2D_versus_3D}).

Under the assumption of spherical symmetry and that the Sérsic profile precisely describes the projected light distribution of galaxies, we expect the three-dimensional luminosity defined inside a sphere of radius $R$ to be smaller than the corresponding one projected onto the plane of the sky (i.e., onto a circular cylinder with same radius $R$). This is because each luminosity point in the projected distribution results from the contribution of all the corresponding points in the 3D distribution stacked along the line of sight. To demonstrate this, we present in both figures the median trend of both the 2D light computed from Sérsic model and the one computed directly from the projected distribution of particles. While the 2D luminosity computed from particles is always greater than its 3D counterpart with almost no tilt, it is observed a tilt on the analytical 2D luminosity at the low-luminosity (smaller galaxies) end.

\begin{figure}
\centering
 \includegraphics[width=0.8\columnwidth]{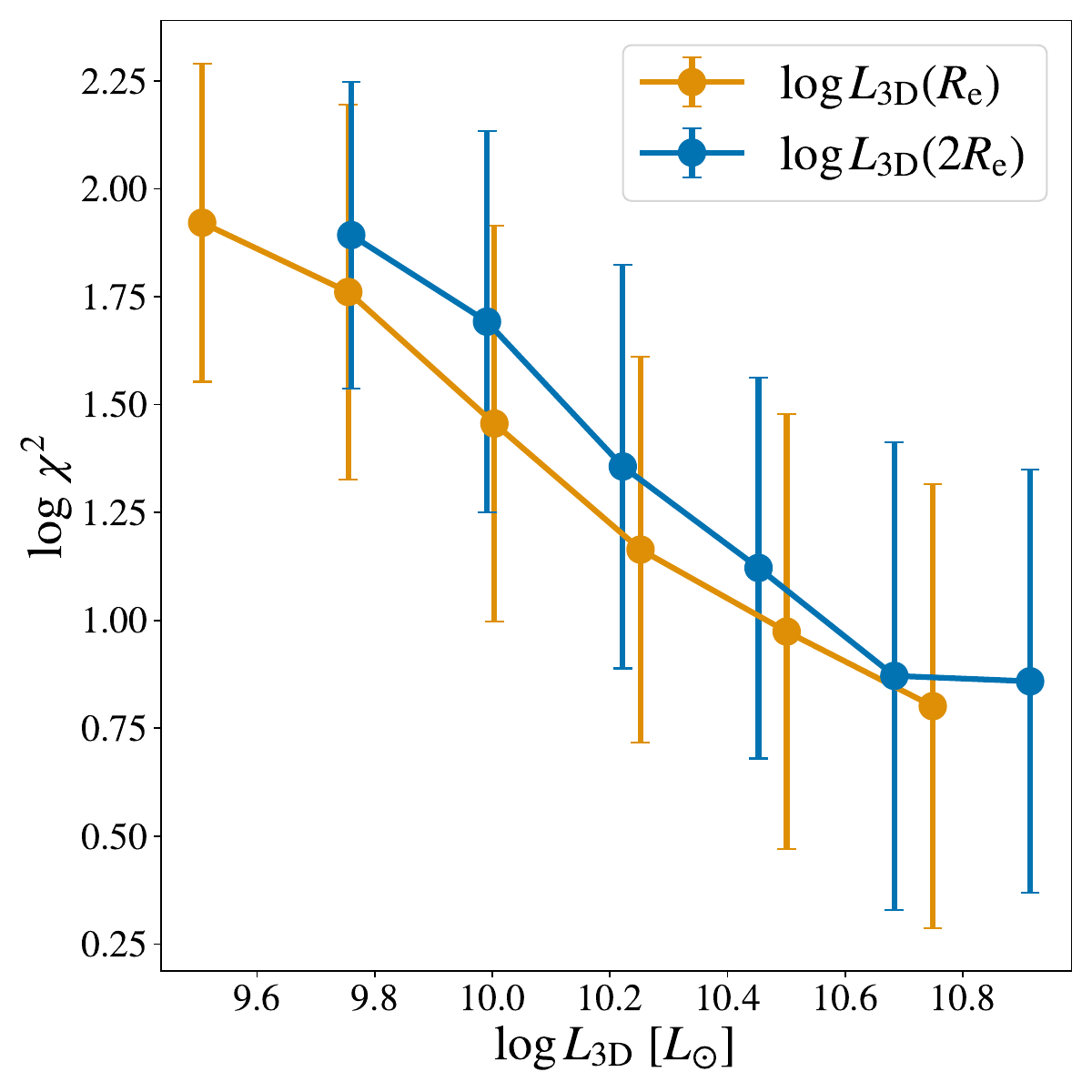}
    \caption{Median behaviour of the (log-scaled) $\chi^2$ for the Sérsic fits versus the 3D luminosity defined at apertures of $1R_{\rm e}$ and $2R_{\rm e}$.}
    \label{chis_versus_lum}
\end{figure}

\begin{figure*}
    \centering
    \includegraphics[width=0.95\columnwidth]{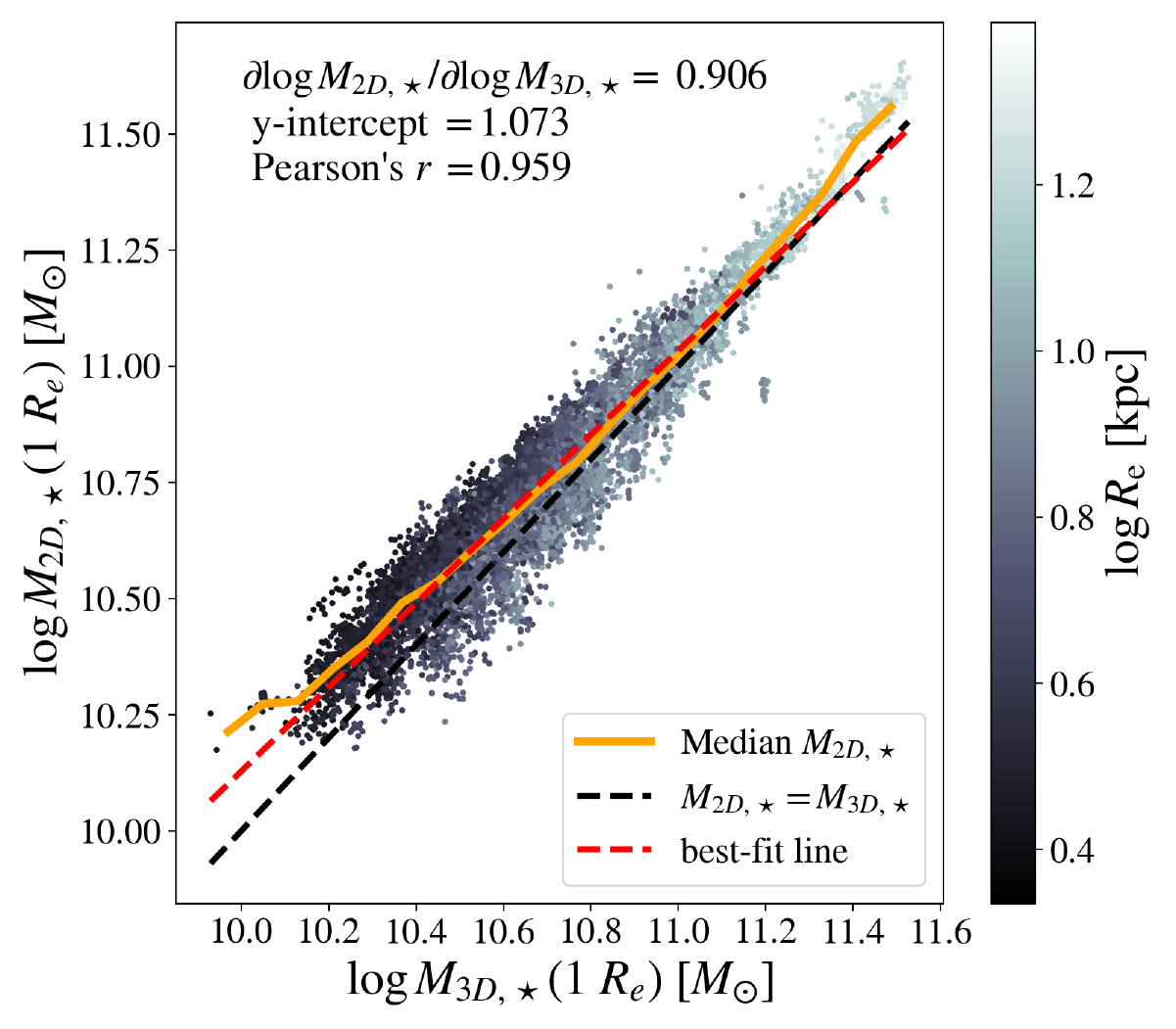}
    \includegraphics[width=0.95\columnwidth]{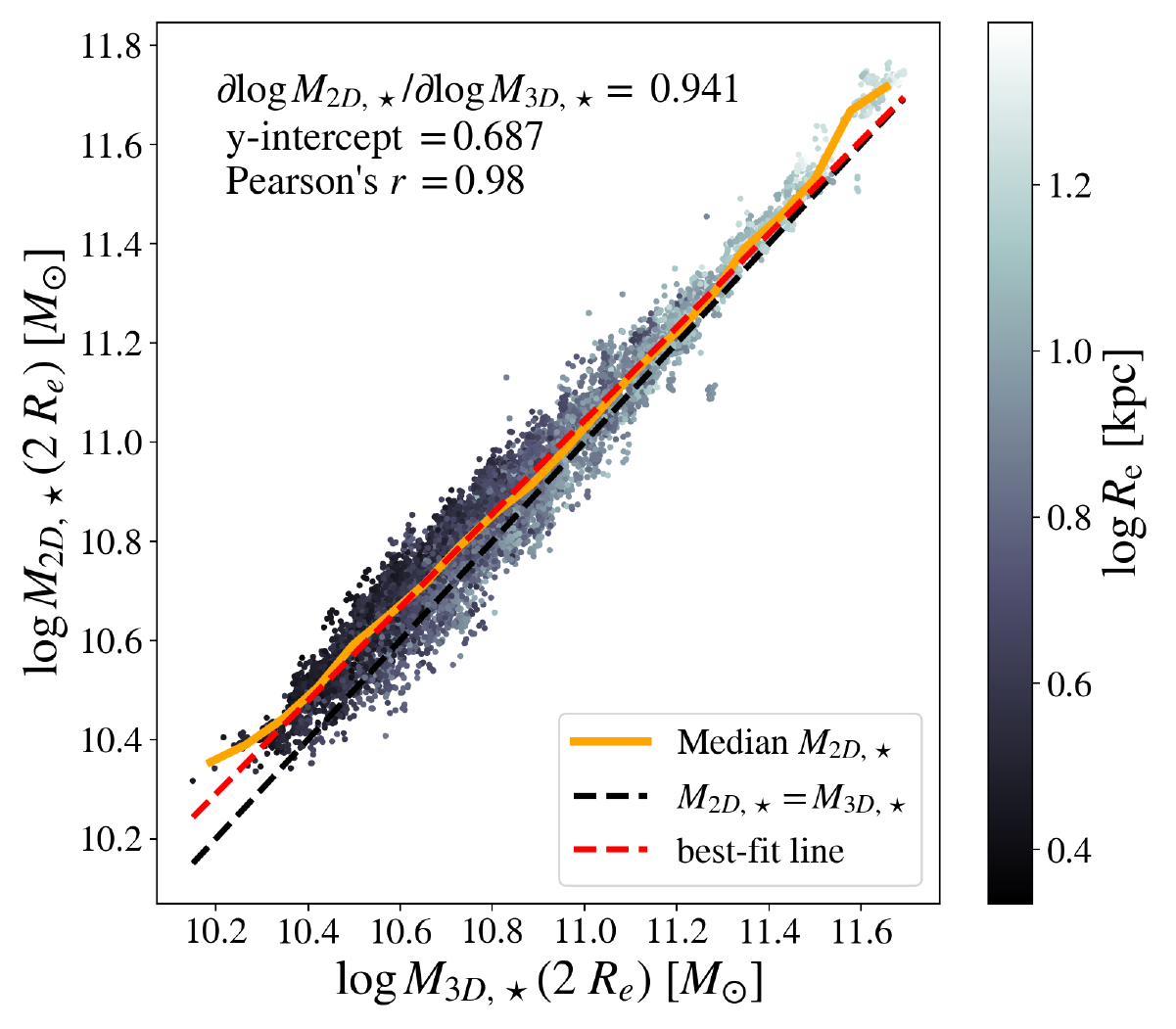}
\caption{Projected stellar inside one effective radii (left) and inside two effective radii (right) computed assuming a Sérsic model for luminosity and a constant stellar mass-to-light ratio versus 3D luminosity inside the same aperture computed from the simulation's particle data. In red is the best-fit line to the relation, in orange is the median trend of the modelled $M_{2D,\star}^{\rm tot}$ (in 20 bins of $M_{3D,\star}$).}
    \label{2D_versus_3D_mass}
\end{figure*}

%In practice, divergences can occur because theoretical predictions are only approximations to the fiducial results. 
Regarding the Fig. \ref{2D_versus_3D}, we believe that differences between the expected behaviour and the theoretical expectations occur because of the properties of objects at the low- and the high-luminosity ends. At the high luminosity end, there is a greater chance that large galaxies  have extra luminous features on their projected light distributions (e.g., diffuse light that were not completely ruled out by the $30$ kpc cut) and hence, are not generally described by the one-component Sérsic model.   {Despite that, as can be observed in Fig. \ref{2D_versus_3D}, these objects represent only a tiny fraction and almost all galaxies have $L_{2D} > L_{3D}$ at this bright end}. At the low luminosity end, there is a greater chance that smaller galaxies have more irregular (noisy) 2D profiles, leading to possible mismatches between the real 2D light and the Sérsic light. This is also the source of the larger scatter observed at the low luminosity end.   {To verify this assertion, we have plotted the median behaviour of  the $\chi^2$ values for the Sérsic fits versus the 3D luminosity at both $1R_{\rm e}$ and $2R_{\rm e}$ apertures, presented in Fig. \ref{chis_versus_lum}. It is clear from this Figure that $\chi^2$ drops as luminosity increases, indicating that fainter (smaller) galaxies have profiles that are not well described by the one-component Sérsic model used here.}

In summary, at the bright end, the result is that the 2D modelled luminosity is   {mildly} underestimated with respect to its fiducial value (the orange solid line is slightly below the blue one for larger galaxies). At the faint end, the Sérsic profiles  are fitted to noisy light distributions, leading to inaccuracies, and  that is the reason we observe in the figures a scatter that decreases with the objects' sizes.   {Because of this larger scatter,} it is also observed that the Sérsic light is overestimated for smaller galaxies. Based on the discussion above, we conclude that the main source of the tilt in the $\log L_{2D} - \log L_{3D}$ relation can be traced to irregularities   {and/or} extra features on the true (projected) light profiles, which are not well described by the assumption of a one-component Sérsic model. Nevertheless,  {it is observed a convergence between} the median trend of the theoretical $L_{2D}$ and the $L_{2D}$ extracted from particle data 
%is approximately the same 
%  {have an approximate convergence}
for objects with $\log L_{3D}(1R_{\rm e})\gtrsim 10.0$ and for $\log L_{3D}(2R_{\rm e}) \gtrsim 10.1$.

\subsubsection{Projected versus three-dimensional stellar 
mass}\label{projected_versus_3d_stellar_mass}
In this section, we discuss how the central stellar mass is affected by its projection. Once it is directly related to the luminosity by a (assumed) constant stellar mass-to-light ratio, the stellar mass inherits the behaviour of the luminosity, either in 2D or 3D, once the only difference between them is a multiplicative constant. Thus, the functional relation between $\log M_{2D,\star}^{\rm tot}$ and $\log M_{3D,\star}$ should be similar to that of the luminosity. To verify this, we compute $M_{2D,\star}^{\rm tot}(xR_{\rm e}) = \Upsilon_{\star}(<30~\mathrm{kpc}) \times L_{2D}(xR_{\rm e})$  and compare this to the 3D stellar mass computed from the particle data, namely $M_{3D,\star}(xR_{\rm e})$ for $x=1,2$. This behaviour is shown in Fig. \ref{2D_versus_3D_mass}, and it indeed has a functional dependence similar to that of the central luminosity.  The best fit lines give the relations:
\begin{equation}
    \log M_{2D,\star}^{\rm tot}(1R_{\rm e}) = 0.906\log M_{3D,\star}(1R_{\rm e}) + 1.073,
\end{equation}
and
\begin{equation}
    \log M_{2D,\star}^{\rm tot}(2R_{\rm e}) = 0.941 \log M_{3D,\star}(2R_{\rm e}) + 0.687.
\end{equation}
Similar to the luminosity, as the aperture radius increases, the slope gets closer to one while the offset decreases. This behaviour is expected, since both values should converge for sufficiently large apertures. However, in our formulation there is an upper limit to the three-dimensional light component, defined as $30$~kpc. Thus, if one wants to compare the light or stellar mass inside a certain aperture proportional to the effective radius $xR_{\rm e}$ and this aperture exceeds $30$~kpc, the Sérsic light (stellar mass) will likely be overestimated with respect to the true projected light (stellar mass).

\section{Conclusions}\label{conclusions}

In this paper, we have derived a full catalogue of ``observational-like'' structural parameters, including total luminosity, effective radius, central surface brightness and S\'ersic index derived by the direct S\'ersic fitting of the Illustris TNG100-1 stellar particles projected along the main axes. We have also computed the total mass density profile slope as well as the slopes of the stellar, dark matter and gas components separately. In a way different from previous literature, we have concentrated on minimizing the impact of the central softening length in 
%the slope of the central luminosity density profiles, the mass density profiles and derived parameters (e.g.,S\'ersic index and effectine radius, and mass density slope) 
the fitting procedures, hence providing the closest quantities to the ones derived from real data. These include also the velocity dispersion from the stellar particles, for which we have provided two ``corrections'': 1) a weighted aperture correction, $\sigma_{\rm e, corr}$, to compensate the central smoothed velocity dispersion profile due to the effect of the softening length; 2) a dynamically motivated velocity dispersion, $\sigma_{\rm e, dyn}$, predicted by the Jeans equation applied to the other structural and mass quantities derived in our catalogue. We argue that this latter quantity 
%the $\sigma_{\rm e, dyn}$ 
is the closest velocity dispersion one would measure in simulations without the effect of the softening dynamics in the very halo centres. 
%the missing velocity dispersion  
%of Using the data product from  the state-of-the-art IllustrisTNG simulations, 

The catalogue includes $10121$ galaxies in the redshift range $0\leq z\leq0.1$ (corresponding to the TNG snapshots $91$ to $99$) and stellar masses  $10^{10.3} \leq M_{\star}/M_{\odot} \leq 10^{11.9}$, compatible with being early-type systems. Potential outliers showing anomalous combinations of structural parameters clearly deviating from the regular scaling relations shown from the bulk of the sample, have been carefully excluded from the sample.
%hence building  
%  {their morphology} and derive   {photometric and spectroscopic} parameters by applying common methods motivated by typical observational procedures, aiming to build a catalogue of ETGs with properties that approximate those of real galaxies. We constructed 
%a clean catalogue  (i.e., no outliers) with $10121$ early-type galaxies  {taht we have dubbed, virtual-ETG}. 
We have dubbed this controlled collection of early-type systems, the {\it virtual-ETG} sample. 

%We examined the fundamental plane ofthe mock galaxies and how well it compares to observations, the mass density slopes, correlations between the ETGs' structural parameters and their central total density slope, and 
%we also investigated how the projection of a galaxy's light and size affects these parameters. 
%  {we also investigated the projection effects on galaxies' light and sizes.}
Apart from the mass density slopes, which have no direct measurement in the real galaxies and has been compared with real galaxy estimates from complex dynamical or lensing modeling, we have conducted a series of tests using the ``luminous'' projected quantities to check the consistency with observational results. Below, we provide a detailed description of the main results presented in this paper:
%and relevant observations regarding the applied methodology and sample. In particular, we have found that:

\begin{enumerate}
  %  \item The fundamental plane of the ETGs is steeper in both $\log\sigma_{\rm e}$ and $\log I_e$ directions, and using a “discriminant” factor $\delta = 4b + a + c$, it is verified that the parameters of the fundamental plane in our sample TNG ETGs is not consistent with $\delta = 0$, in contrast to most observational results \citep{Lu+2020}. %Thus, taking into account the applied methodology, we restate that the TNG FP is not observationally plausible.   {Even with the introduction of galaxy parameters whose derivations account for the effects of the simulation's softening length, we could not reconcile the TNG FP with the observational expectations. }Thus, the tension between the TNG fundamental plane and the observational one is traced to   {a lack of proper mixing between dark matter and baryons, and the absence of correct star formation efficiencies at the right mass scales}, as concluded by \citet{Lu+2020}.
    \item To science validate the virtual-ETG catalogue, we have derived a series of standard scaling relations involving the luminous quantities (namely the $L_{\rm tot,r}-R_{\rm e}$, the $M_{2D,\star}^{\rm tot}-R_{\rm e}$, the $I_{\rm e}-R_{\rm e}$, and the $M_{2D,\star}^{\rm tot}-\sigma$ relations) and found a good agreement with the same scaling relations of a variety of observational sample. This is among the most important results of our paper, especially considering the excellent match of the $\sigma_{\rm e}-M_\star$ relation obtained with the $\sigma_{\rm e,dyn}$, showing, for the fist time, that the overall baryonic physics behind the Illustris-TNG does not miss much of the observations, at least in the massive end of the ETG population, covered by our virtual-ETG sample. {\it This is in contrast with previous claims about the limitations of the baryonic model from the TNG simulations as an explanation of the failure of the scaling relation from TNG galaxies (see, e.g., \citealt{Y.Wang}) but rather tracks the discrepancies to 1) pure numerical effects due to the softening and 2) the lack of full ``observational realism'' in simulations}. This does not imply that the physics in TNG100 is complete. Instead, it suggests that the missing physics may be identified in the residual discrepancies between simulated and observed galaxy catalogues, after full realism is applied. Indeed, most of the discrepancies are linked to differences in the measurements of key physical parameters that “encode” the missing physics. By constructing a sample like our virtual-ETG, where simulated galaxies are processed to mimic observational measurements, we aim to highlight which aspects of the physical model require refinement. 
    A demonstration that we are on the right track  is provided by the fundamental plane results that also nicely match the literature (see de Araujo Ferreira et al., in preparation).
    
    \item The central density slopes values of the early-type galaxies in our catalogue are consistent with those found in the literature for observational samples using galaxy dynamics and strong gravitational lensing modeling. We trace the main source of divergence to the radial interval where the slopes are defined. For instance, using the range $r \in [0.4r_{1/2,} 2r_{1/2}]$, instead of the standard $r \in [0.4r_{1/2,} 4r_{1/2}]$ (see \citealt{C.Wang}) we have found a better agreement of the total mass density slope with observations. In particular, we found (mean $\pm$ scatter) $\langle \gamma_{\rm tot,2}^{\rm PL} \rangle = 2.055 \pm 0.203$ for the total mass, $\langle \gamma_{\star,2}^{\rm PL}\rangle = 2.960 \pm 0.1863$ for the stellar mass, and $\langle \gamma_{\rm DM,2}^{\rm PL} \rangle = 1.717 \pm 0.135$ for the dark matter. 
    %for the stellar component, $\langle \gamma_{\star}\rangle = 3.038 \pm 0.113$, for the dark matter $\langle \gamma_{\rm DM} \rangle = 1.742 \pm 0.103$ and, for the total mass, $\langle \gamma_{\rm tot} \rangle = 1.998 \pm 0.148$, all defined within the radial range $r \in [0.4r_{1/2,} 4r_{1/2}]$. 
    The gas component, that is marginally significant for ETGs, has a mean value $\langle \gamma_{\rm gas,2}^{\rm PL} \rangle =1.113$ with a scatter $\sigma_{\rm gas} = 1.720$, that is $\sim 10$ times greater than the scatter of the other slopes. 
    %As commented, this result for the gas component is a consequence of its low fraction in ETGs compared to the other two components. Hence, in the $\log\rho_{\rm gas}-\log r$ plane, the gas is not as smoothly distributed as the stellar or dark components, leading to the greater scatter.

    \item We have studied scaling relations between the total density slope $\gamma_{\rm tot}$ and the galaxy parameters $n$, $R_{\rm e}$, $\sigma_{\rm e}$ and $M_{\star}$ (see \S\ref{corr_stellar_mass}). %The results are according to the expected behaviour and, 
    Also, for these correlations, we have found a good agreement with observational samples. In particular, the $\gamma_{\rm tot}-\sigma_{\rm e}$ has been found, for the first time, fully consistently recovering observations, thanks to the newly introduced corrected $\sigma_{\rm e}$ discussed above, solving the systematics introduced by the softening dynamics of the centres. In particular, the $\gamma_{\rm tot}-\sigma_{\rm e, dyn}$, turns out to be fully consistent with DynPop \citep[][free and generalized NFW models]{Zhu} and SPIDER ETGs \citep{Tortora+2014}, and more marginally with the steeper trends from the strong lensing samples from SL2S \citep{Sonnenfeld_2013} and SLACS \citep{Auger+2010}. 
    %suggest that the total density slope steepens with increasing $\sigma_{\rm e}$. 
    %This divergence, rather than methodological, can be traced to limitations of the baryonic model from the TNG simulations, as pointed by \citet{Y.Wang}. 
    
    \item The expected behaviour of the projected quantities with respect to their three-dimensional counterparts  is satisfied. In particular, we obtained $\langle r_{1/2}/R_{\rm e}\rangle = 1.39 \pm 0.23$, demonstrating that the relation between the 2D effective radius $R_{\rm e}$ and its 3D definition (where we used the stellar half-mass radius $r_{1/2}$ as an approximation to the latter) are in agreement with dynamically motivated predictions (e.g., \citealt{Ciotti,Wolf}), as anticipated by \citet{wu2023total}. The behaviour of the central projected  luminosity and 3D luminosity is validated from the simulated data, and the divergences are likely due to one-component Sérsic model limitations. Assuming a constant stellar mass-to-light ratio for each galaxy, the comparison between the projected and 3D stellar mass (defined within the same apertures as the luminosity) is also consistent with the expected trend, i.e., because of their strong correlation, the behaviour of the stellar mass closely follows the behaviour of its light. As a possible future upgrade to our catalogue, investigating the impact of colour and stellar mass-to-light ratio (M/L) gradients on the results could provide additional insights.
\end{enumerate}

This work has been thought with the main aim of  developing more observationally motivated strategies for analysing simulated galaxies in a way to produce results that are unbiasedly comparable to real samples. These methods need to minimize the methodological differences in the measurement of structural quantities of simulated and real galaxies. As such, the {\it virtual-ETG} catalogue is the most realistic dataset to use for standard comparisons with similar observed samples, and, most importantly, to the application of Machine Learning (ML) techniques to retrieve more detailed predictions of the mass densities of ETGs based on their photometric and kinematic features (de Araujo Ferreira et al. in preparation). We have already shown that ML can be very effective to predict the total mass inside the effective radius for real galaxies with only a few input features (see \citealt{wu2023total}, but see also \citealt{vonMarttens+2022} for previous tests on mock samples). 
%In subsequent works utilizing ML in the IllustrisTNG simulations, it was demonstrated that these approaches can reliably retrieve the dark and total mass of galaxies, regardless of their morphological type \citep{vonMarttens+2022} and that these techniques can also be applicable to observational samples under the assumption that the distribution of the parameters overlap sufficiently and the overall correlation between them is conserved \citep{wu2023total}. 
%In the next steps of our project,  we plan to investigate the impact of projected quantities on such predictions and how reliable they are to infer the central density trend of ETGs.   {Also comment on the fundamental plane paper}

Overall, the catalogue constructed here is a publicly available resource of local (low redshift) early-type galaxies extracted from the IllustrisTNG100-1 simulation. In a near future, we plan to expand this to all redshifts available from the IllustrisTNG100 database and all galaxy types. This will be an important piece of data that can be used in future researches concerning galaxy formation and evolution to develop and test methodologies aimed at probing structural properties and dynamics of ETGs in both simulated and observed universes.

\section*{Acknowledgements}

Pedro de A. Ferreira gratefully acknowledges the financial support provided by the Coordination of Superior Level Staff Improvement — CAPES. %grant No. 88882.461650/2019-01.

%%%%%%%%%%%%%%%%%%%%%%%%%%%%%%%%%%%%%%%%%%%%%%%%%%
\section*{Data Availability}

The virtual-ETG catalogue is publicly available as a .csv file on \url{https://github.com/PedroFerreirAstro/TNGLDP_catalogue_ETGs}.



\bibliographystyle{mnras}
\bibliography{example} % if your bibtex file is called example.bib

%\begin{thebibliography}{99}
%\end{thebibliography}

%%%%%%%%%%%%%%%%%%%%%%%%%%%%%%%%%%%%%%%%%%%%%%%%%%

%%%%%%%%%%%%%%%%% APPENDICES %%%%%%%%%%%%%%%%%%%%%

\appendix

\section{Scatter matrix before and after outlier removal}\label{Appendix_scatter_matrix}

In Fig. \ref{scatter_matrix}, we provide the scatter matrix for the set of parameters $\{n, \log R_{\rm e}, \log I_e, K,  \gamma_{\rm star}, \gamma_{\rm DM}, \gamma_{\rm tot}\}$ before and after removing the outliers, respectively represented by the light blue and the dark blue colours. The outlier removal followed the criterion of the robust Mahalanobis distance, described in \S\ref{data_cleaning}. Although removing points causes the parameter's range to shrink and, of course, the dataset size decreases, the removal also helps to unveil some correlation that were previously hidden by the outliers, for example the $\gamma_{\rm tot} - \log R_{\rm e}$ relation, which initially was quite weak, but after removing outliers, the positive linear correlation is clearly shown. 

\begin{figure*}
    \centering
    \includegraphics[width=\textwidth]{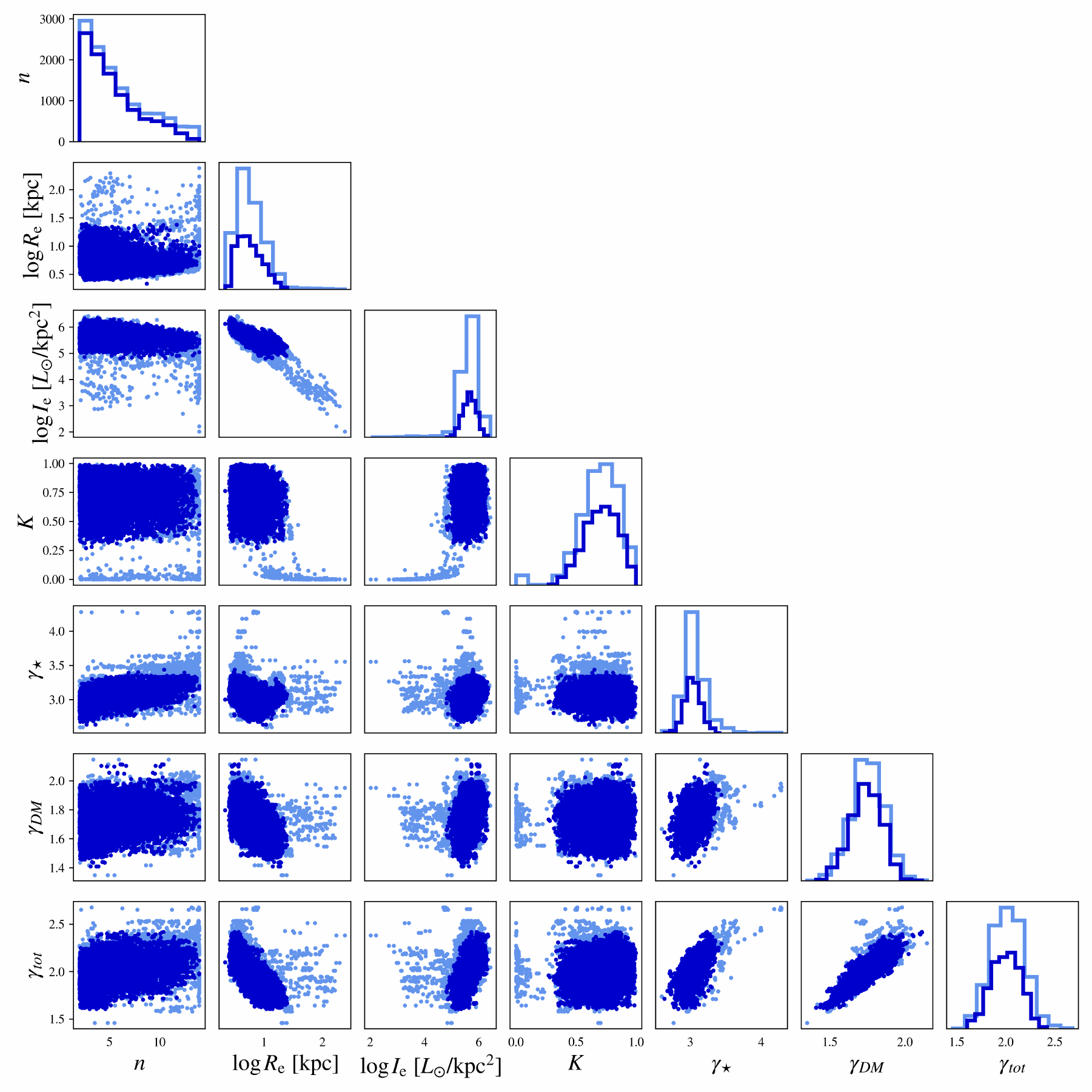}
    \caption{Scatter matrix for the 7-dimensional space of parameters $\{n, \log R_{\rm e}, \log I_e, K, \gamma_{\rm star}, \gamma_{\rm DM}, \gamma_{\rm tot}\}$. In light blue are the points before the outlier removal, and in dark blue the points after removing the outliers.}
    \label{scatter_matrix}
\end{figure*}

\section{Parameters and statistical properties of the catalogue}
\label{appendix_catalogue_stat_properties}
In this appendix, we present a table containing all the parameters in the catalogue (see Table~\ref{catalogue_stat_properties}). Some parameters as the specific star formation rate, are not included because they can be easily derived from the quantities already included. 

\begin{table*}
    \centering
    \caption{Summary of statistical properties and general description of the main parameters in the virtual-ETG catalogue. All the values are computed after the outlier removal task, as described in \S\ref{data_cleaning}. Fields with “$-$” are those where the numerical value is not applicable or is irrelevant.   {Column (1) is the name of these parameters as they are stored in the catalogue, columns (2), (3), (4), (5), and (6) are, respectively, the mean, median, standard deviation, minimum and maximum value of each parameter distribution, and lastly, column (7) gives a brief description of each parameter.}}
    \begin{tabular}[hbt!]{lcccccl} % six columns, alignment for each
        \hline
       \makecell{Stored as \\ (1)} & \makecell{Mean \\ (2)} & \makecell{Median \\ (3)} & \makecell{Std \\ (4)} & \makecell{Min \\ (5)} & \makecell{Max \\ (6)} & \makecell{Short description \\ (7)} \\
        \hline
        \texttt{REDSHIFT} & 0.05 & 0.05 & 0.03 & 0.00 & 0.10 & Redshift of this object \\
        \texttt{TNG\_ID} & - & - & - & - & - & Subfind ID of this object \\
        \texttt{PROJECTION} & - & - & - & - & - & Projection axis ($X = 0$, $Y = 1$, $Z = 2$) \\
        \texttt{SERSIC\_IDX} & 5.31 & 4.58 & 2.68 & 2.00 & 14.00 & Sérsic index \\
        \texttt{HALFLIGHT\_RAD} & 0.76 & 0.73 & 0.21 & 0.34 & 1.40 & 2D effective (halflight) radius \\
        \texttt{HALFLIGHT\_BRIGHTNESS} & 5.65 & 5.66 & 0.25 & 4.80 & 6.34 & Surface brightness at $R_{\rm e}$ \\
        \texttt{AXIS\_RATIO} & 0.71 & 0.71 & 0.14 & 0.27 & 1.00 & Axis ratio $b/a$ of this object \\
        \texttt{SERSIC\_LUM} & 10.42 & 10.38 & 0.24 & 9.85 & 11.32 & Total luminosity from Sérsic model \\
        \texttt{LUM\_HALFLIGHT\_RAD} & 10.12 & 10.08 & 0.24 & 9.55 & 11.02 & 2D luminosity within $1R_{\rm e}$ \\
        \texttt{LUM\_TWO\_HALFLIGHT\_RAD} & 10.25 & 10.21 & 0.25 & 9.72 & 11.13 & 2D luminosity within $2R_{\rm e}$ \\
        \texttt{SERSIC\_MASS\_STAR} & 11.32 & 11.00 & 0.26 & 10.47 & 11.95 & Total stellar mass estimated as $\Upsilon_{\star} \times L_{2D}^{\rm tot}$ \\
        \texttt{MASS\_STAR\_HALFLIGHT\_RAD} & 10.74 & 10.70 & 0.26 & 10.17 & 11.53 & 2D stellar mass within $1R_{\rm e}$ \\
        \texttt{MASS\_STAR\_TWO\_HALFLIGHT\_RAD} & 10.87 & 10.83 & 0.27 & 10.32 & 11.69 & 2D stellar mass within $2R_{\rm e}$ \\
        
        \texttt{SIGMA\_EFF} & 2.16 & 2.15 & 0.10 & 1.88 & 2.55 & Aperture velocity dispersion within $1R_{\rm e}$ \\
        \texttt{SIGMA\_CORR} & 2.18 & 2.17 & 0.09 & 1.91 & 2.57 & Corrected aperture velocity dispersion within $1R_{\rm e}$ \\
        \texttt{SIGMA\_DYN} & 2.22 & 2.21 & 0.07 & 1.98 & 2.51 & Dynamical inferred aperture velocity dispersion within $1R_{\rm e}$ \\
        
        \texttt{MLR} & 4.17 & 4.22 & 0.47 & 2.68 & 5.43 & Stellar mass-to-light ratio \\
        \texttt{SFR} & 0.19 & 0.02 & 0.40 & 0.00 & 4.03 & Star formation rate \\
        \texttt{SPS\_LUM} & 10.33 & 10.30 & 0.22 & 9.84 & 11.05 & 3D total luminosity within $30$ kpc \\
        \texttt{SPS\_LUM\_HALFLIGHT\_RAD} & 10.03 & 9.98 & 0.27 & 9.38 & 10.88 & 3D luminosity within $1R_{\rm e}$ \\
        \texttt{SPS\_LUM\_TWO\_HALFLIGHT\_RAD} & 10.18 & 10.14 & 0.27 & 9.64 & 11.04 & 3D luminosity within $2R_{\rm e}$ \\
        \texttt{MASS\_GAS} & 11.26 & 11.22 & 0.61 & 9.21 & 12.95 & Total gas mass \\
        \texttt{MASS\_GAS\_HMR} & 7.36 & 7.60 & 1.91 & 0.0 & 10.38 & Gas mass within a sphere of $1r_{1/2}$ radius \\  
        \texttt{TNG\_MASS\_STAR} & 11.04 & 10.99 & 0.28 & 10.40 & 11.90 & Total 3D stellar mass \\
        \texttt{TNG\_MASS\_STAR\_HMR} & 10.84 & 10.68 & 0.28 & 10.09 & 11.60 & 3D stellar mass inside a sphere of $1r_{1/2}$ radius \\ 
        \texttt{TNG\_MASS\_STAR\_30KPC} & 10.95 & 10.91 & 0.24 & 10.39 & 11.70 & 3D stellar mass inside $30$ kpc \\      
        \texttt{TNG\_MASS\_STAR\_HALFLIGHT\_RAD} & 10.67 & 10.63 & 0.28 & 9.93 & 11.53 & 3D stellar mass inside $1R_{\rm e}$ \\   \texttt{TNG\_MASS\_STAR\_TWO\_HALFLIGHT\_RAD} & 10.81 & 10.76 & 0.28 & 10.15 & 11.69 & 3D stellar mass inside $2R_{\rm e}$ \\
        \texttt{HMRAD\_STAR} & 0.90 & 0.86 & 0.21 & 0.45 & 1.57 & 3D stellar half-mass radius \\
        \texttt{MASS\_DM} & 12.65 & 12.60 & 0.36 & 11.62 & 13.76 & Total dark mass \\
        \texttt{MASS\_DM\_HMR} & 10.80 & 10.73 & 0.38 & 10.06 & 12.08 & Dark matter mass within a sphere of $1r_{1/2}$ radius \\
        \texttt{MASS\_ALL} & 12.68 & 12.63 & 0.37 & 11.65 & 13.83 & Total mass (dm+gas+stars) \\
        \texttt{MASS\_ALL\_HMR} & 11.08 & 11.01 & 0.33 & 10.43 & 12.21 & Total mass within a sphere of $1r_{1/2}$ radius\\
        
        \texttt{SLOPE\_STAR} & 3.04 & 3.04 & 0.11 & 2.64 & 3.44 & Stellar density slope within $[0.4, 4]r_{1/2}$ \\
        \texttt{SLOPE\_GAS} & 0.67 & 0.74 & 0.98 & -12.95 & 3.33 & Gas density slopes within $[0.4, 4]r_{1/2}$ \\
        \texttt{SLOPE\_DM} & 1.74 & 1.74 & 0.10 & 1.41 & 2.11 & Dark matter density slope within $[0.4, 4]r_{1/2}$ \\
        \texttt{SLOPE\_ALL} & 2.00 & 2.00 & 0.15 & 1.61 & 2.42 & Total density slope within $[0.4, 4]r_{1/2}$ \\
        
        \texttt{SLOPE\_STAR\_2} & 2.96 & 2.96 & 0.16 & 2.35 & 3.63 & Stellar density slope within $[0.4,2]r_{1/2}$ \\
        \texttt{SLOPE\_GAS\_2} & 1.11 & 1.30 & 1.72 & -31.80 & 4.27 & Gas density slopes within $[0.4,2]r_{1/2}$ \\
        \texttt{SLOPE\_DM\_2} & 1.76 & 1.76 & 0.12 & 1.37 & 2.23 & Dark matter density slope within $[0.4,2]r_{1/2}$ \\
        \texttt{SLOPE\_ALL\_2} & 2.07 & 2.08 & 0.18 & 1.54 & 2.74 & Total density slope within $[0.4,2]r_{1/2}$ \\

        \texttt{AV\_SLOPE\_STAR} &  2.96 & 2.93 & 0.17 & 2.51 & 3.62 & Average star density slopes within $[0.4,2]r_{1/2}$ \\
        \texttt{AV\_SLOPE\_GAS} &  1.27 & 1.38 & 1.65 & -28.27 & 17.71 & Average gas density slopes within $[0.4,2]r_{1/2}$ \\
        \texttt{AV\_SLOPE\_DM} & 1.71 & 1.72 & 1.72 & 1.72 & 2.32 & Average dark matter density slope within $[0.4,2]r_{1/2}$ \\
        \texttt{AV\_SLOPE\_ALL} & 2.06 & 2.06 & 2.06 & 1.43 & 2.62 & Average total density slope within $[0.4,2]r_{1/2}$ \\

        \texttt{INTERCEPT\_STAR} & 9.33 & 9.29 & 0.28 & 8.68 & 10.46 &  Stellar density intercept for $r$ within $[0.4,4]r_{1/2}$ \\
        \texttt{INTERCEPT\_GAS} & 5.21 & 5.32 & 0.95 & -9.24 &  8.10 &  Gas density intercept for $r$ within $[0.4,4]r_{1/2}$ \\
        \texttt{INTERCEPT\_DM} & 8.71 & 8.70 & 0.13 & 8.35 & 9.20 & Dark matter density intercept for $r$ within $[0.4,4]r_{1/2}$ \\
        \texttt{INTERCEPT\_ALL} & 9.07 & 9.07 & 0.13 & 8.62 & 9.61 & Total density intercept for $r$ within $[0.4,4]r_{1/2}$ \\

        \texttt{INTERCEPT\_STAR\_2} & 9.25 & 9.23 & 0.23 & 8.48 & 10.19  &  Stellar density intercept for $r$ within $[0.4,2]r_{1/2}$ \\
        \texttt{INTERCEPT\_GAS\_2} & 5.57 & 5.86 & 1.61 & -26.87 & 22.06  &  Gas density intercept for $r$ within $[0.4,2]r_{1/2}$ \\
        \texttt{INTERCEPT\_DM\_2} & 8.68 & 8.68 & 0.13 & 8.23 & 9.20 & Dark matter density intercept for $r$ within $[0.4,2]r_{1/2}$ \\
        \texttt{INTERCEPT\_ALL\_2} & 9.10 & 9.09 & 0.14 & 8.51 & 9.74 & Total density intercept for $r$ within $[0.4,2]r_{1/2}$ \\
        
        \hline
    \end{tabular}
    \label{catalogue_stat_properties}
\end{table*}

\section{TNG100-1 velocity dispersion profiles}
\label{app:veldisp_profiles}
In Fig. \ref{fig:veldisp_profiles}, we show $142$ velocity dispersion profiles for the ETGs in our sample. For the innermost regions, i.e., below the softening length ($\epsilon_{\star} = 0.74$ kpc; see the dashed black vertical line in the Figure), a nearly constant behaviour or a drop on the profiles instead of a central peak is observed, and galaxies with lower stellar mass are affected the most. As commented on the main text, this effect, is artificially induced by the smoothing of the gravitational field for scales below the simulation's softening length (``softened velocity dispersion'' for short) and contrasts the usually peaked velocity dieprsion profiles of observed massive ETGs (see, e.g., \citealt{Coccato+2009,Cappellari+2013}).

\begin{figure}
    \centering
    \includegraphics[width=\columnwidth]{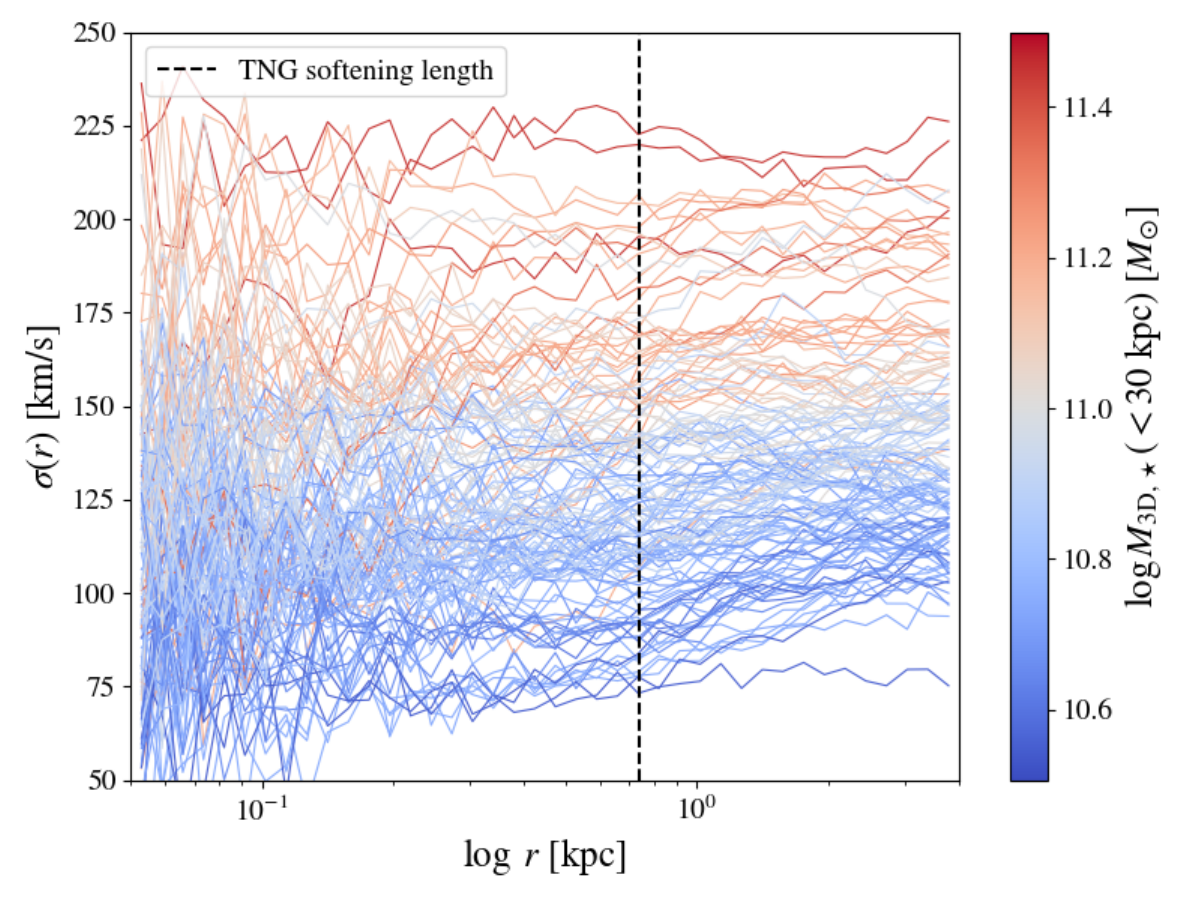}
    \caption{Velocity dispersion profiles for $\sim 140$ ETGs in the stellar mass range $10^{10.5}\leq M_{3D,\star}(< 30~\rm kpc)\leq10^{11.5}$. The profiles were computed for 40 bins over the radial range $ 0.05\leq r/\rm{kpc} \leq 4$.}
    \label{fig:veldisp_profiles}
\end{figure}

\section{Correction to the central velocity dispersion, accounting for the regions below the softening length}
\label{app:sigma_corr}
%However this is, by definition, different from the line-of-sight velocity dispersion of observed galaxies, that instead consider all stars down to the very galaxy centers, where the velocity dispersion profile of ETGs, usually show their maximum value. 
As commented before (see, e.g., Appendix \ref{app:veldisp_profiles}), the line-of-sight velocity dispersion of TNG tend to be underestimated because the simulated galaxies miss the typical central peaks seen in the observed velocity dispersion profiles. To try to account for this missing ``central component'' we introduce here an heuristic correction based on standard aperture corrections applied to observed galaxies (see, e.g., \citealt{Jorgensen+1995, Ziegler-Bender_apcorr, Cappellari+2006}). 

The luminosity-weighted velocity dispersion along a given line of sight defined within an aperture $R_{\rm AP}$ can be obtained in terms of the observed surface brightness and luminosity in the following manner (see, e.g., \citealt{Tortora+2022}):
\begin{equation}
    \sigma_{\rm AP}^2(R_{\rm AP}) = \frac{2\pi}{L(<R_{\rm AP})}\int_{0}^{R_{\rm AP}} sI(s)\sigma_{\rm LOS}^2(s)ds.\label{eq:lw_aperture_veldsip}
\end{equation}
Using this fact, we then define the corrected aperture velocity dispersion inside $R_{\rm e}$ as:
\begin{equation}
\begin{aligned}
    \sigma_{\rm e}^2(R_{e}) =& \frac{1}{L(<R_{\rm e})}\bigg[\int_{0}^{\epsilon_{\star}} 2\pi sI(s)\sigma_{\rm LOS}^2(s)ds\\
    +& \int_{\epsilon_{\star}}^{R_{\rm e}} 2\pi sI(s)\sigma_{\rm LOS}^2(s)ds\bigg].
\end{aligned}
\end{equation}
From Eq.~(\ref{eq:lw_aperture_veldsip}), the first integral in the square parentheses above is equivalent to $L(<\epsilon_{\star}) \sigma^2_{\epsilon_\star}$, and analogously, the second integral is $L(\epsilon_\star < r < R_{\rm e})\sigma_{[\epsilon_{\star}, R_{\rm e}]}^2$. Therefore,
\begin{equation}
\sigma_{\rm e}^2 = \left[\sigma^2_{\epsilon_\star}\frac{L(<\epsilon_\star)}{L(<R_{\rm e})} + \sigma_{[\epsilon_{\star}, R_{\rm e}]}^2\frac{L(\epsilon_\star < r < R_{\rm e})}{L(<R_{\rm e})}\right]. \label{eq:corrected_sigma_0}
\end{equation}
The second term of this equation is obtained from the TNG stellar particle data (see \S\ref{catalogue_section} for more details). About the first term, it cannot be measured from simulations due to the artificial softening of the potential anticipated in \S\ref{catalogue_section} (point vi).
%for the represents an unrealistic value of the velocity dispersion near the very centre of galaxies. 
To overcome this problem, we introduce a correction similar to the one usually adopted in spectroscopy to derive the integrated velocity dispersion for a given aperture starting from the measured dispersion
%in the literature, which is particularly useful when measurements of the velocity dispersion are taken 
through a fiber with a fixed aperture. The adjustment of these measurements to reflect the velocity dispersion within the effective radius is usually obtained through a power-law formula:
\begin{equation}
    \left(\frac{\sigma_{\rm AP}}{\sigma_{\rm e}}\right) = \left( \frac{R_{\rm AP}}{R_{\rm e}}\right)^{\alpha},\label{eq:sigma_ap_corr}
\end{equation}
where $R_{\rm AP}$ is the radius of the aperture and the parameter $\alpha$ can assume several values, e.g., $-0.04$ \citep{Jorgensen+1995}, $-0.066$ \citep{Cappellari+2006}; $-0.055$ \citep{Falcon_barroso+2017}; and, more recently, \citet{Zhu_2023} have suggested a variation of the parameter depending on galaxy properties. In particular, we used the correction $\alpha = -0.066$ from \citet{Cappellari+2006}, as a standard value, but being aware that this might be just a lower limit value, especially for large $\sigma_e$ which might be slightly overestimated. 
%Note that there is not consensus about the value of this $\alpha$ parameter as different dataset/analyses have suggested different values and even a variation depending on galaxy properties (see, e.g.,), with the -0.0 

\begin{figure}
    \centering
    \includegraphics[width=0.9\columnwidth]{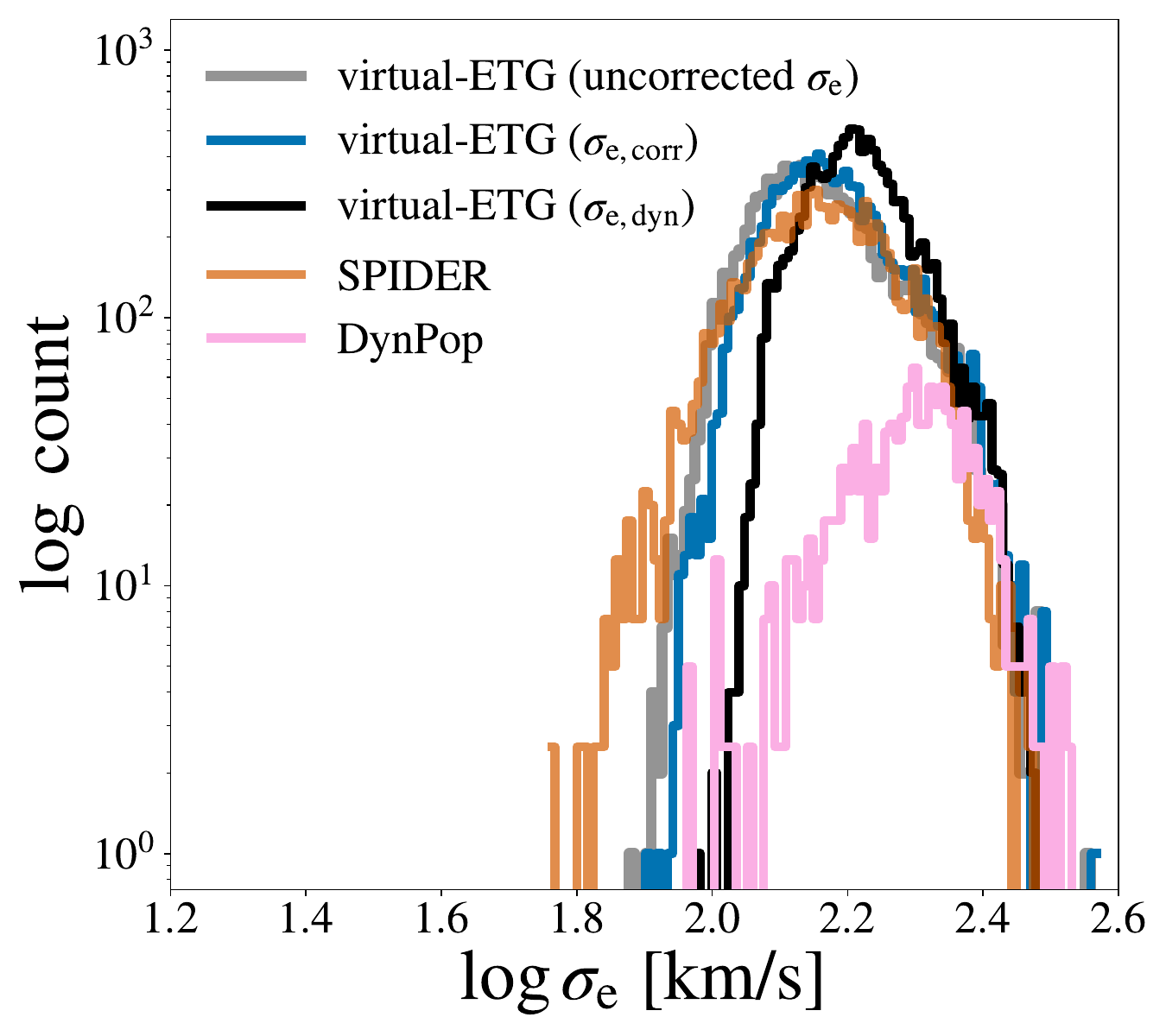}
    \caption{Comparison of the virtual-ETG sample velocity dispersions and observed galaxy sample from SPIDER \citep{Tortora+2014} and Dynpop \citep{Zhu}.   {For better visualization, we normalized the samples by their area on the sky, i.e., we augmented the counts of SPIDER and DynPop by a factor of $2.5$. %(which corresponds to roughly $2$ times the volume of TNG100, i.e., about $3000~\rm{deg}^2$ at $z\lesssim 0.02$) %and, similarly, the DynPop counts are multiplied by $0.86$.}}
    }}
    \label{fig:sig_comp}
\end{figure}

Therefore, from Eq.~(\ref{eq:sigma_ap_corr}), we estimate the velocity dispersion below the softening length $\sigma_{\epsilon_\star}$ in Eq.~(\ref{eq:corrected_sigma}) as:
\begin{equation}
    \sigma_{\epsilon_\star} = \left( \frac{\epsilon_\star}{R_{\rm e}}\right)^{-0.066}\sigma_{\rm e}.
    %\sigma_{[\epsilon_{\star}, R_{\rm e}]}.
    \label{eq:ap_corr}
\end{equation}
This can be substituted in Eq. (\ref{eq:corrected_sigma}) to obtain
\begin{equation}
\sigma_{\rm e}^2 = \left[\left( \frac{\epsilon_\star}{R_{\rm e}}\right)^{-0.132}\sigma_{\rm e}^2~\frac{L(<\epsilon_\star)}{L(<R_{\rm e})} + \sigma_{[\epsilon_{\star}, R_{\rm e}]}^2\frac{L(\epsilon_\star < r < R_{\rm e})}{L(<R_{\rm e})}\right]
\label{eq:corrected_sigma}
\end{equation}
and, by isolating the $\sigma_e$,
\begin{equation}
\sigma_{\rm e,corr}^2 = \sigma_{[\epsilon_{\star}, R_{\rm e}]}^2\frac{L(\epsilon_\star < r < R_{\rm e})}{L(<R_{\rm e})}\left( 1 - \frac{L(<\epsilon_\star)}{L(<R_{\rm e})} \left(\frac{\epsilon_\star}{R_{\rm e}}\right)^{-0.132} \right)^{-1},
\label{eq:sig_e_corr_app}
\end{equation}
%Notice that we are using $\sigma_{[\epsilon_{\star}, R_{\rm e}]}$ to estimate the sigma below $\epsilon_\star$. In an ideal case one should use $\sigma$ inside $R_{\rm e}$ and not in a "mega bin" $r \in [\epsilon_{\star}, R_{\rm e}]$, which, by itself, still leads to an underestimation of the overall velocity dispersion. 
where we have used the suffix ``corr'' to highlight that this includes the correction term as in Eq. (\ref{eq:ap_corr}). 
In Fig. \ref{fig:sig_comp} we compare the distribution of the $\sigma_{\rm e, corr}$, defined in Eq. (\ref{eq:sig_e_corr_app}) and the corresponding $\sigma_e$ from two galaxy samples, introduced in \S\ref{sec:intro}, that we have used in this paper: the SPIDER \citep{Tortora+2014} 
%ATLAS$^{\rm 3D}$ \citep{Cappellari+2013}
and the DynPop \citep{Zhu} samples as they are a better observational match to our mock ETG sample. We also show the distribution of dynamically derived $\sigma_{\rm e, dyn}$ of each virtual-ETGs as introduced in \S\ref{pararaph:projected_quantities}.

As seen in Fig. \ref{fig: sig_corr-dyn} of this latter section, the $\sigma_{\rm e, dyn}$ aligns rather well with the corrected $\sigma_{\rm e, corr}$ for $\log L_{3D}(R_{\rm e})/L_{\odot}>10$, although it starts deviating for small $\sigma_{\rm e}$. This becomes more dramatic if we plot the full virtual-ETG sample shown in Fig. \ref{fig:sigma_corr_sigma_app}. This deviation can have different origins. From the dynamical inference side, one reason can be some unaccounted anisotropy. However, as discussed in \citet{Tortora+2014}, the inclusion of $\beta\sim0.2$ anisotropy parameter (i.e., radial anisotropy) would produce a $\sim0.04$ relative error in the total mass inside $R_{\rm e}$ corresponding to $\sim0.02$ relative errors in $\sigma_{\rm e}$. Considering the lower $\sigma_{\rm e}$ (i.e., lower mass and luminosity systems) mostly affected by radial anisotropy, this eventually would shift the trend by 0.02 dex, which is not sufficient to explain the discrepancies around $\log \sigma_e\sim2-2.1$. From the correction to the particle velocity dispersion side, one might guess some systematics to arise from the aperture correction in Eq. (\ref{eq:sig_e_corr}). However, we have checked that, accounting for a more detailed variation of the $\alpha$ parameter (e.g., from \citealt{Zhu_2023}), it would eventually adjust better to the high $\sigma_{\rm e}$, rather than the lower side. Rather, a final and possible more realistic explanation is that the correction might be just not appropriate for the smaller systems (either in luminosity, i.e., $\log L_{{\rm 3D}}(R_{\rm e})/L_\odot<10$, and size). Here, the motivation very likely resides on the fact that these systems are the ones holding the steepest total density profiles (see the correlations of the $\gamma$ with $M_\star$ and $R_{\rm e}$ in Fig. \ref{fig:gamma_correlations}, upper left and upper right panels, respectively) and also seen by the color-code in Fig. \ref{fig:sigma_corr_sigma_app} where the most deviating systems below $\log \sigma_{\rm e}<2.2$ are the ones with $\gamma>2$. In these cases, the effect of the softened dynamics might be larger than the one with shallower slopes, hence impacting the velocity distribution stronger and causing the $\sigma$ to be diluted more even outside the softening length. On the other hand, the $\sigma_{\rm e, dyn}$ seems to capture the correct physics based on the particle distribution on the position space despite they have been derived starting from the 2D quantities by the S\'ersic fitting and the modeled total mass profile, that can still deviate from the true intrinsic distribution. In other words, the dynamically derived $\sigma_{\rm e, dyn}$ is the closer representation of what the kinematics of the stellar particles would be measured in the TNG100-1 simulation in absence of the softening length. Interestingly, if using these inferred  $\sigma_{\rm e, dyn}$, all the scaling relations involving the $\sigma_e$ for the virtual-TNG nicely re-align with the observations as discussed in \S\ref{results_discussion}, including the fundamental plane (de Araujo Ferreira et al., in preparation). 
%Indeed, this might be the case as one can see by the color code showing that smaller systems (in luminosity and size) are the ones mostly deviating from the one-to-one relation, as already hinted by the sample with $\log L_{{\rm tot}, r}/L_\odot>10$ shown in Fig. \ref{fig: sig_corr-dyn}.  

%with other inferences for the virtual-ETG sample in smaller systems and tends to be lower than the other inferences in larger systems (see Fig. \ref{fig:sigma_corr_sigma_inf}). %The definition of $\sigma_{\rm e, inf}$ using the SB cut is also motivated by the fact the it better reproduces the observed $M_\star-\sigma$ relation, as demonstrated in Fig.~\ref{fig:sigma_Mstar_corr.pdf}, where we compare with  its definition using the 3D stellar mass $M_{\star,3D}(<30 \rm kpc)$ instead of $M_{\star,2D}$ from SB cut.

\begin{figure}
    \centering
    \includegraphics[width=\linewidth]{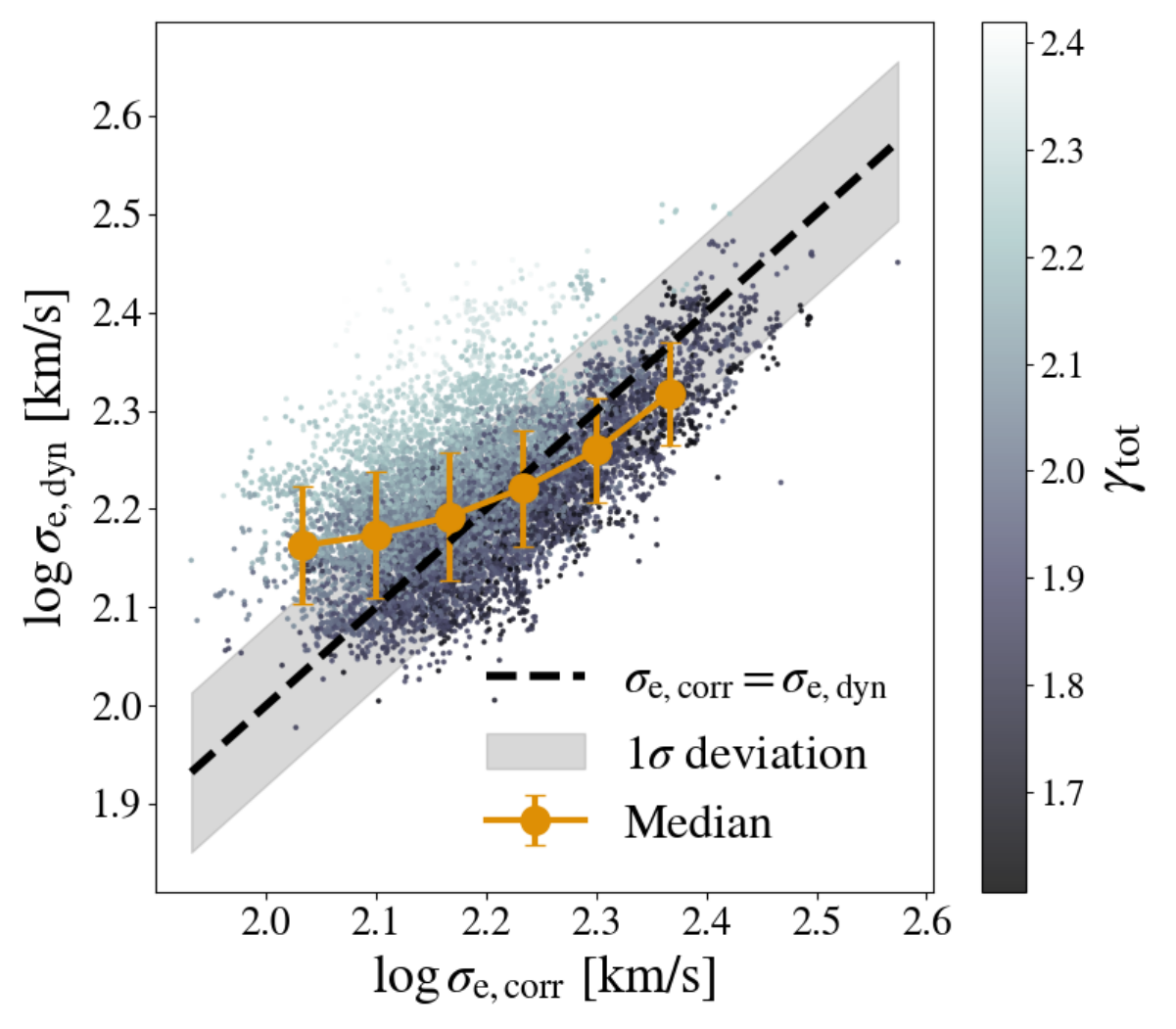}
    \caption{Comparison between the two aperture velocity dispersions introduced in this work: the aperture corrected $\sigma_{\rm e, corr}$ derived according to Eq.~(\ref{eq:sig_e_corr_app}) and dynamically derived $\sigma_{\rm e, dyn}$ of each virtual-ETGs as introduced in  \S\ref{pararaph:projected_quantities}.}
    \label{fig:sigma_corr_sigma_app}
\end{figure}

%\begin{figure}
%    \centering
%    \includegraphics[width=\linewidth]{figures/sigma_Mstar.pdf}
%    \caption{  {there is an equivalent figure for this one in the section about the scaling relations} $M_{\star}-\sigma$ relation for our virtual-ETG sample assuming two }
 %   \label{fig:sigma_Mstar_corr.pdf}
%\end{figure}

%\onecolumn % Switch to single-column layout for the appendix

%%%%%%%%%%%%%%%%%%%%%%%%%%%%%%%%%%%%%%%%%%%%%%%%%%

% Don't change these lines
\bsp	% typesetting comment
\label{lastpage}
\end{document}